\documentclass[twocolumn]{aastex62}
\usepackage{graphicx,ifthen,url,float,color,ulem,amsmath,ragged2e}

\newcommand {\mmpz}{{\sc MMpz}}

\newcommand {\etal}{et.~al.}

\newcommand {\cof}{$^{12}$CO($J\!=$6$\rightarrow$5)}
\newcommand {\htwoo}{p-H$_{2}$O(2$_{1,1}\!\rightarrow$2$_{0,2}$)}
\newcommand {\kms}{km s$^{-1}$}

\def\ltsima{$\; \buildrel < \over \sim \;$}
\def\simlt{\lower.5ex\hbox{\ltsima}}
\def\gtsima{$\; \buildrel > \over \sim \;$}
\def\simgt{\lower.5ex\hbox{\gtsima}}
\newcommand {\uJy}{$\mu$Jy}
\newcommand {\um}{$\mu$m}

\newcommand{\msun}{{\rm\,M$_\odot$}}
\newcommand{\sfr}{{\rm\,M$_\odot$\,yr$^{-1}$}}
\newcommand{\lsun}{{\rm\,L$_\odot$}}

\newcommand{\lpeak}{$\lambda_{\rm peak}$}
\newcommand{\lir}{L$_{\rm IR}$}
\newcommand{\mism}{M$_{\rm ISM}$}
\shorttitle{ALMA 2mm-selected DSFGs}
\shortauthors{Casey et al.}

\begin{document}

\title{\sc Mapping Obscuration to Reionization with ALMA
  (MORA):\\ 2\,mm Efficiently Selects the
  Highest-Redshift Obscured Galaxies}

\correspondingauthor{Caitlin M. Casey}
\email{cmcasey@utexas.edu}

\author[0000-0002-0930-6466]{Caitlin M. Casey}
\affil{Department of Astronomy, The University of Texas at Austin, 2515 Speedway Blvd Stop C1400, Austin, TX 78712, USA}

\author[0000-0002-7051-1100]{Jorge A. Zavala}
\affil{Department of Astronomy, The University of Texas at Austin, 2515 Speedway Blvd Stop C1400, Austin, TX 78712, USA}
\affil{National Astronomical Observatory of Japan, 2-21-1 Osawa, Mitaka, Tokyo 181-8588, Japan}

\author[0000-0003-0415-0121]{Sinclaire M. Manning}
\altaffiliation{NHFP Hubble Fellow}
\affil{Department of Astronomy, The University of Texas at Austin, 2515 Speedway Blvd Stop C1400, Austin, TX 78712, USA}
\affil{Department of Astronomy, University of Massachusetts Amherst, 710 N. Pleasant Street, Amherst, MA 01003, USA}

%
\author{Manuel Aravena}
\affil{N\'{u}cleo de Astronomi\'{i}a, Facultad de Ingenier\'{i}a y Ciencias, Universidad Diego Portales, Av. Ej\'{e}rcito 441, Santiago, Chile}
\author{Matthieu B\'{e}thermin}
\affil{Aix-Marseille Universit\'{e}, CNRS, CNES, LAM, Marseille, France}
\author[0000-0001-8183-1460]{Karina I. Caputi}
\affil{Kapteyn Astronomical Institute, University of Groningen, P.O. Box 800, 9700AV Groningen, The Netherlands}
\affil{Cosmic Dawn Center (DAWN)}
\author{Jaclyn B. Champagne}
\affil{Department of Astronomy, The University of Texas at Austin, 2515 Speedway Blvd Stop C1400, Austin, TX 78712, USA}
\author{David L. Clements}
\affil{Imperial College London, Blackett Laboratory, Prince Consort Road, London, SW7 2AZ, UK}
%
%
\author[0000-0003-3627-7485]{Patrick Drew}
\affil{Department of Astronomy, The University of Texas at Austin, 2515 Speedway Blvd Stop C1400, Austin, TX 78712, USA}
\author[0000-0001-8519-1130]{Steven L. Finkelstein}
\affil{Department of Astronomy, The University of Texas at Austin, 2515 Speedway Blvd Stop C1400, Austin, TX 78712, USA}

\author{Seiji Fujimoto}
\affil{Cosmic Dawn Center (DAWN)}
\affil{Niels Bohr Institute, University of Copenhagen, Lyngbyvej 2, DK-2200 Copenhagen, Denmark}
\author[0000-0003-4073-3236]{Christopher C. Hayward}
\affil{Center for Computational Astrophysics, Flatiron Institute, 162 Fifth Avenue, New York, NY 10010, USA}
%
%
%
\author[0000-0002-6610-2048]{Anton M. Koekemoer}
\affiliation{Space Telescope Science Institute, 3700 San Martin Dr., Baltimore, MD 21218, USA}

\author{Vasily Kokorev}
\affil{Cosmic Dawn Center (DAWN)}
\affil{Niels Bohr Institute, University of Copenhagen, Lyngbyvej 2, DK-2200 Copenhagen, Denmark}
\author{Claudia del P. Lagos}
\affil{International Centre for Radio Astronomy Research (ICRAR), M468, University of Western Australia, 35 Stirling Hwy, Crawley, WA 6009, Australia}
\affil{ARC Centre of Excellence for All Sky Astrophysics in 3 Dimensions (ASTRO 3D)}
\affil{Cosmic Dawn Center (DAWN)}

\author[0000-0002-7530-8857]{Arianna S. Long}
\affil{Department of Physics and Astronomy, University of California, Irvine, CA 92697, USA}

\author{Georgios E. Magdis}
\affil{Cosmic Dawn Center (DAWN)}
\affil{DTU-Space, Technical University of Denmark, Elektrovej 327, DK-2800 Kgs. Lyngby, Denmark}
\affil{Niels Bohr Institute, University of Copenhagen, Lyngbyvej 2, DK-2200 Copenhagen, Denmark}
\author[0000-0003-2475-124X]{Allison W.S. Man}
\affil{Department of Physics \& Astronomy, University of British Columbia, 6224 Agricultural Road, Vancouver, BC V6T 1Z1, Canada}
%
%
\author{Ikki Mitsuhashi}
\affil{Department of Astronomy, The University of Tokyo, 7-3-1 Hongo, Bunkyo, Tokyo 113-0033, Japan}
\affil{National Astronomical Observatory of Japan, 2-21-1 Osawa, Mitaka, Tokyo 181-8588, Japan}

\author{Gerg\"{o} Popping}
\affil{European Southern Observatory, Karl-Schwarzschild-Strasse 2, D-85748, Garching, Germany}

%
%
\author{Justin Spilker}
\altaffiliation{NHFP Hubble Fellow}
\affil{Department of Astronomy, The University of Texas at Austin, 2515 Speedway Blvd Stop C1400, Austin, TX 78712, USA}
\author{Johannes Staguhn}
\affil{The Henry A. Rowland Department of Physics and Astronomy, Johns Hopkins University, 3400 North Charles Street, Baltimore, MD 21218, USA}
\affil{Observational Cosmology Lab, Code 665, NASA Goddard Space Flight Center, Greenbelt, MD 20771, USA}
\author[0000-0003-4352-2063]{Margherita Talia}
\affil{Dipartimento di Fisica e Astronomia, Universit\`{a} di Bologna, Via Gobetti 93/2, I-40129, Bologna, Italy}
%
%
\author{Sune Toft}
\affil{Cosmic Dawn Center (DAWN)}
\affil{Niels Bohr Institute, University of Copenhagen, Lyngbyvej 2, DK-2200 Copenhagen, Denmark}
\author{Ezequiel Treister}
\affil{Instituto de Astrof\'{i}sica and Centro de Astroingenier\'{i}a, Facultad de F\'{i}sica, Pontificia Universidad Cat\'{o}lica de Chile, Casila 306, Santiago 22, Chile}
\author[0000-0003-1614-196X]{John R. Weaver}
\affil{Cosmic Dawn Center (DAWN)}
\affil{Niels Bohr Institute, University of Copenhagen, Lyngbyvej 2, DK-2200 Copenhagen, Denmark}

\author{Min Yun}
\affil{Department of Astronomy, University of Massachusetts Amherst, 710 N. Pleasant Street, Amherst, MA 01003, USA}


%

%



\begin{abstract}
We present the characteristics of 2\,mm-selected sources from the
largest Atacama Large Millimeter and submillimeter Array (ALMA)
blank-field contiguous survey conducted to-date, the Mapping
Obscuration to Reionization with ALMA (MORA) survey covering
184\,arcmin$^2$ at 2\,mm.  Twelve of thirteen detections above
5$\sigma$ are attributed to emission from galaxies, eleven of which
are dominated by cold dust emission.  These sources have a median
redshift of $\langle z_{\rm 2mm}\rangle=3.6^{+0.4}_{-0.3}$ primarily
based on optical/near-infrared (OIR) photometric redshifts with some
spectroscopic redshifts, with 77$\pm$11\%\ of sources at $z>3$ and
38$\pm$12\%\ of sources at $z>4$. This implies
that 2\,mm selection is an efficient method for identifying the
highest redshift dusty star-forming galaxies (DSFGs). Lower redshift
DSFGs ($z<3$) are far more numerous than those at $z>3$ yet
 likely to drop out at 2\,mm.
MORA shows that DSFGs with star-formation rates in
excess of 300\,\sfr\ and relative rarity of $\sim10^{-5}$\,Mpc$^{-3}$
contribute $\sim$30\%\ to the integrated star-formation rate density
between $3<z<6$.
The volume density of 2\,mm-selected DSFGs is consistent with
predictions from some cosmological simulations and is similar to the
volume density of their hypothesized descendants: massive, quiescent
galaxies at $z>2$.
Analysis of MORA
sources' spectral energy distributions hint at steeper
empirically-measured dust emissivity indices than typical literature
studies, with
$\langle\beta\rangle=2.2^{+0.5}_{-0.4}$.  The MORA survey represents an
important
step in taking census of obscured star-formation in the
Universe's first few billion years, but larger area 2\,mm surveys are
needed to more fully characterize this rare population and push to the
detection of the Universe's first dusty galaxies.
\end{abstract}


\section{Introduction} \label{sec:intro}

Half of all extragalactic radiation is absorbed by dust and re-emitted
at long wavelengths \citep[e.g.][]{fixsen98a}.  Decades of progress,
both technological and observational, have taught us that the obscured
emission emanates from very different galaxies than unobscured light;
the former is largely from massive, star-forming galaxies while the
latter is from lower mass galaxies \citep{whitaker17a}.  So while the
need to take census of star-formation has been a key focus of
extragalactic astrophysics for some time \citep[e.g.][]{madau14a}, it
has been clear that the very deep surveys of cosmic star-formation
conducted in the rest-frame ultraviolet and optical may not adequately
capture the full picture.  Surveys of submillimeter-luminous dusty
star-forming galaxies \citep[DSFGs;
  e.g.][]{smail97a,blain02a,casey14a} -- star-forming galaxies with
SFR$\simgt100\,$\sfr\ whose stellar emission is over
$\simgt$95\%\ obscured by dust -- have been the primary method of
unveiling the Universe's obscured contribution to the star-formation
rate density (SFRD).  This approach is in direct contrast with the
strategy of measuring the total SFRD by taking a census of UV-selected
galaxies and correcting estimates for dust attenuation \citep[e.g. as
  in][]{bouwens20a}.

A key limitation in all surveys of distant, dust-obscured galaxies is
the difficulty in identifying their redshifts.  Unlike Lyman Break
Galaxies (LBGs), whose selection method directly indicates their
redshifts, DSFGs' spectral shape in the (sub)mm regime is highly
degenerate with redshift solutions spanning $1\simlt z\simlt12$.
Similar efforts to characterize obscured emission indirectly via
synchrotron radio emission are faced with similar challenges, though
primarily limited to $1\simlt z\simlt4$ \citep{novak17a}.
Following up obscured sources in the optical or near-infrared for
characteristic emission lines present in star-forming galaxies is
difficult due to significant obscuration
\citep{chapman03a,chapman05a,swinbank04a}. Pursuing millimeter
spectroscopy has been prohibitive for large samples until recently due
to technological and sensitivity limitations in available
instrumentation\footnote{And recently, large samples are still quite
  challenging to confirm via millimeter spectral scans, as it requires
  anywhere from 30\,minutes -- 3 hours of integration time per source,
  even for luminous (unlensed) DSFGs \citep[c.f.][]{vieira13a}.}.  In
addition, the large beamsizes of many single-dish (sub)mm facilities
can further obfuscate redshift identification via uncertain astrometry
and source confusion, requiring another intermediate stage of
interferometric follow-up to constrain positions
\citep[e.g.][]{karim13a,hodge13a}.

The complexities in identifying DSFGs' redshifts has led to great
difficulty in measuring the volume density of highly obscured galaxies
beyond $z\simgt3-4$.  A small error in redshift measurement for a
small fraction of a uniformly-selected DSFG sample (typically selected
at $\lambda\le1\,$mm) can result in very different inferred volume
densities for the population at these redshifts \citep[e.g. see the
  wide variety of high-$z$ volume density measurements
  in][]{rowan-robinson16a,koprowski17a,gruppioni20a,dudzeviciute20a,loiacono21a,khusanova21a}.
This is primarily because the peak in the redshift distribution for
850\um-selected DSFGs (as well as 1\,mm-selected DSFGs) is between
$2<z<3$, and sources at higher redshifts are quite rare per unit solid
angle on the sky in comparison.  This has been demonstrated throughout
the literature via the measurement of the redshift distribution of
DSFGs selected at $\sim$850\um--1.2\,mm
\citep[e.g.][]{smolcic12a,brisbin17a,hatsukade13a}.  Recent work by
\citet{dudzeviciute20a} points out that only 6\%\ of 850\um-selected
galaxies sit at $z>4$.  Thus, the accurate identification of such
systems -- often having degenerate submillimeter colors with sources
at lower redshifts -- is effectively equivalent to searching for a
needle in a haystack.

Some efforts have focused on selecting the highest redshift DSFGs
using submillimeter colors, identifying characteristically `red' SEDs
across {\it Herschel} bands
\citep[e.g.][]{dowell14a,ivison16a,donevski18a,duivenvoorden18a,bakx18a,yan20a}.
However, {\it Herschel} bands do not benefit from the negative
$K$-correction because they probe emission near the peak of the dust
SED rather than the Rayleigh-Jeans tail \citep*{blain02a,casey14a},
thus {\it Herschel} datasets tend to have reduced sensitivity to
unlensed DSFGs beyond $z\sim2-3$.  Furthermore, this technique may
select against high-redshift DSFGs with atypically warm SEDs, due to
the degeneracy between dust temperature (or SED peak wavelength,
$\lambda_{\rm peak}$) and redshift.  
There have also been some
attempts to constrain the $z\simgt3$ DSFG population through
measurements of anisotropies in the Cosmic Infrared Background
\citep[CIB;][]{maniyar18a,maniyar21a} though {\it Herschel} is largely
insensitive to the high redshift tail.

This paper presents data from a new large ALMA mosaic conducted at
2\,mm (band 4), whose aim is to efficiently select DSFGs at
$z\simgt3-4$ and measure their volume density at these epochs with
more precision than has been done before.  This program is called the
Mapping Obscuration to Reionization with ALMA (MORA) Survey, for its
focus on this especially early epoch of DSFG formation.

The MORA survey is based on the hypothesis that 2\,mm dust continuum
is an efficient `filter' for $z\simgt3-4$ systems.  Surveys conducted
at 850\um--1.2\,mm benefit from the very negative $K-$correction,
meaning that their expected flux density is not redshift dependent for
a given fixed IR luminosity (i.e. at fixed \lir, $S_{850}\approx C$,
where $C$ is a constant without redshift dependence). Moreover,
surveys at 2\,mm have an even more extreme negative $K-$correction,
such that 2\,mm flux density {\it increases} with redshift (i.e. at
fixed \lir, $S_{2\,mm}\propto (1+z)^{\eta}$, where $\eta\approx0.5-1.0$).
As a result of this extreme negative $K-$correction, $z\simgt3$
galaxies of matched luminosity should appear {\it brighter} at 2\,mm
than those at $z\sim1-3$.  This contrasts with nearly every other
waveband in which galaxies are observed, from the X-ray through the
radio, where more distant objects are expected to be fainter than
objects closer to us.  If a blank-field 2\,mm survey depth is adjusted
appropriately, it should be sensitive to detecting DSFGs at $z\simgt3$
while the much more common $z\sim1-3$ DSFGs should be undetected
\citep[see detailed modeling work in][]{casey18a,casey18b,zavala18a}.
This is the premise for the design of the MORA survey.

A parallel work to this paper is presented by \citet{zavala21a},
hereafter referred to as Z21, who conduct a number counts analysis of
the MORA Survey 2\,mm mosaics and implications for the integrated
cosmic star-formation rate density at $z\simgt3$.

This paper presents the characteristics of the individual sources
detected in the MORA Survey and what can be inferred about their
redshifts, masses, SEDs, and descendants.  Section~\ref{sec:data}
presents the details of the MORA Survey design and data acquisition.
Section~\ref{sec:identify} then presents the identification of the
robust sample of 2\,mm-selected galaxies, and describes details of
each source from what is known in the literature; this section also
presents a discussion of sources with low signal-to-noise detections
below the formal 5$\sigma$ detection
threshold. Section~\ref{sec:models} describes models of the 2\,mm
universe, including semi-empirical models based in cosmological
simulations and empirical models.  Section~\ref{sec:results} presents
the main calculations and results of our manuscript, including
analysis of the 2\,mm population redshift distribution, SEDs, and
other unresolved physical characteristics.
Section~\ref{sec:discussion} presents a discussion, including results
of the contribution of the MORA sample to the cosmic star-formation
rate density, an extrapolation of what the MORA galaxy sample will
evolve to become, discussion of the measured emissivity spectral
index, and the potential impact of cosmic variance on our
measurements.  Section~\ref{sec:conclusions} then presents our
conclusions.  Throughout we assume a standard $\Lambda$-CDM cosmology
adopting the {\it Planck}-measured parameters, with
$H_{0}=67.4\,$\kms\,Mpc$^{-1}$\ \citep{planck-collaboration20a}, and
where SFRs are mentioned, we assume a Kroupa IMF \citep{kroupa03a} and
scaling relations drawn from \citet{kennicutt12a}.

\begin{figure}
\includegraphics[width=0.99\columnwidth]{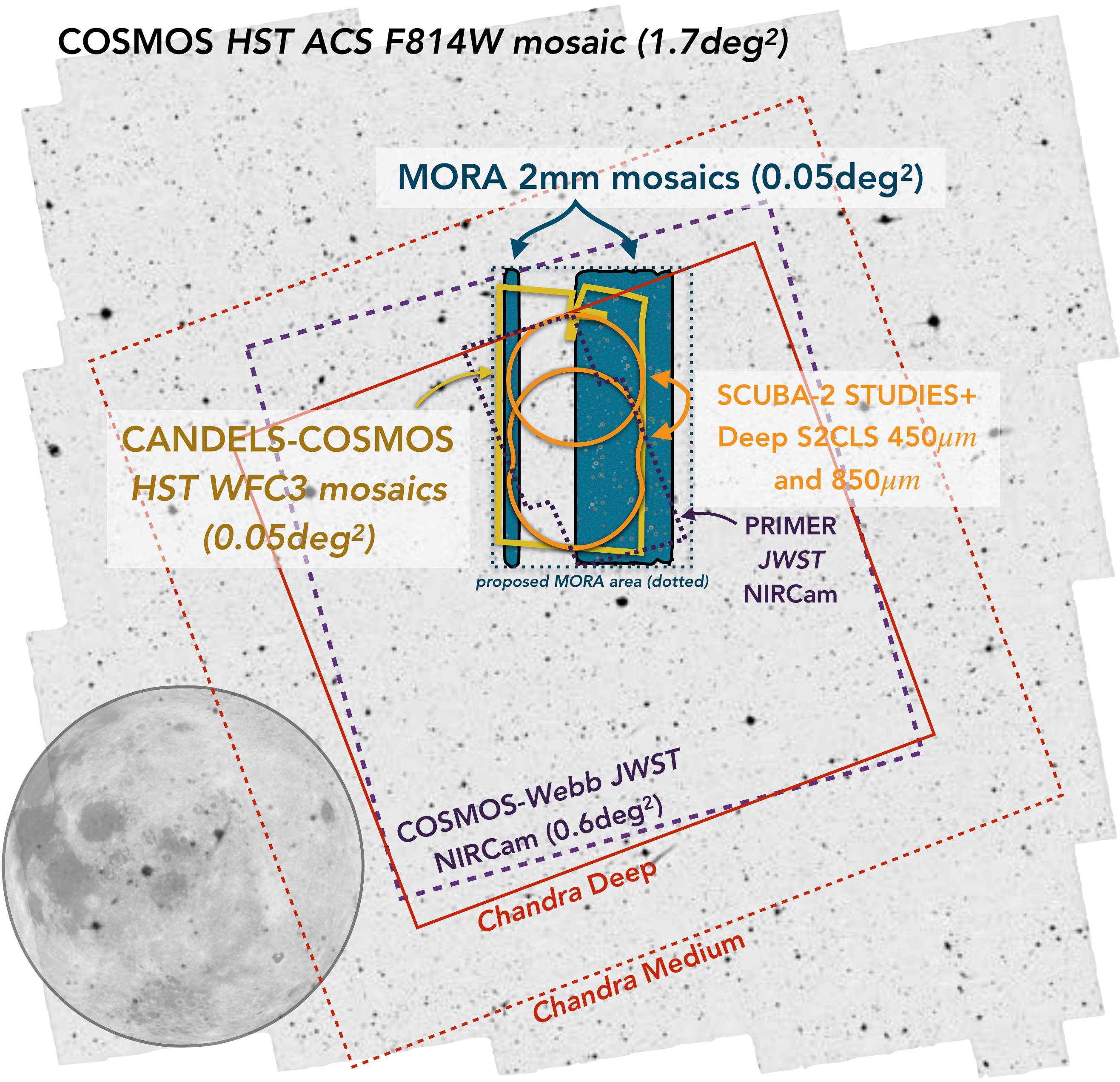}
\caption{The positions of the MORA mosaics shown in teal relative to
  the entire COSMOS field.  The moon is provided for scale in the
  lower left.  Shown are the {\it HST}/ACS F814W imaging
  \citep{scoville07a,koekemoer07a}, the {\it Chandra}-COSMOS deep and
  medium deep survey areas in red \citep{civano16a}, the
  CANDELS-COSMOS area in yellow \citep{grogin11a,koekemoer11a}, the
  deep 450\um\ and 850\um\ SCUBA-2 pointings in orange from the
  STUDIES program \citep{wang17a} and S2CLS
  \citep{roseboom13a,geach17b}, and the MORA mosaics in teal. We also
  show the proposed MORA area (dotted teal square) and the forthcoming
  PRIMER and COSMOS-{\it Webb} deep NIRCam imaging programs (dotted
  purple lines). Other multiwavelength data are not shown in this
  figure for clarity, though many cover the full COSMOS field
  \citep[e.g. deep 3\,GHz radio continuum from][]{smolcic17a}.  Most
  spectroscopy in the field is concentrated within the central
  1\,deg$^2$ around the {\it Chandra}-Deep and COSMOS-{\it Webb}
  footprints.}
\label{fig:cosmos}
\end{figure}

\begin{table*}
\caption{MORA Survey Observed Scheduling Blocks, Weather Conditions and Noise Characteristics}
\centering
\begin{tabular}{ccccccc}
\hline\hline
Position & Tuning & RA & PWV & On-Source & RMS & Synth. \\
& [GHz] & & [mm] & Time [min] & [\uJy/beam] & Beamsize \\
\hline
P03   & 139 & 10:00:43.83 & 5.87 & 44.00 & 62.8 & 1$\farcs$85$\times$1$\farcs$46 79$^o$ \\ 
P03   & 147 & 10:00:42.83 & 4.92 & 48.82 & 62.8 & 2$\farcs$11$\times$1$\farcs$32 72$^o$ \\ 
P10   & 139 & 10:00:29.19 & 6.19 & 43.98 & 98.3 & 1$\farcs$77$\times$1$\farcs$39 68$^o$ \\ 
P11   & 147 & 10:00:25.96 & 5.05 & 48.75 & 83.5 & 1$\farcs$51$\times$1$\farcs$45 30$^o$ \\ 
P12   & 147 & 10:00:23.73 & 4.91 & 48.78 & 88.0 & 1$\farcs$74$\times$1$\farcs$40 54$^o$ \\ 
P13   & 147 & 10:00:21.50 & 5.68 & 49.75 & 87.7 & 1$\farcs$56$\times$1$\farcs$36 86$^o$ \\ 
P14   & 147 & 10:00:19.26 & 5.99 & 49.77 & 90.9 & 1$\farcs$68$\times$1$\farcs$33 67$^o$ \\ 
P15   & 147 & 10:00:17.03 & 2.25 & 48.78 & 68.5 & 2$\farcs$27$\times$1$\farcs$44 72$^o$ \\ 
P16   & 147 & 10:00:14.80 & 4.72 & 48.77 & 112.1 & 2$\farcs$05$\times$1$\farcs$30 60$^o$ \\ 
P17   & 147 & 10:00:12.57 & 3.66 & 48.77 & 69.2 & 1$\farcs$94$\times$1$\farcs$44 68$^o$ \\ 
P18   & 139 & 10:00:10.34 & 5.98 & 43.98 & 60.4 & 2$\farcs$03$\times$1$\farcs$42 68$^o$ \\
P18   & 147 & 10:00:10.34 & 2.98 & 48.75 & 60.4 & 2$\farcs$44$\times$1$\farcs$65 61$^o$ \\
P20   & 147 & 10:00:05.87 & 5.31 & 48.83 & 91.1 & 1$\farcs$82$\times$1$\farcs$36 77$^o$ \\ 
\hline
P03 {\sc Mosaic}      & $-$ & $-$ & $-$ & 92.82 & 62.8 & 1$\farcs$91$\times$1$\farcs$41 83$^o$ \\
P10--P20 {\sc Mosaic} & $-$ & $-$ & $-$ & 528.91 & varies & 1$\farcs$83$\times$1$\farcs$43 88$^o$ \\
\hline\hline
\label{tab:observations}
\end{tabular}\\
\begin{minipage}{\textwidth}
{\bf Table Notes.} The range of declinations of pointing centers in
each scheduling block is uniform across all SBs and is +02:11:04.16 to
+02:33:55.82.  The positions (``P{\it XX}'') correspond to distinct
RAs of the mosaic.  Two positions were observed at both frequencies:
P03 and P18, while all other positions were only observed with one of
the two tunings.  Only P10--P20 are spatially adjacent such that they
have been stitched together in one contiguous mosaic, and the P03 data
is a separate mosaic of its own.  This table states the final
continuum RMS achieved in each position of the final mosaic product,
yet the synthesized beam is measured for a representative subsample of
pointings in each individual dataset.  The last two rows state the
resulting synthesized beams and total on-source time spent for the two
end-product mosaics: the P03 mosaic and the P10--P20 mosaic.
\end{minipage}
\end{table*}

\section{Data \&\ Observations} \label{sec:data}

MORA Survey observations were originally designed to cover
230\,arcmin$^2$ in two tunings, both in ALMA band 4 at 2\,mm in the
center of the Cosmic Evolution Survey (COSMOS) field
\citep{scoville07a,capak07a,koekemoer07a}.  Figure~\ref{fig:cosmos}
shows the context of the proposed and observed MORA mosaics in the
larger COSMOS field relative to other key datasets in the field.  The
COSMOS field was chosen as the location of the mosaic due to its rich
multiwavelength data (discussed more in \S~\ref{sec:ancillary}) and,
specifically, the CANDELS portion of the COSMOS field
\citep{grogin11a,koekemoer11a} was chosen for its even deeper
near-infrared imaging with {\it Hubble/WFC3}.  The near-infrared depth
will be significantly enhanced with the addition of {\it James Webb
  Space Telescope (JWST)} data from the PRIMER and COSMOS-{\it Webb}
surveys.

The target continuum RMS for the entire MORA mosaic was 90\,\uJy/beam
at 1$\sigma$.  This exact tuning configuration in band 4 was chosen
based on the program's secondary goal: to search the known $z\sim2.5$
protocluster structure, ``Hyperion,'' which spatially overlaps with
this map \citep{chiang15a,diener15a,casey15a,cucciati18a}, for a blind
search of molecular and neutral gas emitters.  The first tuning was
centered on a local oscillator (LO) frequency of 147.28\,GHz (referred
to as `Tune147') and covered the frequency ranges 139.5--143.2\,GHz
and 151.4--155.2\,GHz.  This tuning is sensitive to the detection of
CI(1-0) at $2.44<z<2.52$.  The second tuning is centered on a LO
frequency of 139.03\,GHz and covered the frequency ranges
131.2--134.9\,GHz and 143.2--146.9\,GHz (referred to as `Tune139').
It is tuned to enable the detection of CO(4-3) at $2.42<z<2.51$.  This
resulted in 21 scheduling blocks of 149 pointings each for each
tuning, resulting in 42 total scheduling blocks.  Each scheduling
block (SB) was spatially distributed as 2 columns of pointings with
fixed right ascension and 74--75 pointings in declination spanning a
declination range $+$02:11:04 to +02:33:56. Each SB's fixed
R.A. position is referred to as a position and a number in this text,
e.g. ``P{\it XX}'' where {\it XX} ranges from 03 -- 20.  As proposed,
the mosaic would have covered a total of 3129 pointings at two
frequency settings each.
The individual pointings of the mosaic were spaced by 19.3\arcsec,
which is 0.47 times the primary beam FWHM at the highest frequency of
data acquisition, 155.2\,GHz.  This spacing leads to a slightly more
compact mosaic than the default Nyquist spacing for mosaics; the same
spacing was used for both tunings to make data processing more
straightforward.  This consequently resulted in a greater depth of
observations than proposed (as Nyquist sampling was used to derive the
on-source time).

Observations with the Atacama Large Millimeter and submillimeter Array
(ALMA) were carried out under program 2018.1.00231.S from
27-March-2019 through 3-April-2019 in the C43-3 configuration for a
total of 14.6 hours including overheads and calibrations.
Data were acquired under an average precipitable water vapor of
PWV\,=\,5\,mm with conditions ranging from
2\,mm\,$<$\,PWV\,$<$\,6.5\,mm.

The program was observed in part only: 14 of the 42 SBs were executed,
11 of which were at the higher frequency tuning, Tune147, and 3 of
which at the lower frequency tuning, Tune139.  One of the higher
frequency SBs, Tune147 for position P16, did not pass QA0 due to poor
weather conditions, but was processed after the fact manually as
``semi-pass'' data and folded into the final mosaic after flagging
problematic antennae.  Table~\ref{tab:observations} lists the
observational conditions and data characteristics of each observed SB
and the final mosaics.

There are two final mosaics produced from these data that are
spatially distinct: one elongated mosaic represents observations taken
in the `P03' position (two pointings wide) with both tunings over a
total area of 28\,arcmin$^2$. P03 is too far spatially offset from the
rest of the data to be joined in one mosaic. The other mosaic
represents all other data from the spatially adjacent positions
`P10--P20' over a total area of 156\,arcmin$^2$ We refer to these as
the `P03' and the `P10--P20' mosaics, respectively.  The reason there
are two mosaics rather than one is because the program was only
partially completed and not all data were taken.

We imaged these data using natural weighting (Briggs weighting with
robust=2) to optimize source signal-to-noise. The synthesized beam
across all observations was broadly consistent with the beamsize of
the final mosaics: 1$\farcs$91$\times$1$\farcs$41 for P03 and
1$\farcs$83$\times$1$\farcs$43 for P10--P20.  This beamsize is larger
than the characteristic scale of dust in high redshift galaxies
\citep[$\sim$0$\farcs$6, e.g.][]{hodge16a} but smaller than the minimum
anticipated scale of source confusion at 2\,mm
\citep[$>$20$''$;][]{staguhn14a}\footnote{While source confusion is
  often discussed in the context of single-dish (sub)mm maps, it
  should be noted that the density of 2mm sources on the sky is much
  lower than at 850\um\ or shorter wavelengths, and therefore
  confusion would only set in for very low resolution ($\sim$1$'$ beam),
  deep ($<$1\,mJy RMS) 2mm maps.}.  The synthesized beam is ideally
matched to our science goals allowing for the detection of unresolved
point sources; therefore, no tapering or alternative data weighting
was needed.

Several channels covering a 140\,MHz wide frequency range centered on
an atmospheric absorption feature at 142.2\,GHz were flagged for
removal in the Tune147 datasets, while no channels were flagged in the
Tune139 datasets.  We determined that the channel flagging in the
Tune147 datasets improved the RMS depth of the map by 3\%\ on average \citep[see also][]{zavala21a}.

To analyze the computational time required to produce the full mosaic,
we tested time binning the data by 5\,s, 10\,s, and 30\,s.  Time
binning did substantially speed up the process of combining the
visibilities with {\tt tclean} and the resulting mosaic images were
consistent with one another.  For our final analysis, we use the 10\,s
time averaged maps.

Our final mosaics\footnote{Mosaics and Measurement Sets available to
  download at \href{www.as.utexas.edu/~cmcasey/downloads/mora.html}{www.as.utexas.edu/$\sim$cmcasey/downloads/mora.html}.}
cover 184\,arcmin$^2$ of the proposed 230\,arcmin$^2$.  Most of the
area is in the P10--P20 mosaic (156\,arcmin$^2$) while the remaining
28\,arcmin$^2$ is in the P03 mosaic.  Of the full area,
101\,arcmin$^2$ is covered at or below the proposed depth of
90\,\uJy/beam (with the deepest part of the map reaching
60\,\uJy/beam).  See Z21 for more complete information on the noise
characteristics of the maps.

\begin{figure*}
\centering
\includegraphics[width=1.99\columnwidth]{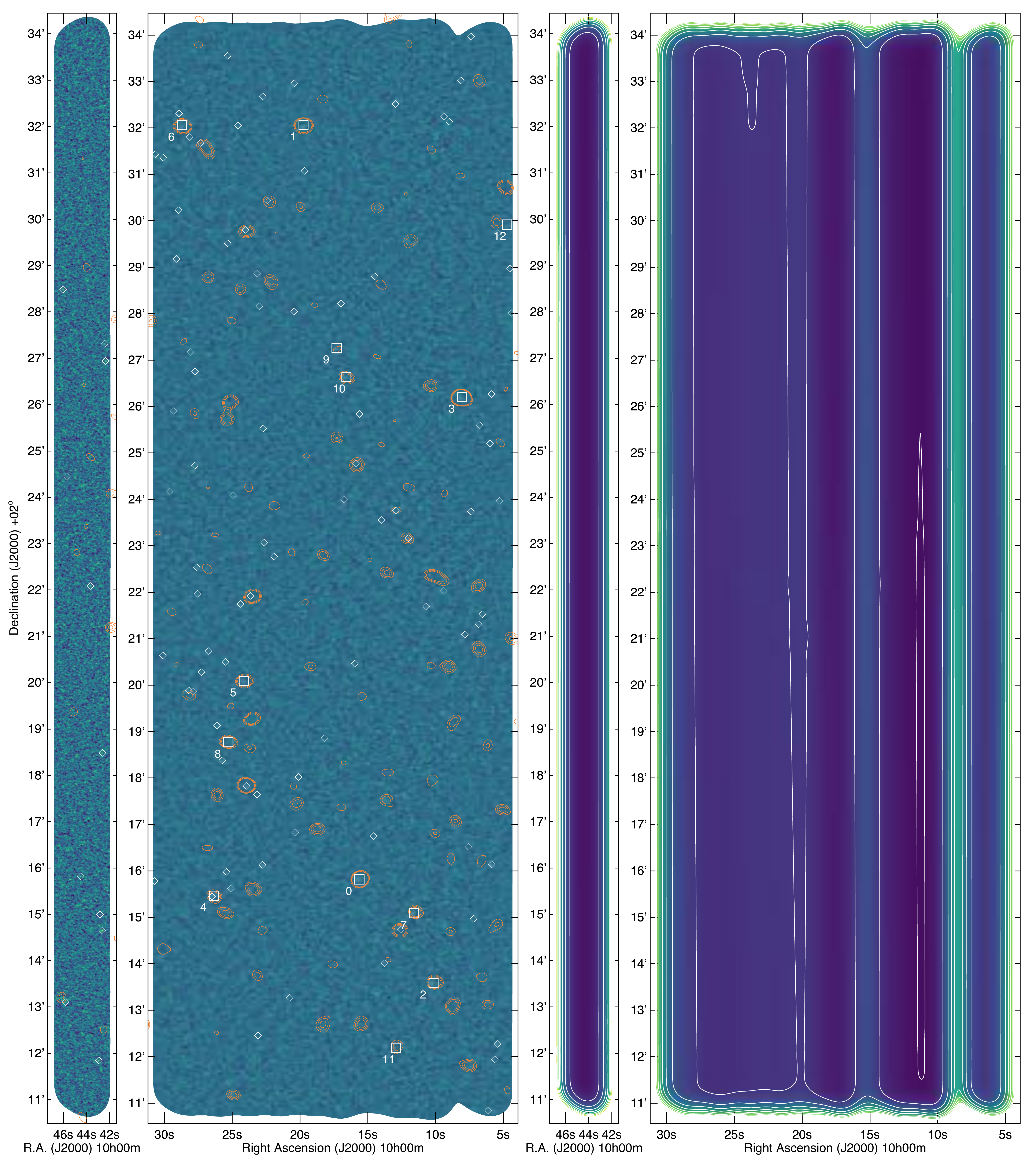}
\caption{Signal-to-noise (SNR; left) and root-mean-square (RMS; right)
  maps of the two MORA mosaics: P03 on the left, and P10-20 on the
  right.  Sources detected at $>$5$\sigma$ significance are enclosed
  in boxes and numbered, corresponding to the sources listed in
  Table~\ref{tab:sources} in decreasing order of SNR.  The SCUBA-2
  850\um\ signal-to-noise map from \citet{simpson19a} is shown in
  orange contours, denoting 3.5\,$\sigma$, 5\,$\sigma$, and
  6.5\,$\sigma$ significance.  Sources detected in the 2\,mm map
  between $4<\sigma<5$ are shown as small diamonds.  The RMS maps are
  overlaid with contours beginning at 60\uJy/beam and increasing in
  30\uJy/beam steps; the distribution of RMS depths across the full
  map is shown in Figure~2 of the accompanying MORA Survey paper, Z21.
  Note that irregularities in the RMS map and discontinuity between
  the two mosaics are due to the lack of completion of the MORA
  program.}
\label{fig:maps}
\end{figure*}

Figure~\ref{fig:maps} shows both signal-to-noise maps and
root-mean-square (RMS) maps of both mosaics.  For context, we have
overlaid (in contours) the S2COSMOS SCUBA-2 850\um\ signal-to-noise map
from \citet{simpson19a}.

\section{The 2\,mm-selected Sample}\label{sec:identify}

Extensive tests of large extragalactic mosaics from ALMA
\citep{umehata15a,dunlop16a,hatsukade16a,walter16a,aravena16a,gonzalez-lopez19a,gonzalez-lopez20a,franco18a,franco20a}
show that ALMA deep fields exhibit Gaussian noise in the absence of
bright (SNR\,$>$\,10) sources.  This is demonstrated for the MORA mosaics
in our accompanying paper, Z21, which analyzed the number counts
and noise characteristics of this dataset in depth.
In \citet{casey18b} we simulated completeness and contamination rates
for mock ALMA datasets using the assumption that the noise is
Gaussian. 
As argued in \S~3.1 of \citet{casey18b}, the measured
contamination rates and completeness of simulated ALMA sources do not
depend on the wavelength or underlying number density of sources in
the map because they are not confusion-limited.  

Z21 tested that confusion is, indeed, not a concern for the MORA
mosaics by masking significantly detected sources (of which there are
few across the large mosaic) and injecting fake sources through Monte
Carlo trials throughout the rest of the map, measuring both
completeness and contamination rates as a function of signal-to-noise
and 2\,mm flux density.  Z21 completed the same procedure for fake
maps which have Gaussian noise and the same heterogeneous RMS
characteristics of our data and find identical rates of contamination
and source completeness.  Furthermore, we find that sources are
recovered at the same rates as simulated in Figure~6 of
\citet{casey18b}, despite the differences in simulated wavelength and
different beamsize of observations.

Note that the synthesized beamsize of our observations,
$\sim$1$\farcs$9$\times$1$\farcs$4, exceeds the expected full width
half maxima of obscured emission for galaxies at $z>1$
\citep[$\sim$1--5\,kpc FWHM, corresponding to scales $<$0$\farcs$6,
  e.g.][]{simpson15a,hodge16a,fujimoto17a}, therefore we do not expect
any sources to be resolved out beyond the scale of one synthesized
beam. Thus the treatment of these sources as point sources rather than
sources with extended emission is appropriate and flux densities are
measured at the sources' peak.  This contrasts with the recent
GOODS-ALMA survey covering 69\,arcmin$^2$ at 1.1\,mm, whose
synthesized beamsize of 0$\farcs$6 is similar to the expected size of
sources; this led to incompleteness with regard to spatially extended
sources in that work and the need to taper observations to recover
extended emission \citep{franco18a,franco20a}.

To further diagnose source contamination, we search the MORA maps for
significant negative peaks and find 106 sources below $-$4$\sigma$
significance (i.e. strong negative peaks), 15 lower than
$-$4.5$\sigma$ and 2 lower than $-$5$\sigma$ significance.  With
$\sim$3$\times$10$^{5}$\,beams in the MORA maps, we expect
48$^{+6}_{-5}$, 7$\pm$3, and 1$\pm$1 negative sources to arise at these
respective significances ($>$\,4, $>$\,4.5 and $>$\,5\,$\sigma$).  The
uncertainties on the expected number of negative noise peaks are
determined by generating several fake noise maps with the same noise
characteristics as the existing data.  The number of negative sources
found in the map skews somewhat higher than expected, despite the
consistencies of the maps' noise characteristics with modeled Gaussian
noise. Z21 demonstrates that the noise characteristics of the MORA
mosaics are indeed Gaussian. We do not suspect that the atypical
number of negative sources is from the sidelobes of nearby bright
sources. Instead the excess of noise peaks could be due to slight
imperfections in modeling the noise.  Z21 highlights that the most
significant negative detection in the map is found at $-6\sigma$, which
only has a $\simlt$0.5\%\ chance of being generated in a map of this
size due to noise.  Whether or not this negative detection is of
genuine astronomical origin (i.e. potentially a decrement in the CMB
caused by inverse compton scattering) is briefly discussed in Z21, but
requires further observational follow-up to refute or confirm.

Our tests are consistent with our previous findings in
\citet{casey18b}: sources identified at $>$5$\sigma$ significance have
little-to-no contamination with false noise peaks; only 1$\pm$1 false
source is expected across both MORA mosaics above this threshold from
Gaussian noise.  Contamination rises sharply at $4.5<\sigma<5$
significance, consisting of $\sim$\,40\,\%\ false noise peaks, and
noise peaks come to dominate the sample detected between
$4<\sigma<4.5$ significance.  In this paper, we present the robust
sample of $>5\sigma$ detected sources and then proceed to analyze the
marginal $4<\sigma<5$ sample in conjunction with prior identification
at other complementary wavelengths.  Thirteen $>$5\,$\sigma$ sources
are identified, one of which is thought to be false, translating to a
$>$5$\sigma$ purity of $12/13\approx92\%$ (a figure that would be
higher if real sources were more common).

\subsection{COSMOS Ancillary Data}\label{sec:ancillary}

We make extensive use of the rich ancillary data available in the
COSMOS field including the two most recent generations of the
optical/near-infrared photometric catalog, COSMOS2015
\citep{laigle16a} and COSMOS2020 (Weaver, Kauffmann \etal, submitted).
There are over 30 bands of optical/near-infrared (OIR) imaging that
make up these photometric databases, from deep broadband coverage
(with depths of 26--28 magnitudes for 5$\sigma$ point sources) to
intermediate and narrow band imaging campaigns (with depths 25--26
magnitudes).  

COSMOS also contains a wealth of multiwavelength data, from the X-ray
\citep{civano12a,civano16a} through the radio
\citep{schinnerer07a,smolcic17a}.  We include details of detections at
these other wavebands where relevant.  Particularly important for this
work are the other millimeter and submillimeter datasets in the field,
including SCUBA-2 at 850\um\ \citep{casey13a,geach17b,simpson19a},
450\um\ \citep{casey13a,roseboom13a}\footnote{Though deeper 450\um\ data exist in a subset of the MORA field from \citet{wang17a}, these data are not publically accessible and are therefore not included in our analysis.}, {\it Herschel} PACS and SPIRE at
100--500\,\um\ \citep{lutz11a,oliver12a}, AzTEC at 1.1\,mm
\citep{scott08a,aretxaga11a}, and {\it Spitzer} at
24\um\ \citep{le-floch05a}.  In addition, a variety of ALMA archival
datasets have built up in the field -- see \citet{liu18a} for details
on the A3COSMOS project -- at a range in observed frequencies, with
the most common tunings in band 6 (1.2\,mm) and band 7 (870\um).  We
make use of these data to refine the spectral energy distribution fits
for galaxies detected in the MORA maps.

\subsection{Photometric Redshift Fitting}

Photometric redshifts are fit to this existing photometry using the
{\sc Lephare}\footnote{\software{{\sc Lephare}
  \citep{arnouts02a,ilbert06a}}} package and \citet{bruzual03a}
templates for star-forming and quiescent galaxies, as in
\citet{ilbert13a}.  Where OIR photometric redshifts exist for
MORA-detected sources, we provide the most recent estimates from
COSMOS2020 when available, or in some instances, we use photometric
redshifts from elsewhere in the literature.  Several photometric
redshift catalogs are presented in the COSMOS2020 compilation; we make
use of the {\sc Classic} {\tt SExtractor}\footnote{\software{{\tt
    SExtractor} \citep{bertin96a}}} photometry and {\sc Lephare}
photometric redshifts, as we find they minimize $\chi^2$ for the
redshift estimates.  Note that the redshifts overall are
indistinguishable from one another for this sample and would not
change our results.  See Weaver, Kauffmann \etal, submitted for more
detail on the differences between these catalogs.

For each source in our sample, we also employ the
\mmpz\ technique\footnote{\software{{\sc MMpz} \citep{casey20a}.}}
for FIR/mm photometric redshift fitting described in \citet{casey20a}
as an independent check on redshift constraints at other wavelengths.
The \mmpz\ technique uses the aggregate FIR/mm flux density
measurements for a source to derive a probability density distribution
in redshift.  Rather than basing the redshift fit on a single
long-wavelength template, \mmpz\ presumes the galaxy is most likely to
lie on the correlation between galaxies' IR luminosities and their
rest-frame peak wavelengths, i.e. in the \lir-\lpeak\ plane \citep[as
  shown in][]{lee13a,strandet16a,casey18a}, and the algorithm
determines the redshift range over which the galaxy's SED is likely to
be most consistent with that empirical relationship.

\subsection{The 5$\sigma$ subsample}\label{sec:sample}

Thirteen sources are identified in our maps above 5$\sigma$
significance. They are numbered in order of decreasing signal-to-noise ratio at
2\,mm and their basic detection characteristics are listed in
Table~\ref{tab:sources}.  Twelve of the thirteen have been detected at
other wavelengths and reported in the literature in various forms; in
particular, those twelve sources are also detected with SCUBA-2 at
850\um\ \citep{simpson19a} and with {\it Spitzer} IRAC at 3.6\um.
Two of the sources in the sample (sources 5 and 9) have particularly
uncertain redshifts and are undetected in deep near-infrared $H$-band
imaging from the CANDELS-COSMOS survey \citep{koekemoer11a}; these two
sources and their properties are described in greater detail in the
accompanying paper, Manning~\etal

\begin{figure*}
\centering
\hspace{0.0788382\columnwidth}\includegraphics[width=1.9\columnwidth]{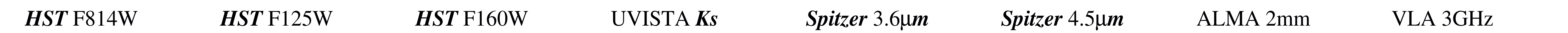}\\
\includegraphics[width=0.0788382\columnwidth]{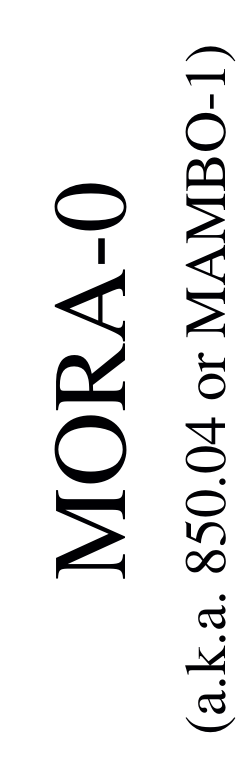}\includegraphics[width=1.9\columnwidth]{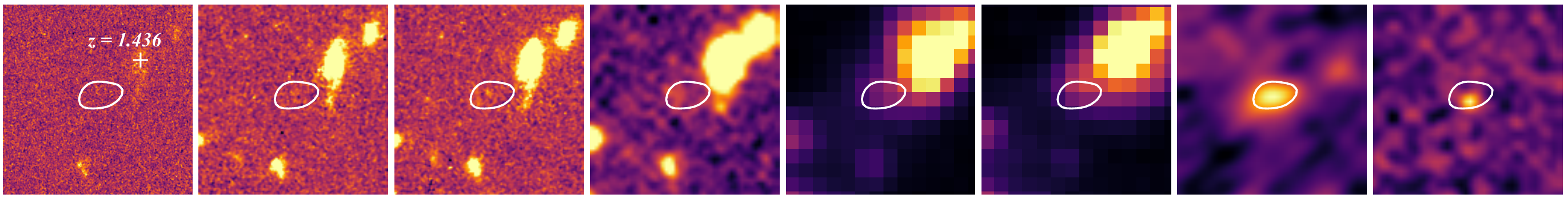}\\
\includegraphics[width=0.0788382\columnwidth]{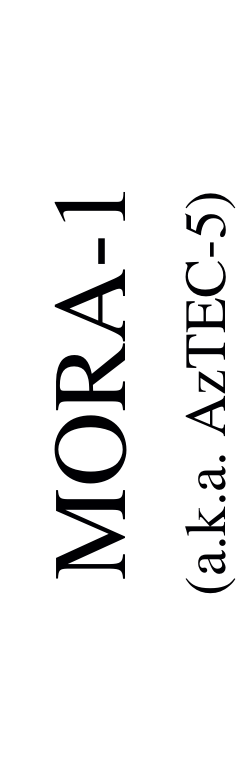}\includegraphics[width=1.9\columnwidth]{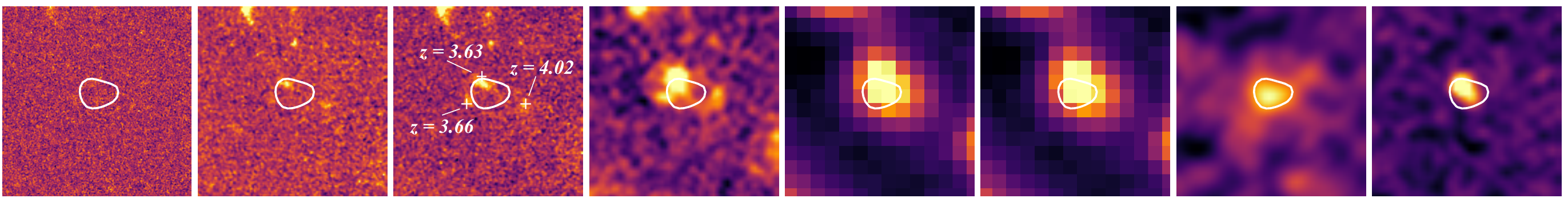}\\
\includegraphics[width=0.0788382\columnwidth]{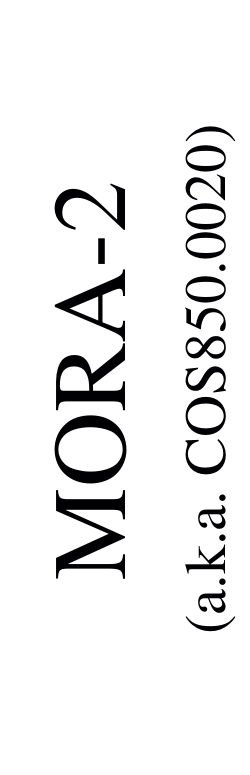}\includegraphics[width=1.9\columnwidth]{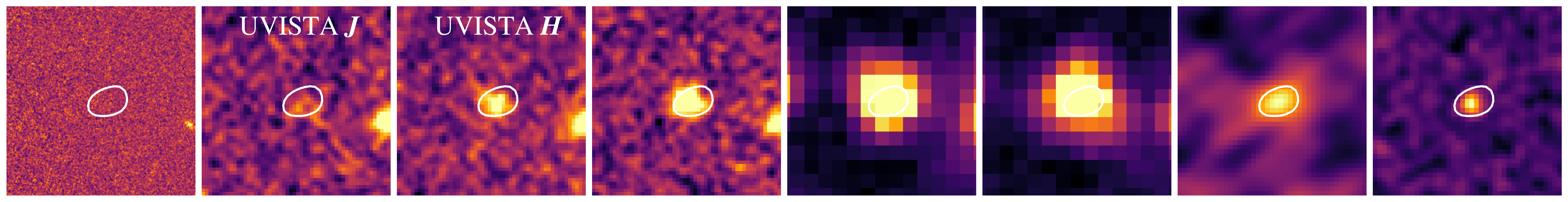}\\
\includegraphics[width=0.0788382\columnwidth]{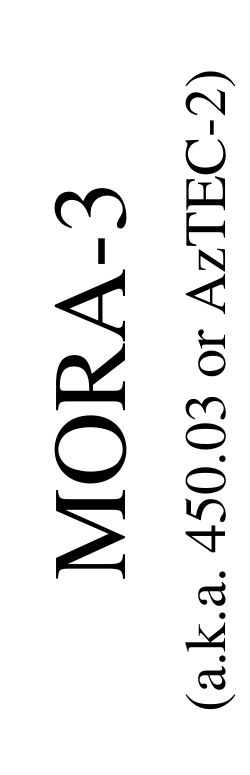}\includegraphics[width=1.9\columnwidth]{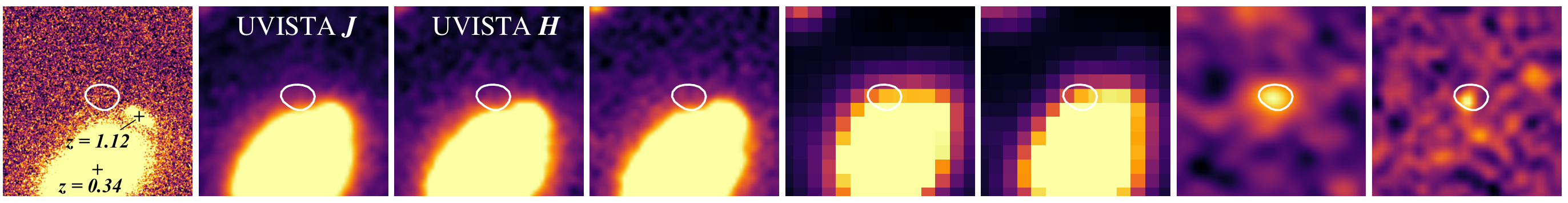}\\
\includegraphics[width=0.0788382\columnwidth]{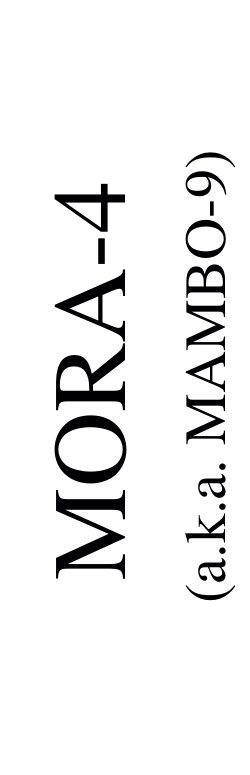}\includegraphics[width=1.9\columnwidth]{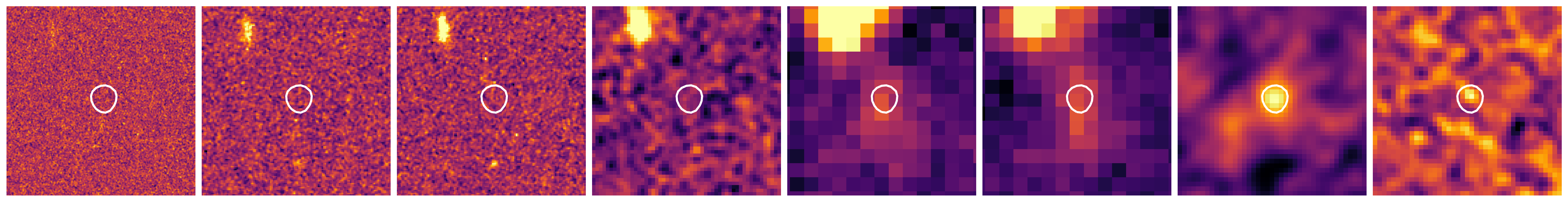}\\
\includegraphics[width=0.0788382\columnwidth]{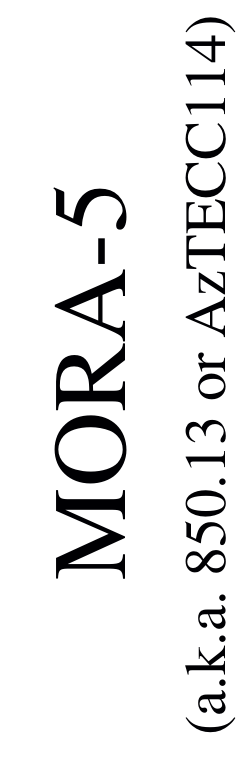}\includegraphics[width=1.9\columnwidth]{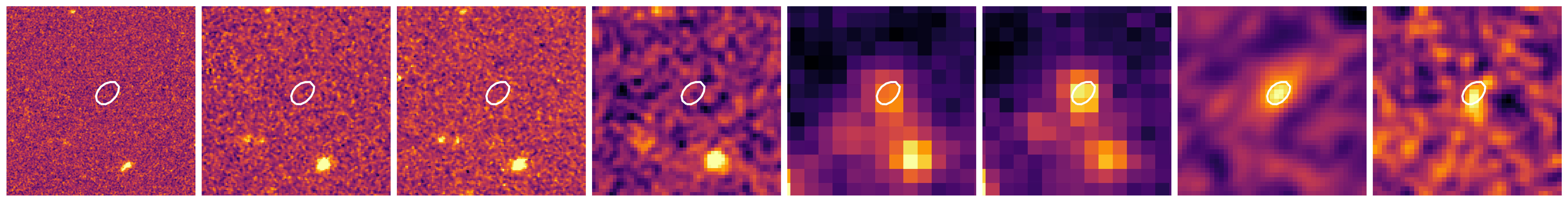}\\
\includegraphics[width=0.0788382\columnwidth]{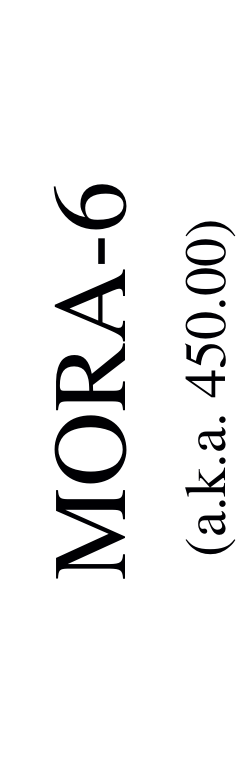}\includegraphics[width=1.9\columnwidth]{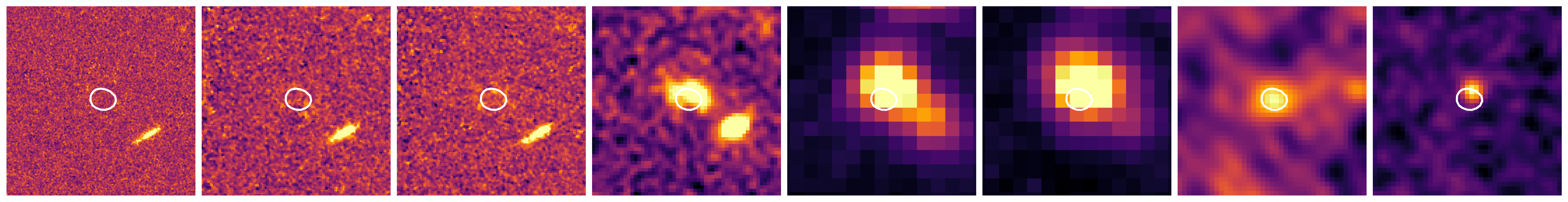}\\
\caption{Multiwavelength cutouts (8$''\times$8$''$, where north is up
  and east is left) of the 13 sources identified at $>$5$\sigma$
  significance at 2\,mm (see continuation of this figure on the next
  page).  Cutouts are from {\it HST} ACS/F814W
  \citep{scoville07a,koekemoer07a}, {\it HST} WFC3/F125W and
  WFC3/F160W \citep{koekemoer11a,grogin11a}, Ultra-VISTA $Ks$-band
  \citep{laigle16a}, {\it Spitzer} 3.6\um\ and
  4.5\um\ \citep{ashby15a}, 2\,mm (this work), and VLA 3\,GHz
  \citep{smolcic17a}.  The sources that sit outside of the CANDELS
  WFC3 coverage area have $J$ and $H$-band cutouts from Ultra-VISTA
  shown instead. The 5$\sigma$ contour from the 2\,mm imaging is
  overlaid in each panel for reference.  Some sources that sit
  adjacent to the 2\,mm-detected galaxies are labeled with redshifts
  with more details given in the text.}
\label{fig:cutouts}
\end{figure*}

Here we provide a brief summary of each of the $>5\sigma$ sources and the
extent to which they have already been characterized in the
literature.  Table~\ref{tab:photometry} in the appendix includes
additional photometric measurements of each sources' integrated flux
density at all available wavelengths.  Figure~\ref{fig:cutouts} shows
multiwavelength cutouts of all thirteen sources from the optical
through the radio.

\begin{figure*}
\centering
\hspace{0.0788382\columnwidth}\includegraphics[width=1.9\columnwidth]{cutout_header.pdf}\\
\includegraphics[width=0.0788382\columnwidth]{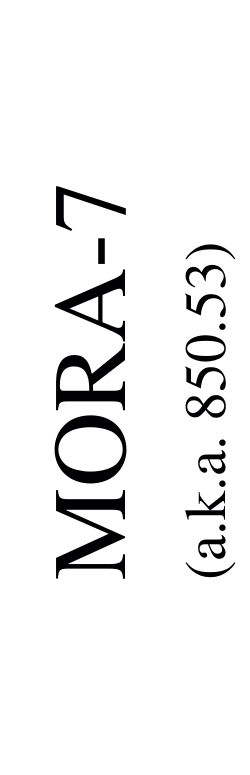}\includegraphics[width=1.9\columnwidth]{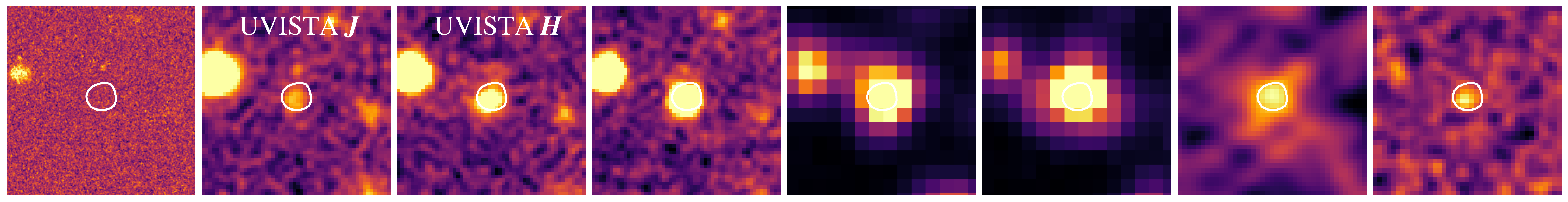}\\
\includegraphics[width=0.0788382\columnwidth]{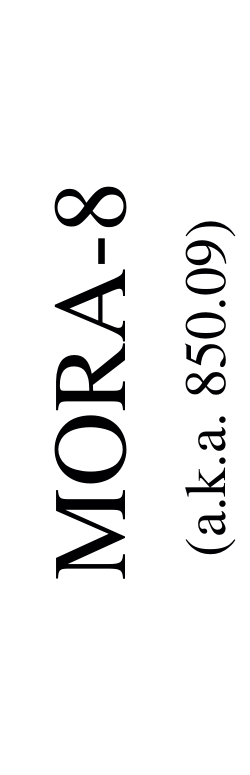}\includegraphics[width=1.9\columnwidth]{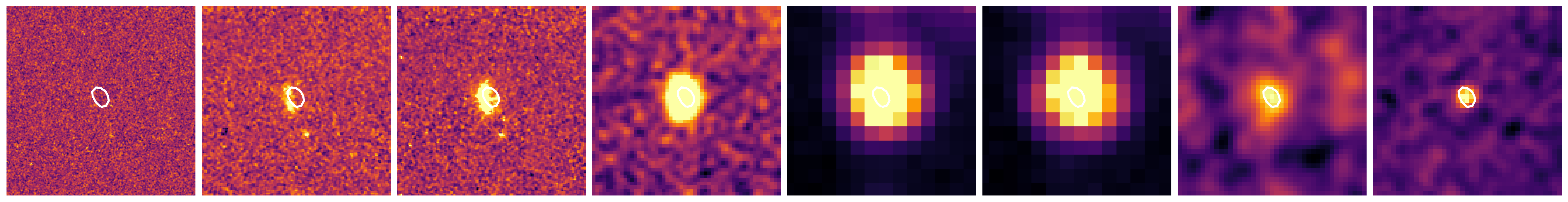}\\
\includegraphics[width=0.0788382\columnwidth]{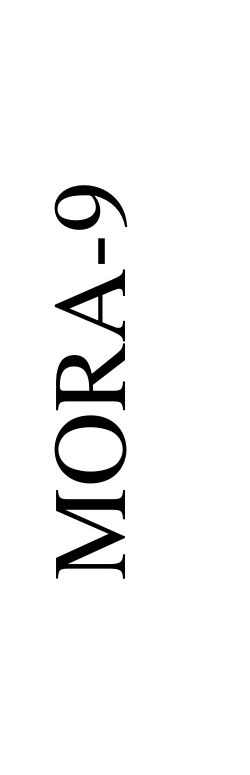}\includegraphics[width=1.9\columnwidth]{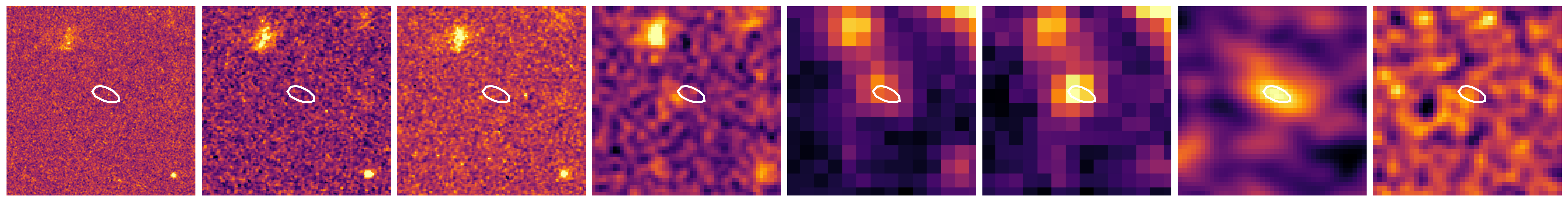}\\
\includegraphics[width=0.0788382\columnwidth]{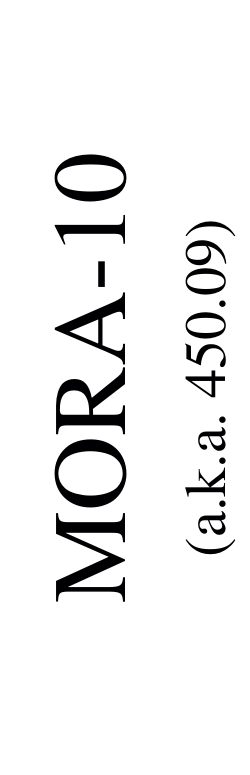}\includegraphics[width=1.9\columnwidth]{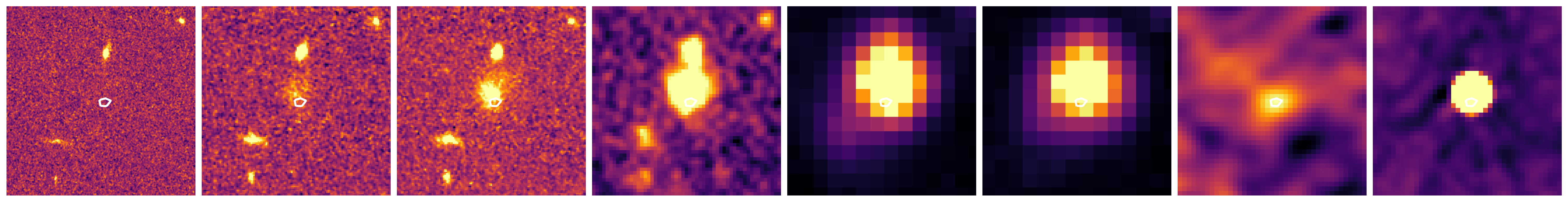}\\
\includegraphics[width=0.0788382\columnwidth]{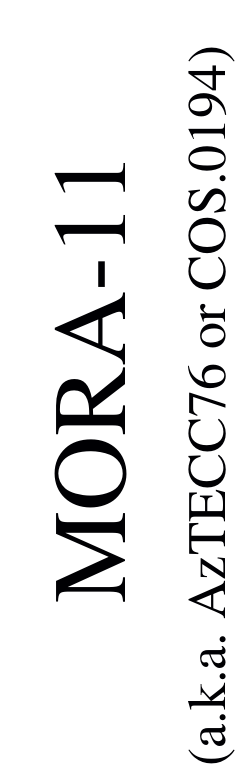}\includegraphics[width=1.9\columnwidth]{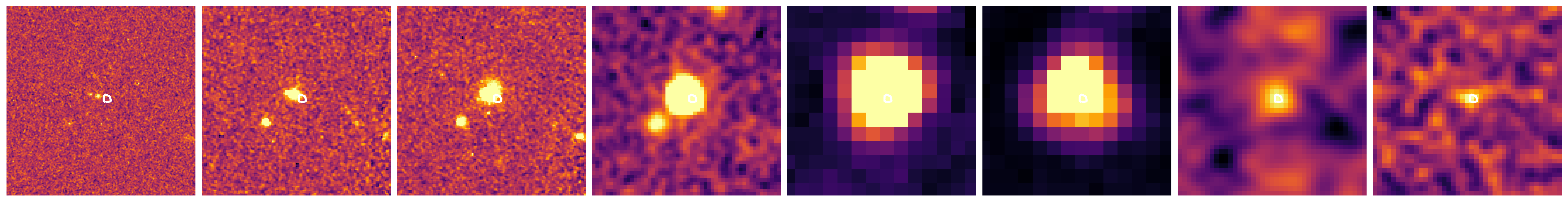}\\
\includegraphics[width=0.0788382\columnwidth]{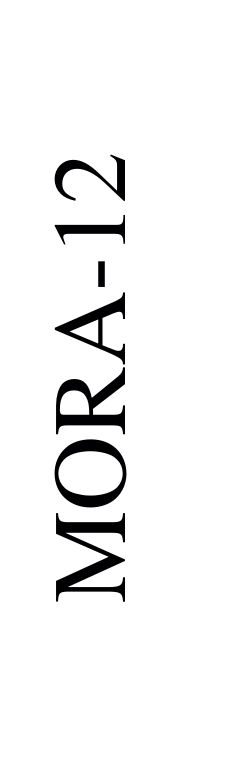}\includegraphics[width=1.9\columnwidth]{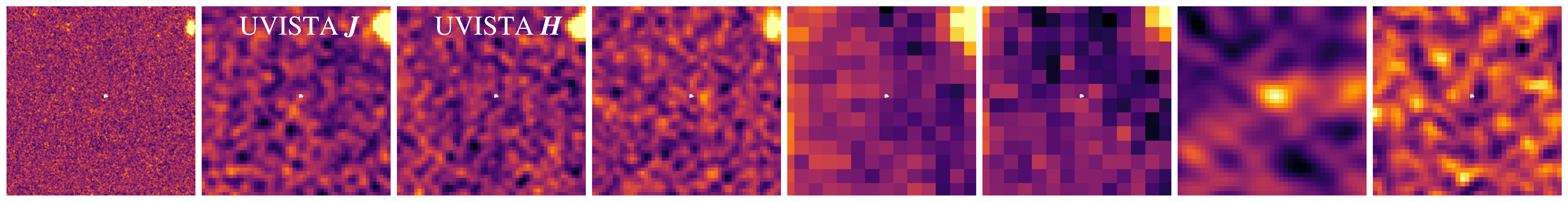}\\
--- Continuation of Figure~\ref{fig:cutouts} --- 
\vspace{3mm}
\end{figure*}

\subsubsection{MORA-0, a.k.a. 850.04 or MAMBO-1}

The highest signal-to-noise source in the MORA maps is detected at
nearly 8$\sigma$, and has been previously identified as a DSFG through
detection at 1.2\,mm \citep[MAMBO-1 in][]{bertoldi07a}, 1.1\,mm
\citep[AzTECC7 in][]{aretxaga11a}, and at 850\um\ \citep[850.04
  in][]{casey13a}.  The source is only marginally detected with {\it
  Herschel} SPIRE and SCUBA-2 at 450\um\ \citep{oliver12a,casey13a}.
To date, it does not have a reliable spectroscopic confirmation
despite being targeted repeatedly in near-infared campaigns; the
H-band MOSFIRE spectrum presented in \citet{casey17a} is spatially
offset from the ALMA source by 1$''$.  The closer (and much fainter)
OIR counterpart, only 0$\farcs$3 offset from the ALMA centroid, has an
OIR-based photometric redshift from the \citet{laigle16a} COSMOS2015
catalog of $z_{\rm phot}=3.31^{+0.76}_{-0.81}$; the source is missing
from the COSMOS2020 photometric catalogs for no obvious reason other
than its marginal detection near the detection limit of the
near-infrared catalogs.

MORA-0 has a FIR photometric redshift from
\citet{brisbin17a} of $z_{\rm FIR}=4.4\pm1.0$. Using the \mmpz\ FIR/mm
photometric redshift technique, we derive a mm-based redshift of
$z_{\rm mm}=3.4^{+0.6}_{-0.6}$; both FIR/mm photometric redshifts are
higher than, though consistent with, the OIR photometric redshift.
The source is detected in both the 1.4\,GHz and 3.0\,GHz radio maps.
There are several ALMA programs that have obtained continuum data on
MORA-0 in band 6 (1287\,\um\ and 1250\,\um) and in band 7 (870\,\um).  While
two radio galaxies were originally thought to be associated with this
source \citep{bertoldi07a}, these are now thought to sit at different
redshifts: one with $z_{\rm spec}=1.436$ \citep[reported in][]{casey17a} and
MORA-0 with a higher photometric redshift.  The source at $z=1.436$ is
not detected in any of the ALMA data, while MORA-0 is detected in all
of bands 4, 6 and 7 where data exists.  No millimeter spectroscopy
exists for MORA-0.

The lack of detection in the COSMOS2020 photometric redshift catalog,
yet detection in COSMOS2015, casts some doubt on the quality of OIR
constraints; however, the broad consistency of the COSMOS2015
constraint with the \mmpz\ redshift is reassuring.  We adopt the
COSMOS2015 photometric redshift throughout the rest of the text.

\subsubsection{MORA-1, a.k.a. AzTEC-5}

The second most significant source, MORA-1, has been studied under
many names, the most widely-used of which is AzTEC-5 for its initial
detection in \citet{scott08a} at 1.1\,mm. \citet{magnelli19a} includes
a nice discussion of its known characteristics and redshift
constraints, which we summarize here for completeness. AzTEC-5 is {\it
  Herschel} SPIRE, SCUBA-2 (450\um\ and 850\um) and ALMA detected,
with continuum measurements in bands 6 and 7.

The literature alludes
to a spectroscopic identification for AzTEC-5 of $z=3.791$ based on
Ly$\alpha$ emission in a DEIMOS spectrum \citep{capak10a,smolcic12a};
however, that solution was revealed to be inaccurate by subsequent
ALMA follow-up that failed to detect emission lines to corroborate the
redshift.

\citet{gomez-guijarro18a} identify four components of AzTEC-5 in deep
near-infrared imaging, which they dubbed AzTEC5-1, 5-2, 5-3, and 5-4.
MORA-1 is coincident with the position of AzTEC5-1, which is the only
component lacking direct redshift constraints.
\citeauthor{gomez-guijarro18a} do not fit a photometric redshift for
AzTEC5-1; it is absent from the COSMOS2020 photometric catalogs.  The
photometric redshifts for the other components are
$z=3.63^{+0.14}_{-0.15}$ for AzTEC5-2, $z=4.02\pm0.08$) for AzTEC5-3,
and $z=3.66^{+0.40}_{-0.43}$ for AzTEC5-4.  The redshift constraints
on existing components is shown in the third panel (second row) of
Figure~\ref{fig:cutouts}.  In addition to AzTEC5-1, AzTEC5-2 also
appears to have associated 870-\um\ emission, separated from AzTEC5-1
by 0$\farcs$7; ATEC5-2 is not detected at $>$5$\sigma$ significance in
our mosaics.  Our millimeter-derived photometric redshift for the
aggregate photometry of all components of AzTEC-5 give $z_{\rm
  mm}=2.6^{+2.2}_{-0.8}$, while AzTEC5-1 alone has $z_{\rm
  mm}=4.8^{+3.7}_{-2.1}$.  Due to the significant uncertainties on
both, they are consistent with the photometric redshift constraints of
the other three components from \citeauthor{gomez-guijarro18a}

Having detected AzTEC-5 at 2\,mm with GISMO, albeit with far worse
spatial resolution than the MORA map, \citet{magnelli19a} argue that
the redshift of the entire system is likely $z\sim3.6$, with
substantial obscuration in AzTEC5-1 making it difficult to
spectroscopically confirm.  Indeed, the close spatial separation
$<$1$''$ between the two sources (AzTEC5-1 and AzTEC5-2) would support
this claim, as the likelihood of identifying two 870\um\ sources with
$>$3\,mJy at different redshifts separated by $<$1$''$ is exceedingly
low \citep[0.02\%\ based on 850\um\ number
  counts;][]{geach17b,simpson19a}.
While spectroscopic confirmation is needed for MORA-1\footnote{We note
  that alternate band 4 ALMA spectral line observations of MORA-1
  exist under the source name `GalD' in program 2018.1.01824.S;
  however, the tuning would only capture CO(6-5) from redshifts
  $z=3.65-3.77$ and $z=4.06-4.20$.  No CO line is detected in the
  dataset to a depth of $1\sigma\sim$0.4\,mJy/beam in 100\,km/s
  channels.}, we determine that it is best, for the purposes of this
work, to adopt the median redshift of the three other components of
the AzTEC-5 system given in \citet{gomez-guijarro18a}, which is
$z_{\rm phot}=3.78^{+0.27}_{-0.32}$.

\subsubsection{MORA-2 a.k.a. COS850.0020}
The third source in the MORA sample, MORA-2, is detected at a
signal-to-noise of 7.6 at 2\,mm.  It has no match in the earliest
SCUBA-2 surveys \citep{roseboom12a,casey13a} but is detected in the
wider-field surveys of COSMOS from \citet{geach17b} and
\citet{simpson19a}.  Both ALMA band 6 and 7 data exist for MORA-2,
which was selected both as a submillimeter source and dropout source
in the ZFOURGE survey \citep{spitler12a}. Its OIR photometry give a
photometric redshift from COSMOS2020 of $z_{\rm
  phot}=3.36^{+0.60}_{-0.28}$.  There are no spectroscopic constraints
for MORA-2; our millimeter-derived photometric redshift for the source
is $z_{\rm mm}=2.8^{+0.6}_{-0.6}$ which is consistent with the OIR
photometric redshift.

\subsubsection{MORA-3, a.k.a. AzTEC-2 or 450.03}

The fourth source, MORA-3, is more widely known as AzTEC-2
\citep{younger07a} detected with the SMA, and at 450\um\ and
850\um\ by SCUBA-2 \citep[where it is known as 450.03;][]{casey13a}.
It was also previously detected at 2\,mm as GISMO-C2
\citep{magnelli19a}.  It is detected by {\it Herschel} SPIRE and has
ALMA continuum data in band 6 and 7.  The interferometric data reveal
two millimeter sources, the primary (coincident with the 2\,mm
emission) is AzTEC2-A and the secondary fainter source (detected at
2.6$\sigma$ significance in the 2\,mm map) is AzTEC2-B.  While a
spectroscopic identification based on a possible OIR counterpart
existed at $z=1.123$ \citep{smolcic12a,casey17a} 1$''$ to the south of
the primary source A, that redshift has since been shown to be
associated with a foreground galaxy.

The spectroscopic redshift of AzTEC-2 is now confirmed through the
detection of [C{\sc ii}] and CO(5-4) at $z=4.63$ by
\citet{jimenez-andrade20a}; this was independently confirmed in
\citet{simpson20a}.  AzTEC2-A has a spectroscopic redshift of
$z=4.625$ while AzTEC2-B is at $z=4.633$. Our millimeter-derived
photometric redshift for this source is consistent with its
spectroscopic identification, $z_{\rm mm}=3.3^{+1.0}_{-0.8}$.  Both
components of AzTEC-2 are undetected in existing {\it HST} deep
imaging, and the {\it Spitzer} imaging is highly confused with two
foreground galaxies: the $z=1.123$ galaxy 1$''$ to the south of
AzTEC2-A and an elliptical galaxy at $z=0.34$ at 1$''$ south of the
$z=1.123$ system.  Both components of AzTEC-2 are detected at 3\,GHz
in radio continuum.  See \citet{jimenez-andrade20a} for a more
thorough discussion of this source.

Given the proximity of the foreground elliptical galaxy at $z=0.34$,
\citet{jimenez-andrade20a} estimate that the luminosity of AzTEC2-A,
or MORA-3, is gravitationally lensed by a magnification factor of
$\mu=1.5$.  We scale physical quantities proportional to luminosity by
this magnification factor for MORA-3 for the rest of this paper.  Note
that this source sits in a large scale overdensity at $z=4.6$
identified and described in \citet{mitsuhashi21a}, further
corroborating the conjecture that high star-formation rate galaxies at
$z>4$ are highly clustered and good signposts for the most massive
overdense structures in the Universe \citep{casey16a,chiang17a}.

\subsubsection{MORA-4, a.k.a. MAMBO-9}

The fifth source of the sample, MORA-4, is known as MAMBO-9.  It was
originally detected at 1.2\,mm by the MAMBO instrument
\citep{bertoldi07a}, and subsequently detected at 1.1\,mm \citep[as
  AzTEC/C148;][]{aretxaga11a} and SCUBA-2 at 850\um\ \citep[as 850.43
  and COS.0059 in][respectively]{casey13a,geach17b}.  Several teams
identified the source as potentially high-$z$ based on being
undetected in the {\it Herschel} SPIRE bands; \citet{jin19a} initially
report a spectroscopic identification of $z=5.850$ based on a low
signal-to-noise 3\,mm spectral scan, confirmed by detection of
\cof\ and \htwoo\ in \citet{casey19a}.  MAMBO-9 is comprised of two
galaxies separated by 6\,kpc (1$''$) and both confirmed at $z=5.850$.
Our millimeter-derived photometric redshift for MAMBO-9 is $z_{\rm
  mm}=5.1^{+0.6}_{-0.8}$, only in slight tension with the measured
spectroscopic redshift.  MAMBO-9 is the most distant unlensed DSFG
found to-date and we refer the reader to \citet{casey19a} for a more
thorough characterization of the MAMBO-9 system.

\subsubsection{MORA-5 a.k.a. 850.13 or AzTEC\,C114}

The sixth source, MORA-5, has been identified at both 850\um\ and
1.1\,mm \citep[][named 850.13 and AzTEC\,C114,
  respectively]{casey13a,aretxaga11a}.  There is no spectroscopic
redshift for MORA-5, and there is no OIR-based photometric redshift
from either the COSMOS2015 or COSMOS2020 catalogs due to a lack of
counterpart in the near-infrared.  \citet{brisbin17a} offer a FIR
photometric redshift of $z_{\rm FIR}=5.3\pm3.2$, while noting that a
radio-FIR-based photometric redshift is consistently lower, based on
the source's detection at both 1.4\,GHz and 3\,GHz (respective
radio-FIR photometric redshifts of $z_{\rm 1.4GHz}=2.9^{+4.2}_{-0.4}$
and $z_{\rm 3.0GHz}=1.9^{+1.1}_{-0.3}$).  Our millimeter-derived
photometric redshift is $z_{\rm mm}=3.4^{+1.1}_{-0.9}$.  This source
is undetected at all wavelengths shortward of 3.6\,\um, including deep
CANDELS $H$-band and Ultra-VISTA $K$-band imaging. Our accompanying
paper, Manning~\etal, present a more detailed analysis of this source
and calculate a hybrid photometric redshift estimate for the source of
$z_{\rm phot}=4.3^{+1.5}_{-1.3}$ combining the \mmpz\ redshift with
direct extraction and refitting of OIR constraints using both {\sc
  eazy}\footnote{\software{{\sc eazy}
  \citep{brammer08a}}} and {\sc magphys}\footnote{\software{{\sc Magphys}
  \citep{da-cunha08a}}} approaches (see Manning~\etal\ for more
details). We adopt the Manning \etal\ hybrid photometric redshift for
the rest of this paper.

\subsubsection{MORA-6 a.k.a. 450.00}

The seventh source, MORA-6, was detected as the brightest
450\um\ source, named 450.00, in the 394\,arcmin$^2$ map in
\citet{casey13a}.  This source is also detected at 850\um\ with
SCUBA-2 and has ALMA continuum follow-up in both bands 6 and 7. It
lacks a spectroscopic redshift but does have a fairly well-constrained
OIR-based photometric redshift of $z_{\rm phot}=3.34^{+0.13}_{-0.12}$
from COSMOS2020.  Our millimeter-derived photometric redshift is
$z_{\rm mm}=2.5^{+1.6}_{-0.7}$, consistent with the OIR photometric
redshift.  

\subsubsection{MORA-7 a.k.a. 850.53}

The eighth source, MORA-7, has been previously identified at 850\um\ in
both \citet{casey13a} as 850.53 (lacking a corresponding detection at
450\um) and \citet{geach17b} as COS850.0035.  This source lacks a
spectroscopic redshift, but is detected with {\it Spitzer} and
Ultra-VISTA, rendering an OIR photometric redshift estimate of $z_{\rm
  phot}=2.85^{+0.24}_{-0.33}$. The source has dust continuum
observations from ALMA in both bands 6 and 7.  Our millimeter
photometric redshift is $z_{\rm mm}=2.3^{+2.7}_{-0.8}$, consistent
with the OIR photometric redshift.

\subsubsection{MORA-8 a.k.a. 850.09}

The ninth source of our sample, MORA-8, has been previously
identified at both 450\um\ and 850\um\ in \citet{casey13a} wherein it
was referred to as 850.09 and in \citet{geach17b} where it was named
COS850.0016.  The source lacks spectroscopic confirmation, but the OIR
photometric redshift from COSMOS2020 is fairly well-constrained as
$z_{\rm phot}=2.29^{+0.12}_{-0.08}$. This source also has ALMA band 7
continuum data, from which our millimeter-derived redshift is $z_{\rm
  mm}=3.0^{+2.2}_{-1.0}$.  Both ALMA 2\,mm and 870\,\um\ sources are
well aligned with the OIR counterpart.

\subsubsection{MORA-9}

The tenth source of the sample, MORA-9, has been detected at
850\um\ in \citet{simpson19a} at 4.4$\sigma$ significance; it had not
been detected in the earlier compilation of \citet{geach17b}.  Aside
from its detection at 2\,mm and 850\um, the source is also detected at
3.6\um\ and 4.5\um\ from {\it Spitzer} and at low significance in
Ultra-VISTA $K$-band imaging.  This source is undetected in all other
datasets, but its $K$-band counterpart renders it present in the
COSMOS2020 photometric catalogs with an OIR photometric constraint of
$z_{\rm OIR}=4.57^{+1.33}_{-1.22}$.  It is one of only two
$>$5$\sigma$ sources not already surveyed by ALMA at other wavelengths
(the other being MORA-12). The ratio of 850\um-to-2\,mm flux density
is highly suggestive of a high-$z$ solution; we derive a millimeter
photometric redshift of $z_{\rm mm}=5.5^{+0.6}_{-0.8}$ for this
source.  Manning~\etal\ describe this source's characteristics and
derive a hybrid photometric redshift for MORA-9 of $z_{\rm
  phot}=4.3^{+1.3}_{-1.0}$ which includes this millimeter photometric
redshift along with OIR constraints using photometric redshift fitting
techniques {\sc eazy}
and {\sc magphys}.  We adopt the Manning
\etal\ hybrid photometric redshift for the rest of this paper.

\subsubsection{MORA-10, a.k.a. 450.09}

The eleventh source of our sample, MORA-10, is well characterized as a
known DSFG detected at both 450\um\ and 850\um\ from SCUBA-2
\citep[named 450.09;][]{casey13a}, with a near-infrared spectroscopic
redshift of $z=2.472$ as reported in \citet{casey15a}.  The source is
one of many DSFGs in the COSMOS field that sits in a protocluster
environment at $z\sim2.5$, the same structure that motivated the dual spectral tunings for the MORA program
\citep{chiang15a,diener15a,casey15a,cucciati18a}. ALMA data exists for
450.09 in both band 7 and band 3, where the band 3 data confirm its
spectroscopic redshift via detection of CO(3-2).  The millimeter
photometric redshift for the source is $z_{\rm mm}=4.5^{+1.4}_{-1.0}$,
which is 2$\sigma$ discrepant with the measured spectroscopic redshift
(this discrepancy originates from the galaxy appearing to be a bit
colder than the average SED).

MORA-10 is unique among the MORA Survey sample for being particularly
luminous in its radio continuum ($S_{\rm 1.4\,GHz}=5.7$\,mJy and
$S_{\rm 3.0\,GHz}=3.2$\,mJy) with a rest-frame radio luminosity of
$L_{\rm 178\,MHz}=1.9\times10^{27}$\,W\,Hz$^{-1}$, nearly bright
enough to fit the local Fanaroff-Riley class II radio-loud AGN
\citep{fanaroff74a} definition at its redshift.  Fitting the existing
radio continuum measurements to a powerlaw, we derive a synchrotron
slope of $\alpha=0.85$; extended to the observed 2\,mm band data, we
estimate a total synchrotron contribution of $S_{\rm
  2mm,synch}=159\pm60$\,\uJy\ toward the total observed $S_{\rm
  2mm}=405\pm78\,$\uJy.  This implies that $\approx$\,39$\pm$17\%\ of
the total measured 2\,mm flux density is due to synchrotron processes.
While this clearly does not dominate the total flux density, an
absence of synchrotron emission in this source would have rendered it
below the 5$\sigma$ detection limit of our sample.  Even if the
synchrotron sloped varied somewhat, the source would have not been
detected at high significance from its dust emission alone once its
synchrotron component is subtracted from the total 2\,mm flux density.
Because the primary goal of this work is to identify thermal dust
emission at 2\,mm, we will exclude MORA-10 from analysis of population
statistics, like its contribution to the 2\,mm-selected galaxy
redshift distribution and star formation rate density.  For MORA-10's
far-infrared/millimeter SED fit, we adjust the 2mm flux density to
only account for the estimated dust continuum component, removing the
synchrotron component, as we are only fitting the dust SEDs in this
paper.

\subsubsection{MORA-11 a.k.a. AzTECC76 or COS.0194}

The penultimate source of the sample, detected at 5.2$\sigma$
significance, is MORA-11.  It is detected at 850\um\ in both
\citet{geach17b}, where it was named COS.0194, and in
\citet{simpson19a}. It is also detected at 1.1\,mm from
\citet{aretxaga11a} and in the {\it Herschel} {\sc SPIRE} bands at
250--500\um.  Broadly detected in the near-infrared through radio,
MORA-11 has a reported medium-band survey redshift of $z=3.17\pm0.12$
from the NEWFIRM Medium Band Survey \citep{whitaker11a}; the source
has additional Keck-NIRSPEC $K$-band spectroscopic observations from
\citet{marsan17a}, though no emission lines were detected.  This
redshift is in agreement with its OIR photometric redshift from
COSMOS2020 of $z_{\rm phot}=3.00^{+0.10}_{-0.13}$.  We measure a
millimeter photometric redshift of $z_{\rm mm}=4.3^{+4.9}_{-1.2}$ for
this source.  We adopt the medium-band photometric redshift for
MORA-11 for the rest of our analysis in lieu of the COSMOS2020
photometric redshift due to the improved precision offered by the
spectro-photometric analysis of \citet{marsan17a}.
Note that spectral analysis of our own MORA mosaic has tentatively
identified an emission line at 140.85\,GHz, which could be CO(5-4) at
$z=3.091$, consistent with the adopted redshift of $z=3.17\pm0.12$;
further analysis of this line identification is in progress
(Mitsuhashi \etal, in prep).

\subsubsection{MORA-12}

The last source in our sample, MORA-12, is detected at 5.02$\sigma$
significance; unlike the rest of the $>$5$\sigma$ sample, it has no
corresponding detection at 850\um.  While there is a 850\um\ source
\citep[with strong 24\um\ emission, named 850.77 in][]{casey13a}
9.6$''$ away from the ALMA 2\,mm position, we think it is unlikely
that the two are associated.  MORA-12 is also not detected in the {\it
  Herschel} COSMOS maps, as well as the GISMO map from
\citet{magnelli19a}. This source is the only source in the
$>$5$\sigma$ sample to lack detection in {\it Spitzer} IRAC at
3.6\um\ or 4.5\um.  There is no additional ALMA data at any other
wavelength at the position of MORA-12. With only one detection at one
wavelength, we are unable to derive a millimeter photometric redshift
for this source. Because it has not been detected at any other
wavelength, sits on the boundary of the P10-P20 mosaic, and given our
expected contamination rate of 1$\pm$1 false source detected above the
5$\sigma$ threshold, we conclude it most likely that MORA-12 is not
real and that it is a positive noise peak.

\subsubsection{Summary Characteristics of $>$5$\sigma$ Sample}

Of the thirteen sources detected at $>$5$\sigma$, we determine that
twelve of them are real 2\,mm-detected galaxies while the last and
least significant (MORA-12) is likely a positive noise peak.  Of the
twelve real sources, all are detected with both {\it Spitzer} IRAC and
SCUBA-2 at 850\,\um.  Despite all being previously detected, not all
sources had been identified as high-$z$ candidates.  One of the twelve
sources (MORA-10) is thought to have a substantial contribution
(39$\pm$17\%) from synchrotron radio emission to its 2\,mm flux
density, rendering the sample of purely dust-selected 2\,mm sources
with only eleven galaxies.

Three of the twelve sources are spectroscopically-confirmed at
$z=2.472$ (MORA-10, a.k.a. 450.09, with the synchrotron component),
$z=4.625$ (MORA-3, a.k.a. AzTEC-2), and $z=5.850$ (MORA-4,
a.k.a. MAMBO-9); of the remaining nine sources, eight have some form
of OIR-based photometric redshift \citep[][Weaver, Kauffmann \etal,
  submitted]{laigle16a,marsan17a,gomez-guijarro18a}.  The last source
(MORA-5) lacks an OIR counterpart and any redshift constraint; MORA-5
along with MORA-9, whose OIR phot-$z$ is highly uncertain, have hybrid
photometric redshift fits provided in our accompanying manuscript,
Manning \etal

Four of the 11 dust-selected sources (36\%) are ``OIR-dark,'' meaning
they lack near-infrared $H$-band counterparts in deep WFC3 imaging,
which in this case is CANDELS-COSMOS data reaching a 5$\sigma$ point
source depth of 27.15 \citep{koekemoer11a}.  These sources are MORA-3
(a.k.a. AzTEC-2, spectroscopically confirmed at $z=4.63$), MORA-4
(a.k.a. MAMBO-9, spectroscopically confirmed at $z=5.85$), MORA-5 and
MORA-9.  As these represent the potential highest redshift subset of
the 2\,mm-selected sample, a more thorough analysis of them is given
in the accompanying paper by Manning~\etal. It appears that MORA-3
(AzTEC-2) is the only galaxy in the sample that is gravitationally
lensed \citep[$\mu=1.5$;][]{jimenez-andrade20a}.

The majority of the MORA 2\,mm-selected galaxy sample is consistent
with relatively little AGN activity, or AGN activity that does not
dominate the galaxies' bolometric luminosities.  We investigate the
sample's AGN content by analyzing X-ray imaging, radio emission, and
mid-infrared emission.  None of the $>5\sigma$ sample is detected in
the deep COSMOS {\it Chandra} or {\it XMM} data, though the
sensitivity of such X-ray surveys is rather shallow at $z>3$.  The
galaxies' radio luminosities measured at 3\,GHz \citep{smolcic17a},
and in the somewhat shallower 1.4\,GHz data, are in line with
expectation for synchrotron emission generated via star-formation
processes instead of AGN \citep[e.g.][]{yun01a,ivison10a,delhaize17a}.
The one source that proves an exception to this is MORA-10, which
appears to be radio-loud and whose 2\,mm emission is partially
dominated by such non-thermal emission mechanisms.  In the
mid-infrared, five of eleven\footnote{MORA-3, a.k.a. AzTEC-2 is too
severely blended with foreground galaxies to discern whether or not it
is 24\um-luminous.}  (5/11=45\%) are 24\um-detected.  Aside from
MORA-10 which clearly has an AGN, three of the other four
24\um-detected galaxies are the lowest redshift sources in the sample
(MORA-6, -7, and -8), which, at those redshifts, could be due to
either AGN or star-formation via PAH emission.  The last
24\um\ detection, MORA-1, is a blend of several sources $z\sim3.6-4.0$
and the centroid of the emission is not precisely constrained to the
2\,mm source in question.  While AGN overall seem non-dominant in this
sample of 2\,mm-selected DSFGs, it should be noted that recent
modeling work has shown that AGN can contribute substantially to the
heating of host galaxy-scale dust, even in IR-luminous galaxies
without prominent AGN \citep{mckinney21a}; such effects are difficult
to account for without constraining observations, thus we do not
account for it directly in this work.

\subsection{Marginal Sources with $<$5$\sigma$ significance}

Below the 5$\sigma$ threshold, contamination from positive noise peaks
becomes a significant concern.  There are 87 sources identified in the
MORA P10-20 mosaic and 11 sources found in the P03 mosaic at
$4<\sigma<5$ significance, or 98 in total.  Their positions and
characteristics are given in Table~\ref{tab:marginal}.  We determine
which of these sources are most likely to be real by testing for
coincident identification at other wavelengths, for example, in the
COSMOS2015 catalog, the COSMOS2020 catalog, the 3\,GHz radio continuum
survey \citep{smolcic17a}, or at 850\um\ from SCUBA-2
\citep{simpson19a}.

Given the high source density of galaxies in the OIR catalogs, there
is a 8\%\ chance of a random point in our map aligning with a
COSMOS2015 counterpart within 1$''$ \citep[which is conservative
  maximum physical scale on which we see spatial offsets of obscured
  and unobscured emission in galaxies, e.g.][ with more characteristic
  scales of 0$\farcs$1--0$\farcs$3, as in
  \citealt{cochrane19a}]{biggs08a}. The probability of a chance
alignment with a source in COSMOS2020 is slightly higher within 1$''$,
$\sim$13\%, given its increased depth at near-infrared wavelengths.
Out of a sample of 98 marginal sources, this would suggest a total of
8$\pm$3 sources matched at random to COSMOS2015 and 13$\pm$4 sources
matched at random to COSMOS2020. We find a total of 12 matches to
COSMOS2015 and 17 matches to COSMOS2020 within the MORA $4<\sigma<5$
2\,mm sample (11/12 overlap between COSMOS2015 and COSMOS2020),
suggesting that there are very few real sources ($<1/3$) in this
marginal sample and none that can be directly identified reliably.

We also tested the correlation between the marginal sample with
850\um-selected DSFGs observed by SCUBA-2 \citep{simpson19a}; we find
that there should be a 2.9\%\ chance of random alignment between a
$>3.5\sigma$ SCUBA-2 source (within the SCUBA-2 15$''$ FWHM beam) and
a marginal source in our 2\,mm catalog. The rate of false positives is
lower for 850\um\ counterparts than for OIR counterparts because the
sky density is much lower for 850\um\ sources. We find twelve sources
that are spatially coincident with a 850\um\ SCUBA-2 source (a total
of 12\%\ of the marginal sample), well in excess of the anticipated
$\sim$3\%.  Contrary to our findings using OIR counterparts, this
demonstrates a true excess of real sources within the marginal sample.
Of these twelve sources, three have secure IRAC 3.6\um\ counterparts
matched to the positions of the 2\,mm emission, corroborating their
identification as real sources.  

We further investigate radio continuum counterparts for the marginal
sample using the 3\,GHz radio continuum map in COSMOS presented in
\citet{smolcic17a}.  Some of the sources are also covered by the
deeper COSMOS-XS survey \citep{van-der-vlugt21a,algera20a}.
We find that there is a 0.6\%\ chance of random alignment between a
marginal MORA source and a 3\,GHz $>5\sigma$ detected source within
1$''$ of one another.  The rate of false positives is lowest for radio
counterparts due to their increased rarity on the sky, in addition to
their precisely measured positions. We find that six MORA 2\,mm
sources are spatially coincident with a 3\,GHz radio continuum source
(a total of 6\%\ of the marginal sample), a factor of ten higher than
the anticipated $\sim$0.6\%.  The strength of this excess is such that
the 3\,GHz-detected subset can be reliably identified as having real
2\,mm emission.  Nevertheless, the sample is somewhat limited in size
to analyze in further detail.

We include an abbreviated table of marginally-detected $4<\sigma<5$
2\,mm sources in Table~\ref{tab:marginal} for reference and note which
sources have counterparts at which wavelengths.  However, due to high
contamination rates and a similarly high incompleteness rate in this
flux density regime, we do not analyze the sample further.  Later in
\S~\ref{sec:scuba2} we analyze the 2\,mm emission properties of any
DSFGs that have been independently observed by ALMA (not a part of
MORA) in the field, some of which overlap with this marginal sample.

The lack of reliability of the marginal sample is further verified by
analyzing the number of negative peaks in the mosaics between
4--5$\sigma$, of which there are a total of 106.  If there were a
significant population of real sources embedded in the
marginally-identified sample, the positive peaks (98) would likely
outnumber the negative peaks (106).

\section{Models of the 2mm Universe}\label{sec:models}

We make use of several cosmological semi-empirical and empirical
models of the 2\,mm-luminous Universe to draw comparisons with the
MORA dataset.  A brief description of each model dataset follows.

First, we compare to the Simulated Infrared Dusty Extragalatic Sky
(SIDES) model \citep{bethermin17a}, which builds galaxies' SEDs from
their stellar masses and star formation rates (assuming a bimodal
population of main sequence galaxies and starbursts) and which updates
the 2SFM (2 Star Formation Modes) galaxy evolution model
\citep{bethermin12a,sargent12a} to analyze the impact of clustering on
IR map analysis.  The 2SFM model posits that there is a bimodal
population of star forming galaxies: those that are on the main
sequence and starbursts that have elevated specific star formation
rates; the 2SFM model uses this framework to model all galaxies. We
make use of the full SIDES 2\,deg$^2$ lightcone in our analysis, both
to sample cosmic variance and understand possible trends for
2\,mm-selected galaxies on angular scales larger than the MORA survey.

Second, we compare our results to the {\sc Shark} semi-analytic model
of galaxy formation \citep{lagos18a}. By using the SED code {\sc
  ProSpect} \citep{robotham20a}\footnote{\software{{\sc ProSpect}
  \citep{robotham20a}}} and the radiative transfer analysis of the
{\sc EAGLE} hydrodynamical simulations of \citet{trayford20a}, the
{\sc Shark} model was successfully able to predict the ultraviolet to
far-infrared emission of galaxies over a wide range assuming a
universal IMF \citep{lagos19a}\footnote{See also \citet{lovell20a} and
\citet{hayward21a} for similar analysis using the {\sc Simba} and {\sc
  Illustris-TNG} hydrodynamical simulations from \citealt{dave19a} and
the radiative transfer code {\sc Powderday} from
\citet{narayanan20a}\footnote{\software{{\sc Powderday}
  \citep{narayanan20a}}}.}.  \citet{lagos20a} presents detailed
predictions for the (sub-)millimeter galaxy population, including
2\,mm number counts and redshift distributions.  We make use of the
full 108\,deg$^2$ {\sc Shark} lightcone for our comparisons.

Third, we compare our work to the semi-empirical model for dust
continuum emission published by \citet{popping20a}; that work
primarily focused on comparison of 850\um\ and 1.1\,mm number counts
and redshift distributions with the ASPECS survey results
\citep{aravena20a,gonzalez-lopez19a,gonzalez-lopez20a}.  Using the
same methodology of \citeauthor{popping20a}, several lightcones from
the {\sc UniverseMachine} framework \citep{behroozi18a}, which is
grounded in the {\it Bolshoi-Planck} dark matter simulation
\citep{klypin16a,rodriguez-puebla16a}, are stitched together to form a
7.7\,deg$^2$ lightcone.  Dark matter halos are populated with galaxies
calibrated to observationally constrained relations (in stellar mass
and SFR distributions), and dust continuum characteristics are applied
using the relations derived in \citet{hayward11a,hayward13a} using
the {\sc SUNRISE} \citep{jonsson06a} dust radiative transfer code
as a function of SFR and dust mass.

Lastly, we compare to the empirical model predictions for the submillimeter sky
presented by \citet{casey18a,casey18b} and expanded on in
\citet{zavala18a} and, finally, in Z21.  The key difference between
the \citet{casey18a} models and the cosmological semi-empirical models is the
built-in flexibility to {\it test} different hypothetical evolving
infrared luminosity functions against data. Not tethering this model
directly to any cosmological simulation renders it a tool to interpret data
that may be discrepant with such simulations.  \citet{casey18a} use
it to present the hypothetical (sub)mm sky in two diametrically
opposed Universes: the `dust poor' Universe (model A) in which there
is steep evolution from $z\sim7$ to $z\sim2$ in the characteristic
number density of the IRLF ($\Phi_\star$), and the `dust rich'
Universe (model B), in which the evolution of the characteristic
number density is much more shallow.  The `dust poor' Universe
effectively translates to a very minor contribution of intense, dusty
starbursts to cosmic star formation beyond $z\simgt5$ ($<$10\%), while
they would dominate cosmic star formation at similar epochs in the
`dust rich' Universe ($>$90\%).  These models are built to capture two
extremes.  \citet{zavala18a} uses 3\,mm number counts from the ALMA
archive to argue for a solution between these two extremes.
Our accompanying paper, \citet{zavala21a} presents an update to
\citet{zavala18a} using MORA 2\,mm number counts, updated 3\,mm
archival counts, as well as deep 1\,mm number counts from the ASPECS
survey \citep{aravena20a,gonzalez-lopez19a,gonzalez-lopez20a}.  The
expanded dataset used to constrain the model in \citet{zavala21a}
relative to \citet{zavala18a} has resulted in a change of the
predicted $z\simgt2$ evolution of $\Phi_\star$, the number density of DSFGs
at early times, from $\Phi_\star\propto(1+z)^{-4.2}$ to
$\Phi_\star\propto(1+z)^{-6.5}$ \citep[see][for more details]{zavala21a}.

\section{Results}\label{sec:results}

\begin{figure}
\includegraphics[width=0.99\columnwidth]{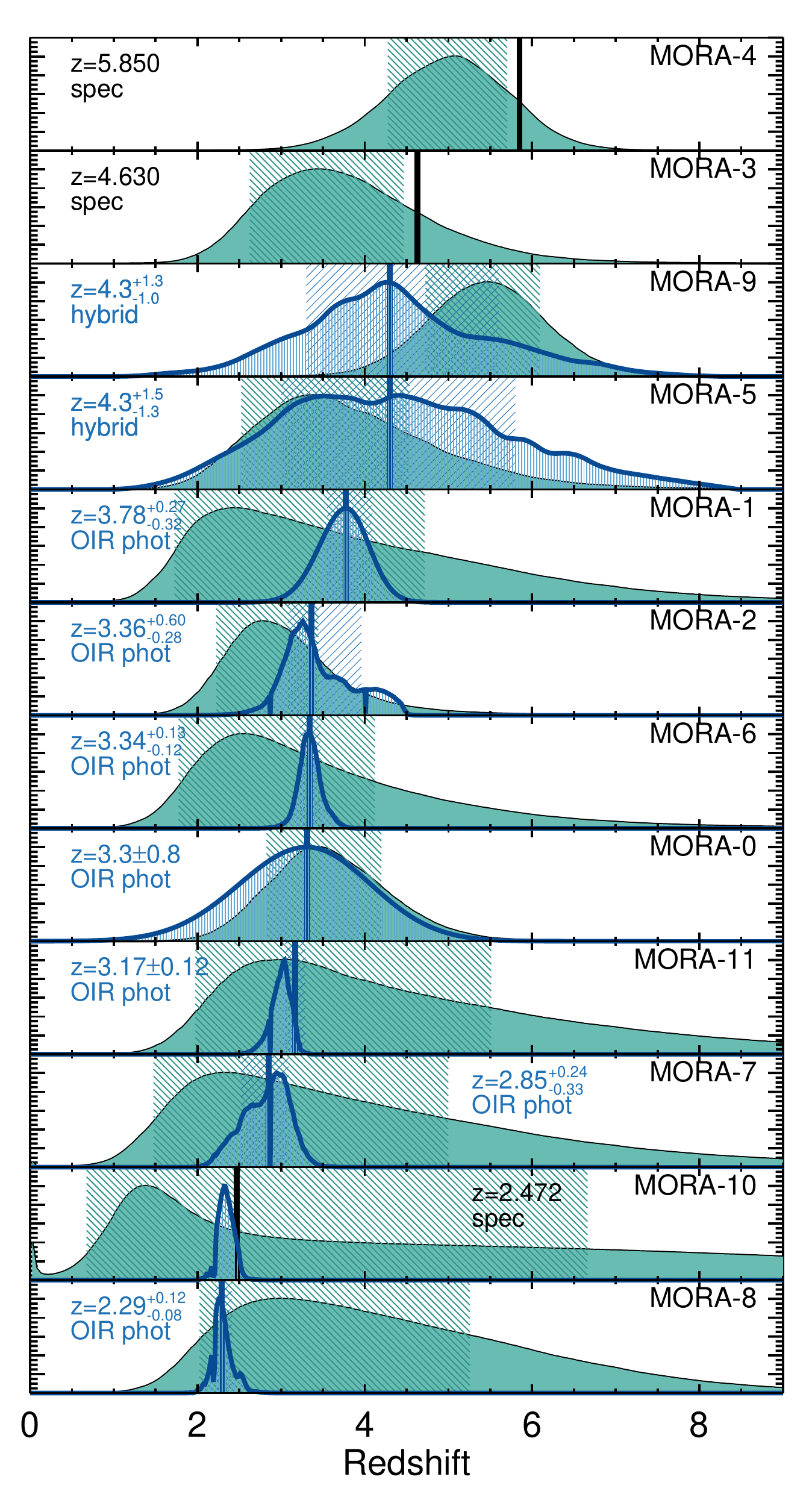}
\caption{Comparison of the redshift probability density distributions
  derived from: the FIR/mm using the \mmpz\ technique (teal
  distributions), OIR photometric redshifts (blue distributions), and
  spectroscopic constraints (black lines).  The hashed regions of
  those same colors indicate the 68 percentile inner confidence
  interval.  Sources are ordered by redshift, from highest (top) to
  lowest (bottom).  The two unconfirmed OIR-dark sources with the most
  uncertain redshift constraints are MORA-5 and MORA-9, described in
  more detail in our accompanying manuscript, Manning \etal.  The
  redshifts for these two galaxies are derived jointly using OIR and
  millimeter constraints, here called hybrid.  }
\label{fig:mmpz}
\end{figure}

\subsection{Redshift Constraints \&\ Distribution}

Redshift constraints for our sample are heterogeneous, ranging from
firm spectroscopic confirmations to limited far-infrared/millimeter
photometric constraints. We exclude MORA-10 (which has $z_{\rm
  spec}=2.472$) from analysis of the sample's redshift distribution
due to the suspected contribution of synchrotron emission to its 2\,mm
flux density, without which it would not have been detected above
5$\sigma$ in our data.

\begin{figure*}
\includegraphics[width=0.99\columnwidth]{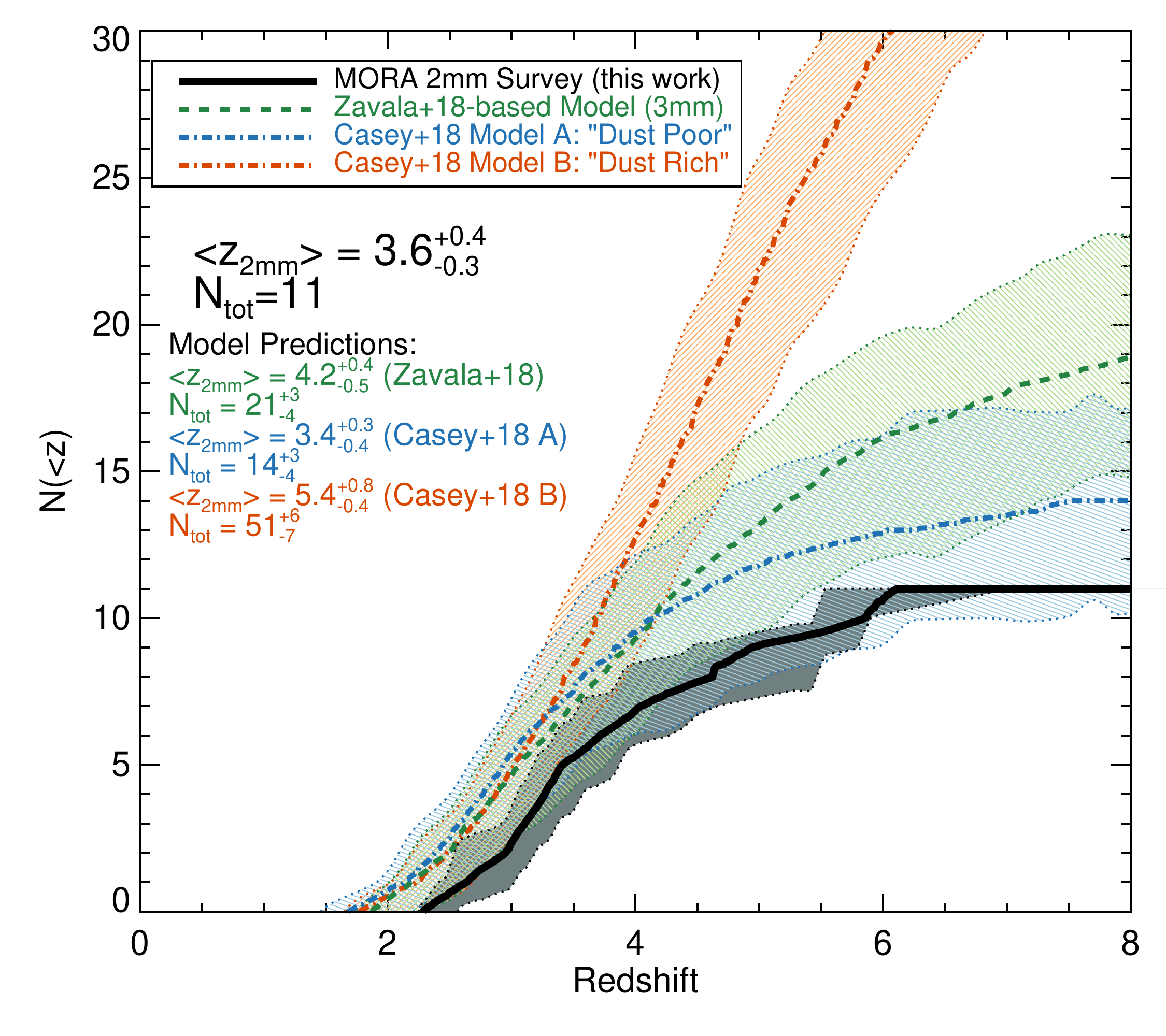}
\includegraphics[width=0.99\columnwidth]{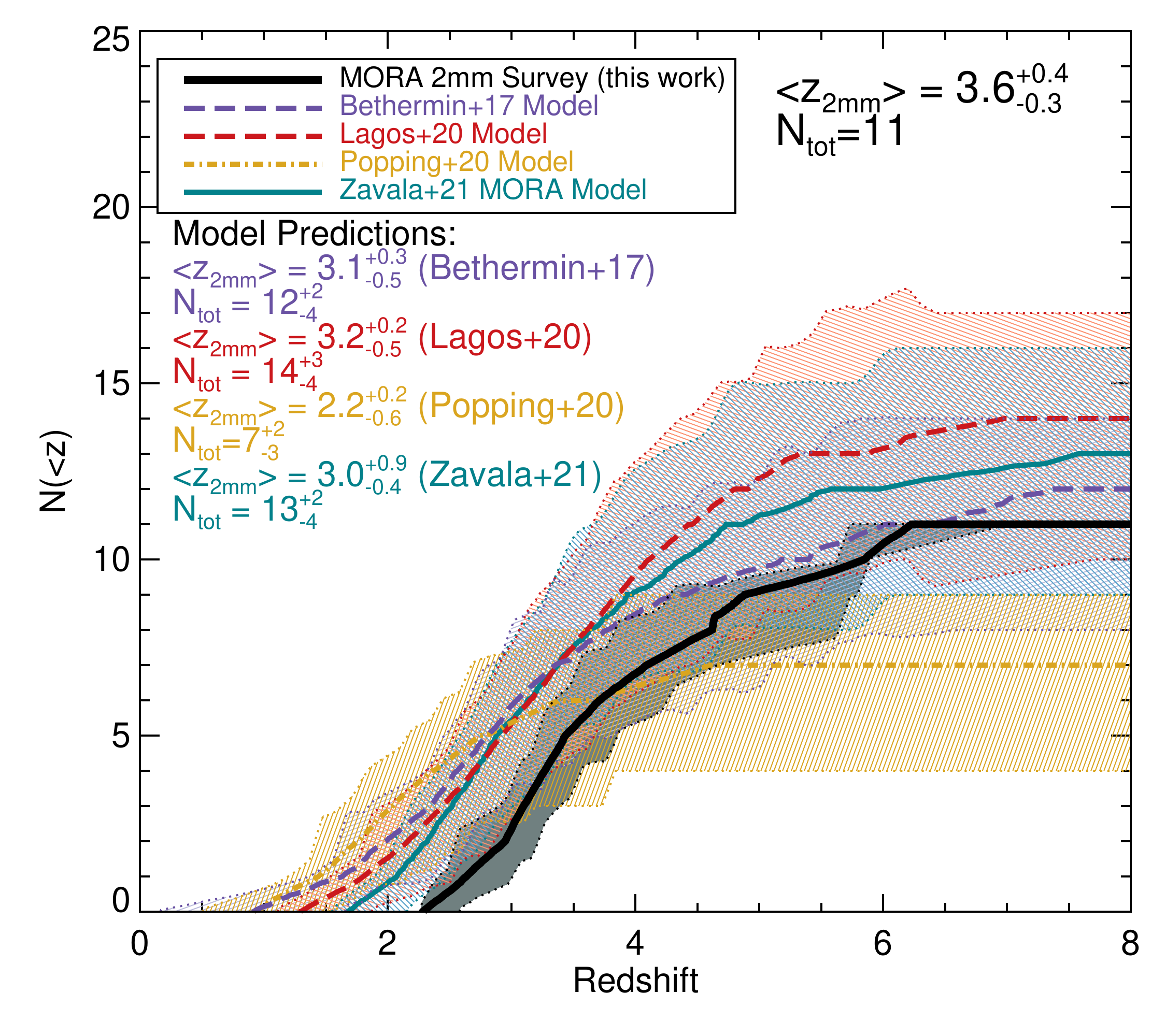}
\caption{The measured cumulative redshift distribution for the eleven
  dust-continuum dominated 2\,mm galaxies detected in the MORA maps
  compared to literature models.  Each galaxy's redshift probability
  density distribution is coadded here to reflect the uncertainty in
  the total CDF; we measure the uncertainty on the median redshift
  estimate, $\langle z_{\rm 2mm}\rangle=3.6^{+0.4}_{-0.3}$, through
  Monte Carlo draws from the CDF.  Literature model redshift CDFs are
  measured by drawing galaxies from the model with matched flux
  densities to our real sources (within uncertainties) and inferring
  the redshift distributions of those sources. The left panel compares
  to three hypothetical universe predictions using the
  \citet{casey18a} model framework: the Casey \etal\ ``dust poor''
  model A Universe (blue), the ``dust rich'' model B Universe
  (orange), and the \citet{zavala18a} 3\,mm-number counts based model
  (green).  The right panel compares with three other models: the
  SIDES model \citep[purple;][]{bethermin17a}, the SHARK model
  \citep[red;][]{lagos20a}, and the most recent 1\,mm/2\,mm/3\,mm
  updated number counts-based model from Z21.  Note that the Z21 model
  is largely based on our MORA maps, but it does not use any source
  redshift information as input, and thus its redshift distribution
  predictions are derived independent of these data.}
\label{fig:nz}
\end{figure*}

From the most to least reliable, two (2/11=18\%) have spectroscopic
redshifts,
seven (7/11$\approx$64\%) have OIR photometric redshifts, and two
(2/11$\approx$18\%) have hybrid FIR/mm and OIR photometric redshift
constraints.  It is somewhat interesting to note that the two
spectroscopic redshifts are the highest redshift sources in the
sample. Figure~\ref{fig:mmpz} shows the redshift constraints for each
source (including MORA-10) in order of increasing redshift from bottom
to top and how consistent existing constraints are with FIR/mm derived
redshifts from the \mmpz\ tool described in \citet{casey20a}. The
millimeter photometric redshifts serve as a sanity check on the
tighter constraints given by other methods.

Figure~\ref{fig:nz} shows the cumulative redshift distribution for the
entire sample in two panels.  Given the heterogeneous constraints,
each source's probability density distribution in $z$ is coadded and
shown as a cumulative distribution function (CDF) to make clear the
relative fraction of the sample below or above a certain redshift
threshold.  Accounting for the uncertainties in the individual
redshift constraints through Monte Carlo draws from the CDF, we
measure the median redshift of the sample as $\langle z_{\rm
  2mm}\rangle=3.6^{+0.4}_{-0.3}$.  The variance in the redshift CDF is
shown in gray and encompasses a 68\%\ ($\pm1\sigma$) minimum
confidence interval.  We find that 77$\pm$11\%\ of the distribution
lies at $z>3$ and 38$\pm$12\%\ lies at $z>4$.

The median redshift of 2\,mm-selected galaxies has been measured twice
before in the literature, using the GISMO instrument on the
single-dish IRAM 30\,m telescope.  \citet{staguhn14a} measure a median
redshift of $\langle z_{\rm 2mm}\rangle=2.9\pm0.9$ for sources in the
GISMO Deep Field in GOODS-N while \citet{magnelli19a} measure a median
redshift of $z\sim4$, though only for five sources and four sources
respectively with S/N$>$4.  Both are in agreement with our measured
median redshift of $\langle z_{\rm 2mm}\rangle=3.6^{+0.4}_{-0.3}$.

Figure~\ref{fig:nz} also shows the predicted cumulative redshift
distributions for the MORA dataset from several models in the
literature, described in \S~\ref{sec:models}.  For all models, we have
generated the cumulative redshift distribution by sampling simulations
over the 184\,arcmin$^2$ area of the MORA survey.  Some of the models
have been simulated over larger areas (e.g. {\sc Shark}, SIDES, and
{\sc UniverseMachine}) while the Casey \etal\ and Zavala
\etal\ empirical models have multiple realizations the same size of
the MORA survey.  Sources in each model dataset then have an RMS noise
assigned to them following the heterogeneous distribution of RMS noise
in the MORA maps to best mimic the data.  We retain simulated sources
that were detected at or above 5$\sigma$ significance.  Sources in
these simulations are assumed to be point sources, as the probability
of them being spatially resolved on scales larger than
1$\farcs$5$\approx$10\,kpc is unlikely at these redshifts.  This
procedure is repeated 1000 times to constrain the uncertainties in the
total number of selected sources and their redshift distribution.

The predictions of the \citet{casey18a} models A and B, as well as the
\citet{zavala18a} model, are shown in the left panel of
Figure~\ref{fig:nz}.  All are an overestimate of the number of sources
in the true data, though the redshift distribution for the model A
from \citeauthor{casey18a} does agree with our data within
uncertainties.  The right panel of Figure~\ref{fig:nz} shows four
models, one of which is the updated empirical model from our
accompanying paper, Z21.  Z21 uses number counts from the MORA survey,
alongside 1.2\,mm number counts from ASPECS and 3\,mm number counts to
derive constraints on the high-$z$ IRLF, even in the absence of direct
redshift measurements of individual galaxies.  Thus it is important to
emphasize that the predicted redshift distribution from Z21
constitutes an independent prediction, despite the use of MORA number
counts to generate the model constraints.

Also shown on the right side of Figure~\ref{fig:nz} are the results
from the three semi-empirical models grounded in cosmological simulations: SIDES
\citep{bethermin17a}, {\sc Shark} \citep{lagos20a}, and {\sc
  UniverseMachine} \citep{popping20a}.
All models, with exception of the \citeauthor{popping20a} model which is only in slight tension,
accurately predict the number of sources to be found in MORA within
uncertainties due to cosmic variance.
Similarly, their redshift distributions are also in broad agreement.
Nevertheless, within the uncertainties on the redshift distributions,
there is a systematic offset where most models predict median
redshifts of $\langle z_{\rm 2mm}\rangle=2.2-3.2$ versus the observed
$\langle z_{\rm 2mm}\rangle=3.6$.  The {\sc Shark} model comes closest
to the measured median redshift, though we note the SIDES model has a
more prominent high-redshift tail to its distribution similar to the
MORA distribution.  Only further data can reduce the uncertainties on
the redshift distribution and discriminate between these models.

In summary, comparing the measured redshift distribution and number
counts of our MORA results with models suggests that, indeed, the
prevalence of dusty star-forming galaxies beyond $z\simgt5$ is
inherently low.  This is well aligned with predictions from
cosmological simulations, which fundamentally limit the build up of
massive star-forming galaxies in the first $\sim$2\,Gyr of the
Universe's history directly from the volume density of massive halos at that epoch.

\begin{figure}
\includegraphics[width=0.99\columnwidth]{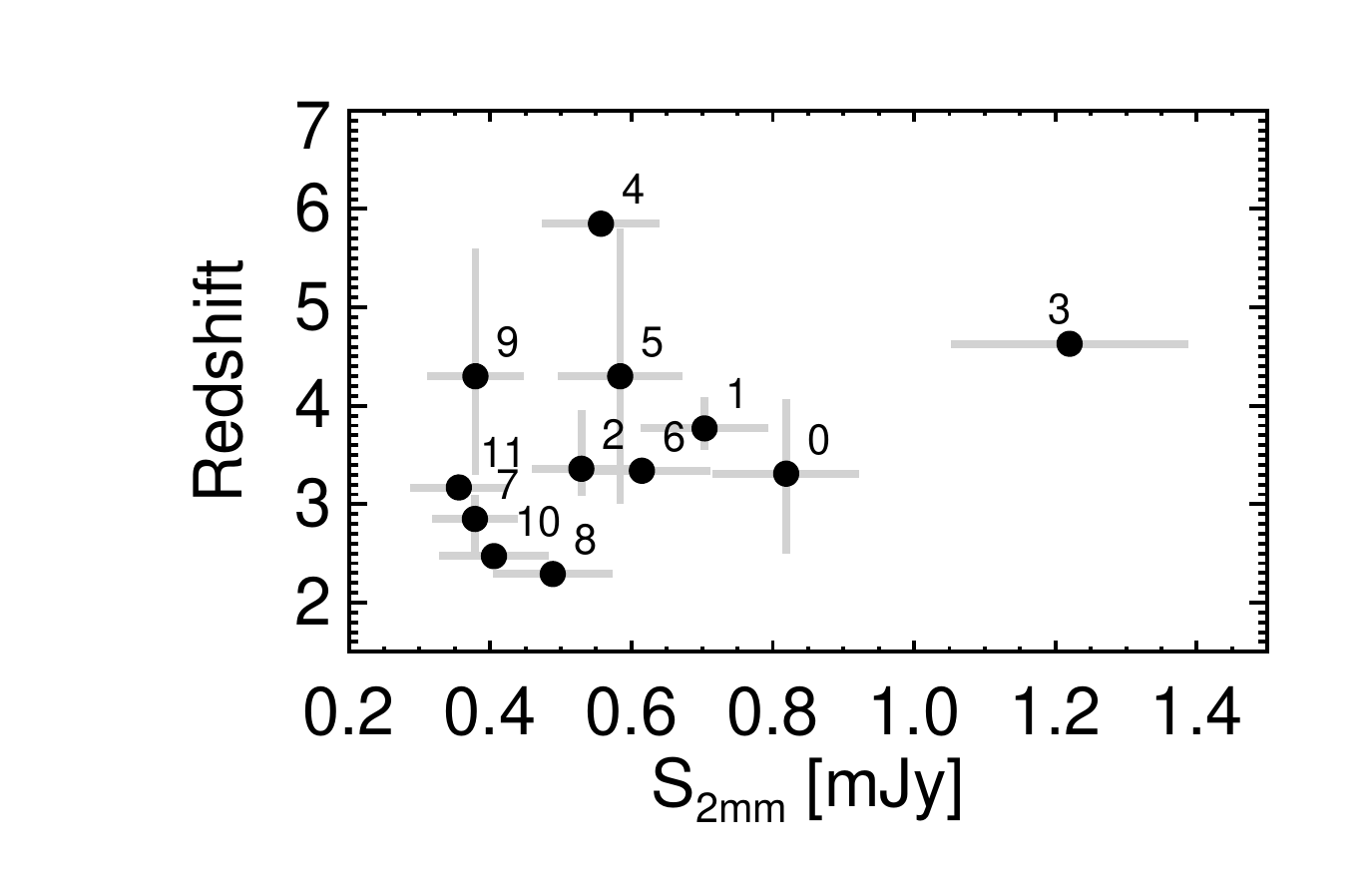}
\includegraphics[width=0.99\columnwidth]{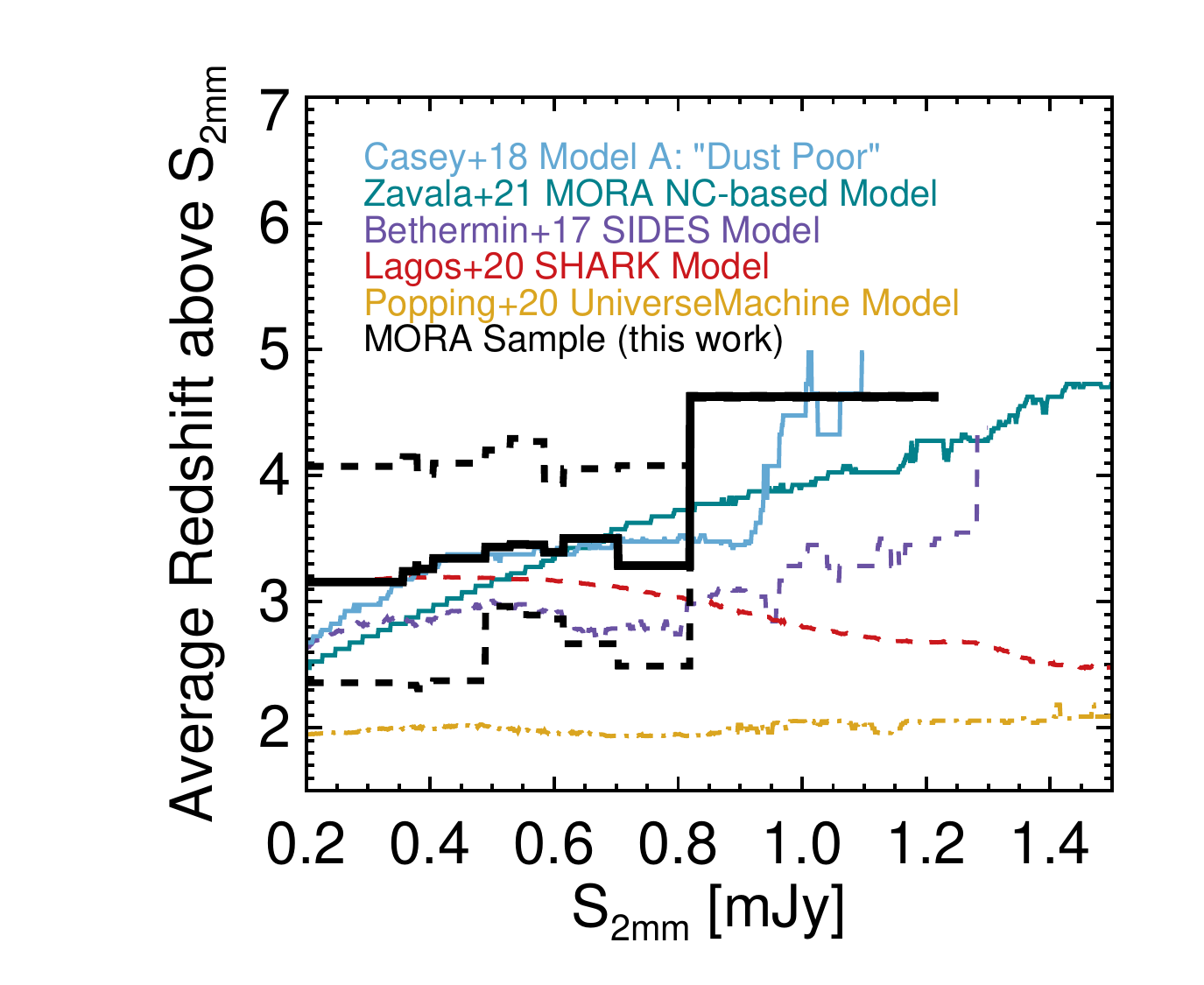}
\caption{{\it Top:} 2\,mm flux density against best redshift
  constraints for the MORA Sample; sources are labeled by their IDs.
  {\it Bottom:} the cumulative median redshift of the MORA sample
  above a given 2\,mm flux density (thick black line, dashed lines
  enclose the inner 68$^{th}$-percentile).  The median redshifts of
  five models are overplotted for comparison: the SIDES model from
  \citet{bethermin17a} in purple (dashed line), the SHARK model from
  \citet{lagos20a} in red (dashed line), the {\sc UniverseMachine}
  model from \citet{popping20a} in yellow (dot-dashed line), the
  dust-poor model from \citet{casey18a} in light blue (solid), and the
  Z21 model in teal (solid).  The dearth of sources $>$1\,mJy in this
  dataset limit our ability to distinguish between models, with
  exception of the \citet{popping20a} model that underestimates the
  redshifts of 2\,mm-selected galaxies of all flux densities.}
\label{fig:s2mmz}
\end{figure}

Despite this verification of models, there are still subtleties to
these measurements worth exploring that could help refine early
Universe DSFG volume densities further.  For example,
Figure~\ref{fig:s2mmz} shows the relationship between 2\,mm flux
density and redshift for the MORA sample and model predictions.  The
\citet{popping20a} model predicts a low average redshift for
2\,mm-luminous sample that does not vary as a function of flux
density. While SIDES and {\sc Shark} accurately predict the redshift
distribution and source density in the MORA map, they offer different
predictions of {\it which} sources are most likely to sit at the
highest redshifts; in other words, while SIDES predicts sources to sit
at higher redshifts with brighter 2\,mm flux density, {\sc Shark}
predicts that the average redshift for bright ($S_{\rm
  2mm}\simgt0.5$\,mJy) 2\,mm sources is relatively low compared to
faint ($S_{\rm 2mm}\simlt0.5$\,mJy) sources.  Our data -- though
limited severely by the small sample size -- suggest that brighter
sources tend to sit at higher redshifts \citep[see
  also][]{koprowski17a}.  This is also in line with the predictions of
both the \citet{casey18a} Model A and Z21 adjusted number counts-based
MORA model, both of which show the highest average redshift for
$S_{\rm 2\,mm}\simgt1$\,mJy sources.

The origins of these second order discrepancies between model
predictions and data will inevitably require larger samples of 2\,mm
galaxies identified over wider areas.  Nevertheless, we discuss
possible origins of such discrepancies (and other potential
degeneracies in our conclusions) later in \S~\ref{sec:varymodel};
possible causes include evolution in the faint-end slope of the IRLF,
possible evolution of galaxies' dust emissivity spectral indices, or
galaxies' bulk luminosity-weighted dust temperatures.

\subsection{Direct SED fits using redshift priors}\label{sec:sedfits}

We fit the FIR through millimeter spectral energy distributions of the
2\,mm-detected sample using a modified blackbody plus a mid-infrared
powerlaw; this procedure is a modified version of the fitting
technique described in \citet{casey12a} and will be described in full
in a forthcoming paper (Drew \etal, in preparation).  The difference
with the \citet{casey12a} analytical approximation is that the
blackbody and mid-infrared powerlaw are added together as a piecewise
function (where the mid-infrared powerlaw is joined at the point where
the slope of the blackbody is equal to the powerlaw index $\alpha_{\rm
  MIR}$).  The functional form of the fit used is:
\begin{equation}
S_\nu(\lambda) = \left\{
\begin{array}{lr}
N_{\rm pl} \lambda^{\alpha} & : \lambda\ |\ \frac{\partial \log S_\nu}{\partial \log \lambda} > \alpha \\
N_{\rm bb} \frac{(1-e^{-(\lambda_0/\lambda)^\beta})\lambda^{-3}}{e^{hc/\lambda kT}-1} & : \lambda\ |\ \frac{\partial \log S_\nu}{\partial \log \lambda} \le \alpha \\
\end{array}
\right.
\label{eq:sed}
\end{equation}
Here $\lambda$ is the rest-frame wavelength, $T$ the modeled dust
temperature, $\lambda_0$ the wavelength where the dust opacity is
$\tau=1$, $\alpha$ is the slope of the mid-infrared powerlaw, and
$\beta$ is the observed integrated emissivity spectral index.  $N_{\rm
  pl}$ and $N_{\rm bb}$ are normalization constants whose value is
tied to one another such that the SED is contiguous at the point of
transition, i.e. where $\partial \log S_\nu/\partial \log
\lambda=\alpha$.  Both $N_{\rm pl}$ and $N_{\rm bb}$ are tied to \lir,
defined as the integral of the SED in $S_{\nu}$ between 8--1000\,\um.

We adopt the general opacity model for all SEDs; for lack of direct
constraints on the dust opacity in these systems, we adopt
$\lambda_0=200$\,\um.  This is broadly consistent with what is seen in
high-\lir\ systems where $\lambda_0$ can be measured
\citep[e.g.][]{conley11a,spilker16a}; even if this presumption is
incorrect in the case of these systems, the exact adopted value does
not impact the results of this work.  Specifically, it only impacts
the functional relationship between $T$ and $\lambda_{\rm peak}$,
where the latter quantity is insensitive to choice of $\lambda_0$; for
example, \citet{cortzen20a} show that the measured dust temperature
for GN20 may vary by 20\,K depending on the adopted $\lambda_0$,
whereas the measured $\lambda_{\rm peak}$ remains unimpacted.  So
while it is known there are degeneracies in measured quantities
$\beta$, $T$, and $\lambda_0$ that cannot be resolved with the limited
data we have on this sample in full, it is worth emphasizing that
measurement of $\lambda_{\rm peak}$ is independent and not sensitive
to choice of opacity model.
For the purposes of SED fitting, the photometric points are allocated
an additional 10\%\ calibration error, representing the degree to
which the absolute value of the flux density across the
(sub)millimeter regime is known.

\begin{figure}
\includegraphics[width=0.99\columnwidth]{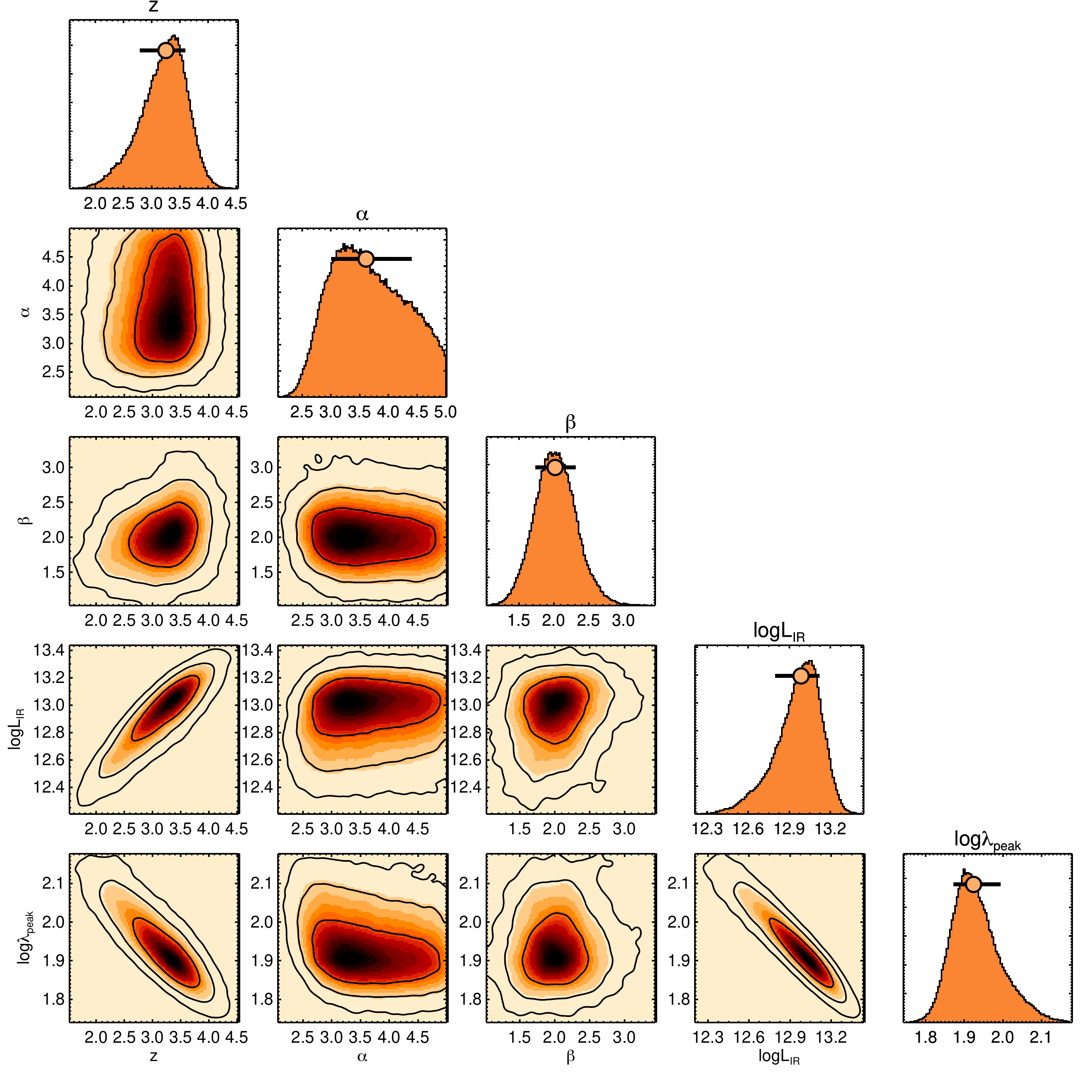}
\caption{An illustrative example of the two-dimensional joint
  posterior distributions for the IR SED fit to source MORA-2. The
  parameters fit are redshift $z$, the mid-infrared powerlaw slope
  $\alpha$, the emissivity spectral index $\beta$, the IR luminosity
  $L_{\rm IR}$, and the rest-frame peak wavelength $\lambda_{\rm
    peak}$.  MORA-2 shows strong covariance between $z$, $L_{\rm IR}$,
  and $\lambda_{\rm peak}$.  The resulting best-fit SED for MORA-2 is
  shown in Figure~\ref{fig:seds}.}
\label{fig:examplefit}
\end{figure}

We converge on best-fit SEDs using a Markov Chain Monte Carlo routine.
In all cases, we use the best available redshift probability
distribution as a prior.  Each SED is handled individually, whereby
the choice to fix a variable or let it vary depends on the degree of
photometric constraints relevant to each variable; for example,
$\alpha_{\rm MIR}$ is fixed to $\alpha_{\rm MIR}=4$ if there are no
detections at rest-frame mid-infrared wavelengths, and $\beta$ is
fixed to $\beta=1.8$ \citep{planck-collaboration11a} if there are no
more than 2-3 photometric points on the Rayleigh-Jeans tail of the
blackbody emission in the millimeter regime.  The choice of fixed
$\alpha_{\rm MIR}=4$ is motivated by fits to lower redshift DSFGs with
constrained mid-IR emission \citep[e.g. at $z\sim2$;][]{pope08a}, while
the choice of $\beta=1.8$ is motivated by well-constrained
Rayleigh-Jeans emission in similar systems
\citep[e.g.][]{scoville16a}.
The free parameters for our least-constrained SED (MORA-9) are
redshift, \lir, and \lpeak\ (where both $\alpha_{\rm MIR}$ and
$\beta$, the emissivity spectral index have been fixed).  In contrast,
the best-fit SEDs constrain up to five free parameters: $z$, \lir,
\lpeak, $\alpha_{\rm MIR}$, and $\beta$.

A graphical representation of the
two-dimensional posterior distributions for each variable are shown in
Figure~\ref{fig:examplefit} for MORA-2.

\begin{figure*}
\centering
\includegraphics[width=0.66\columnwidth]{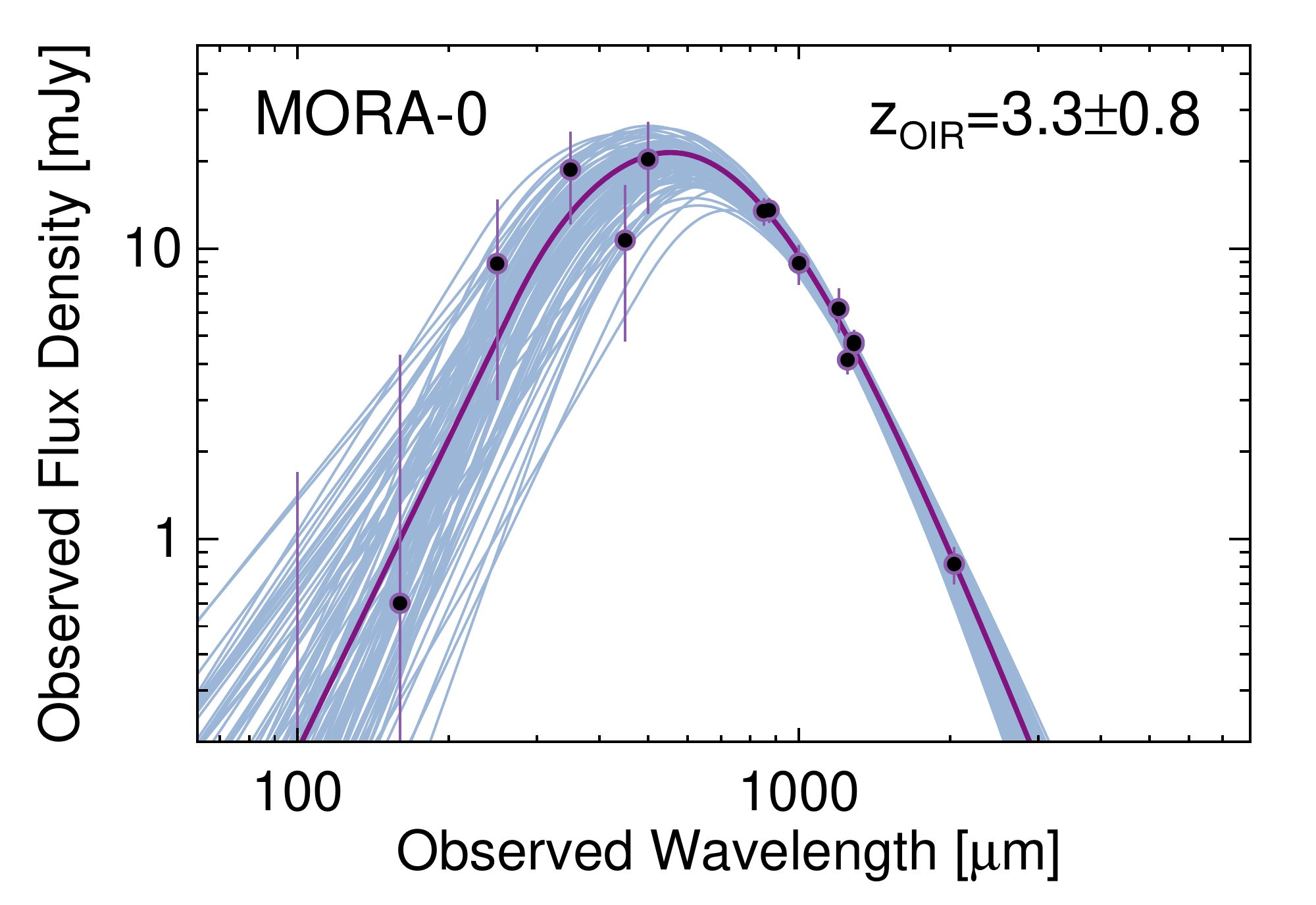}
\includegraphics[width=0.66\columnwidth]{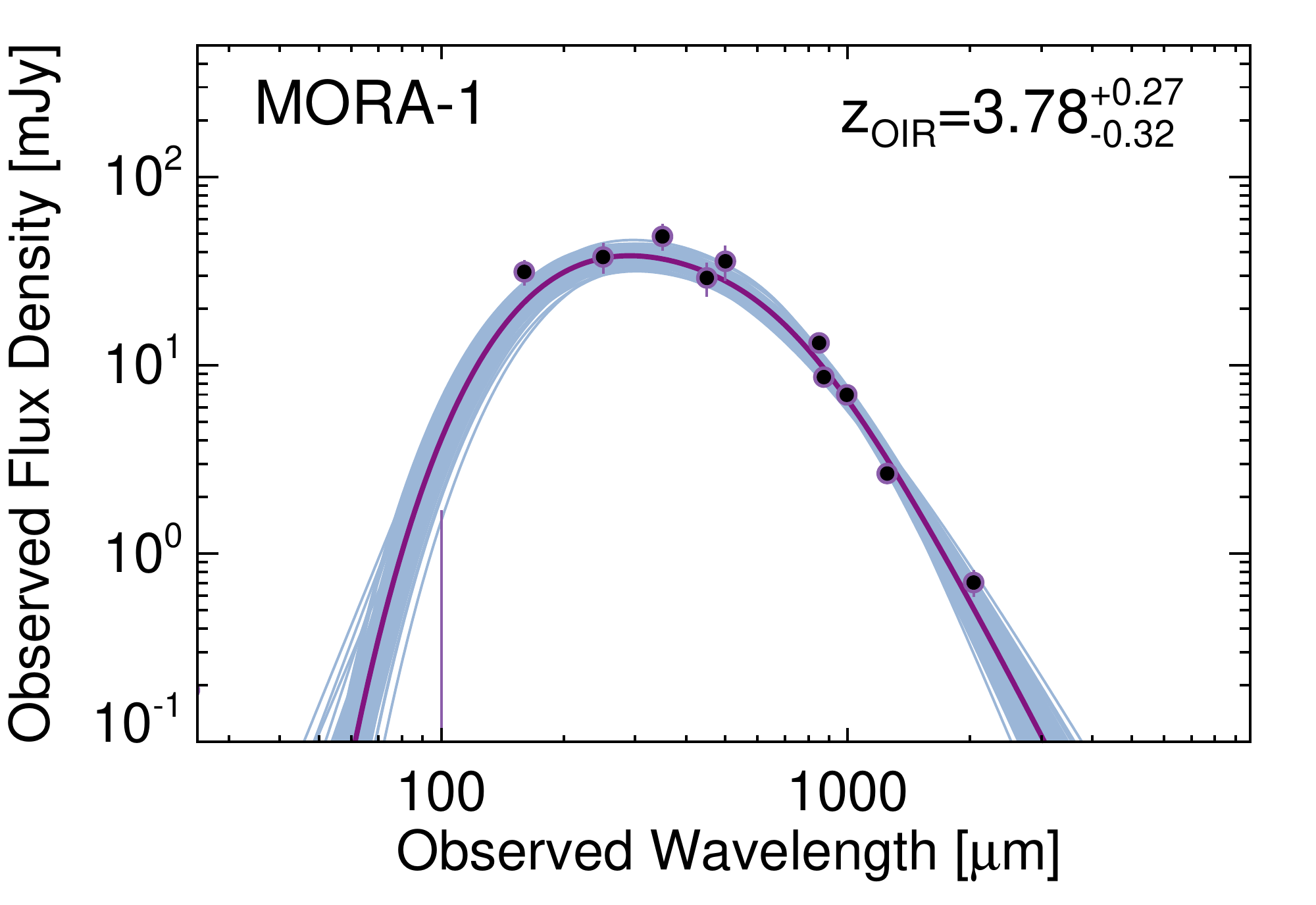}
\includegraphics[width=0.66\columnwidth]{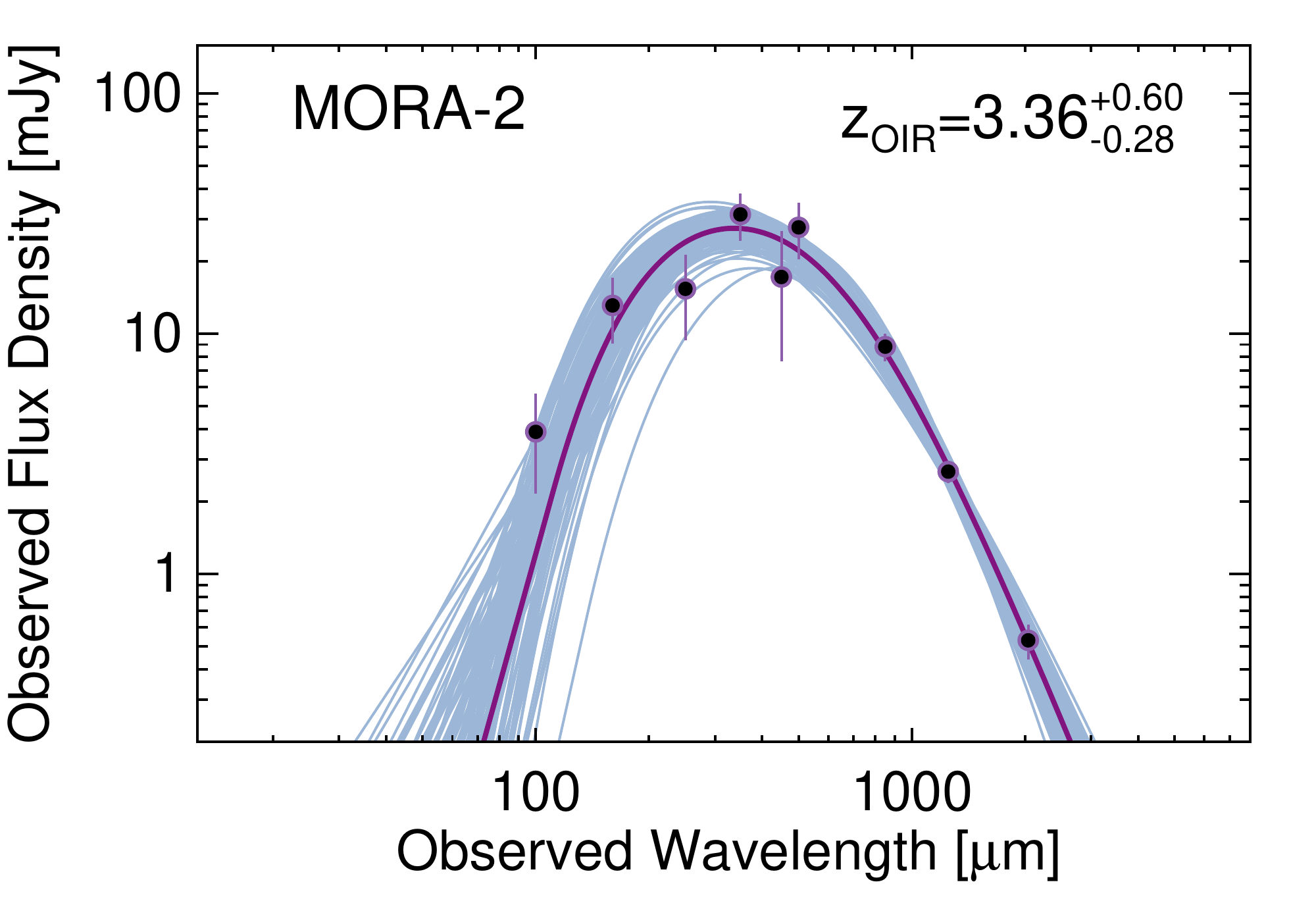}\\

\includegraphics[width=0.66\columnwidth]{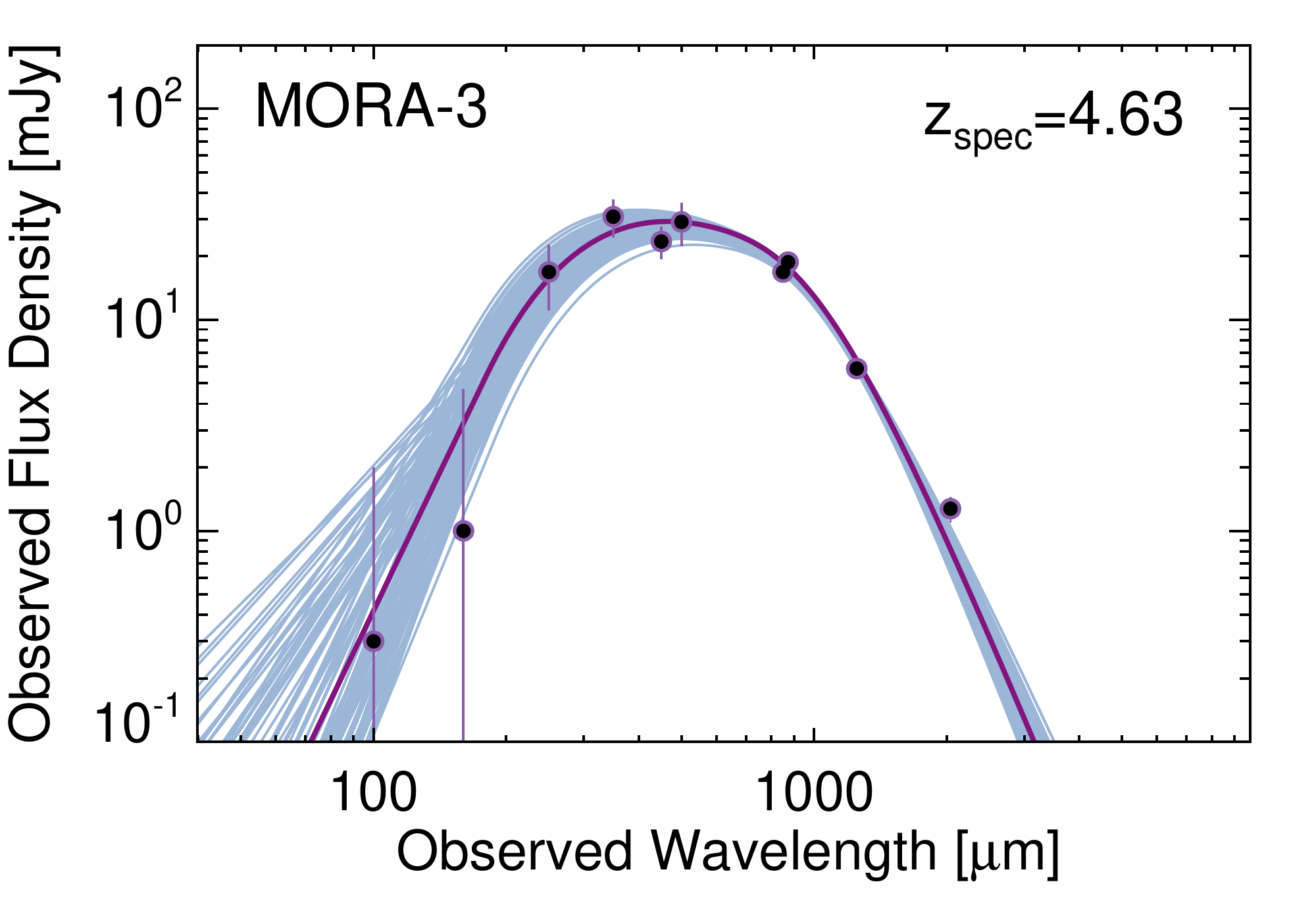}
\includegraphics[width=0.66\columnwidth]{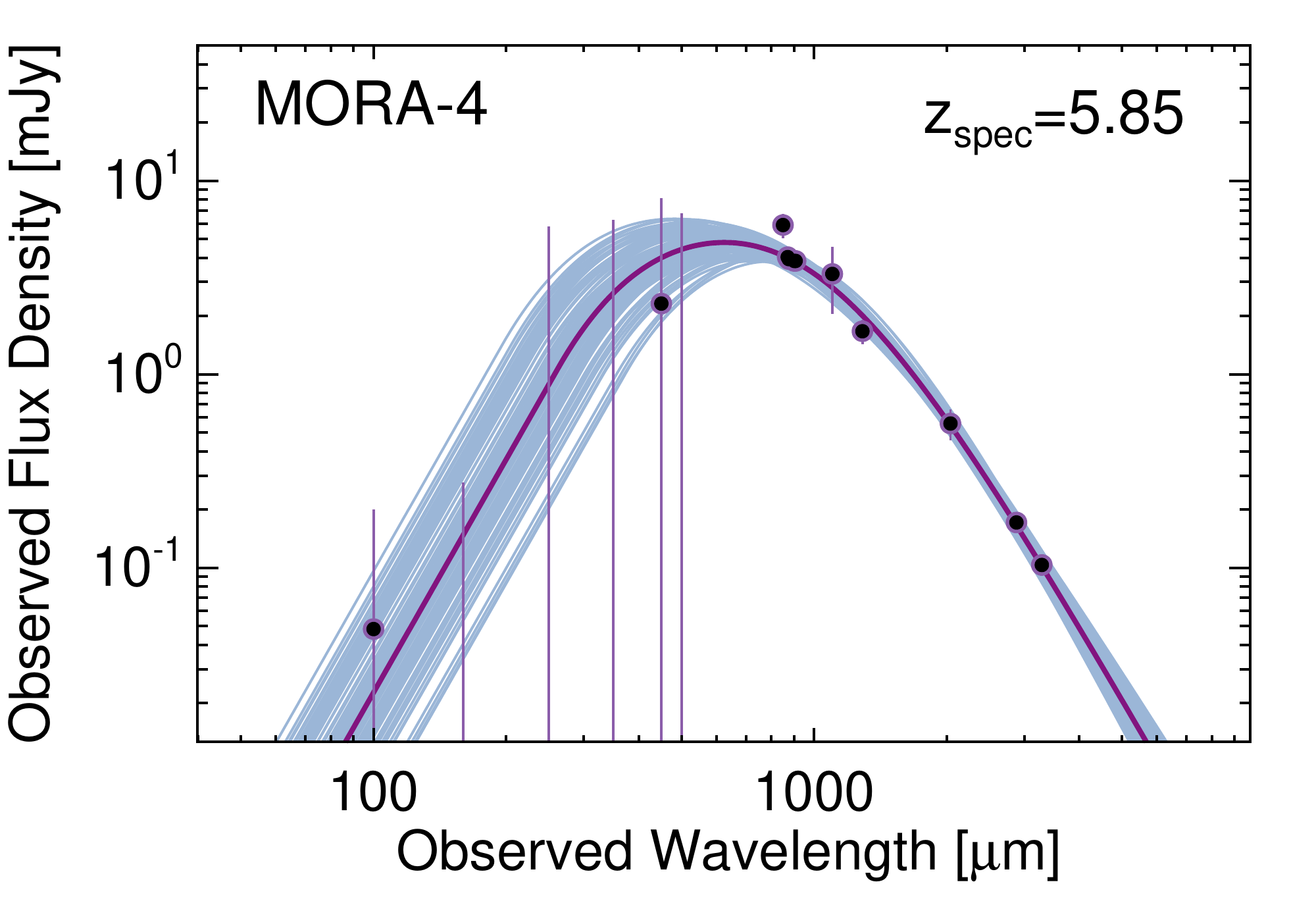}
\includegraphics[width=0.66\columnwidth]{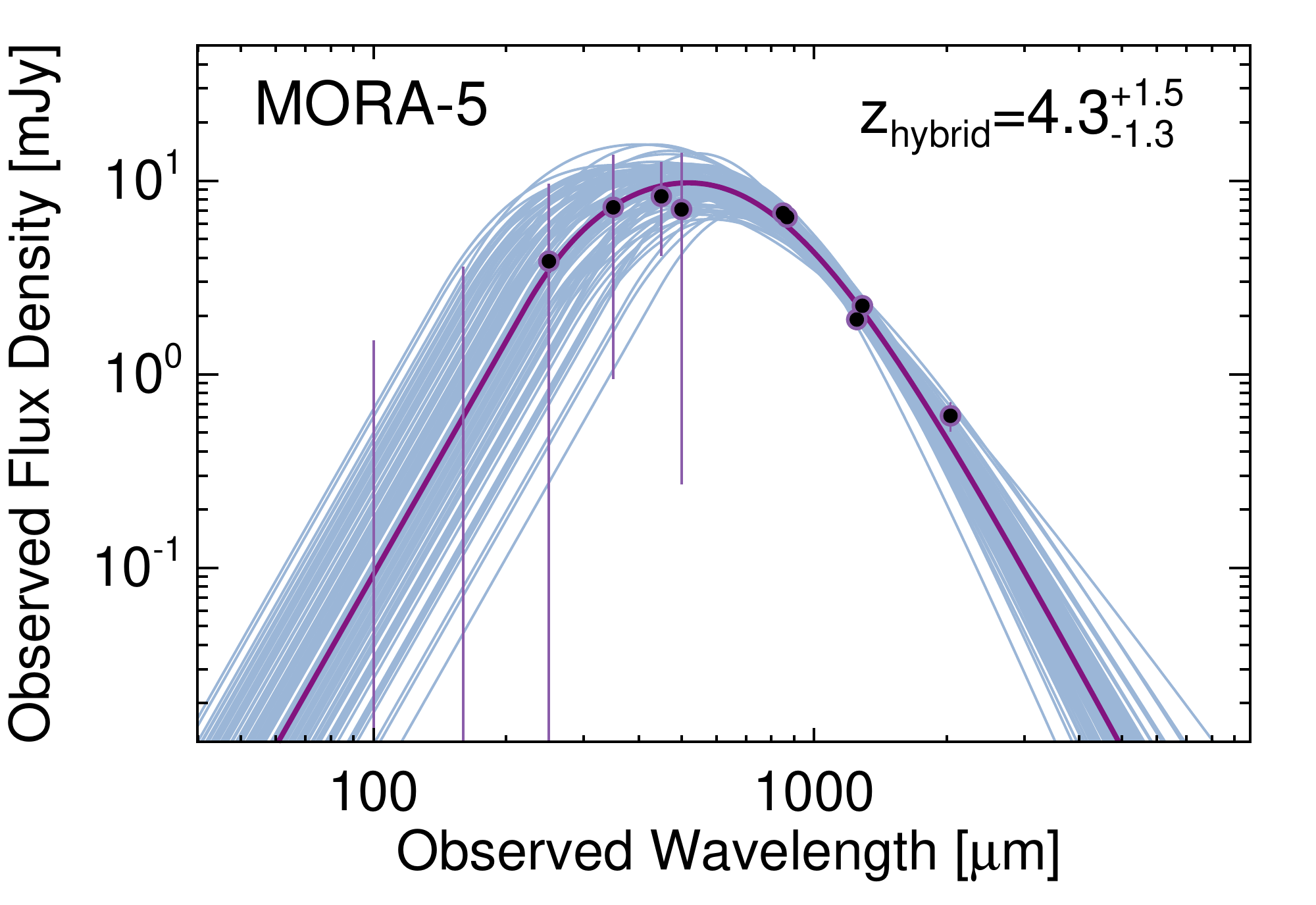}\\

\includegraphics[width=0.66\columnwidth]{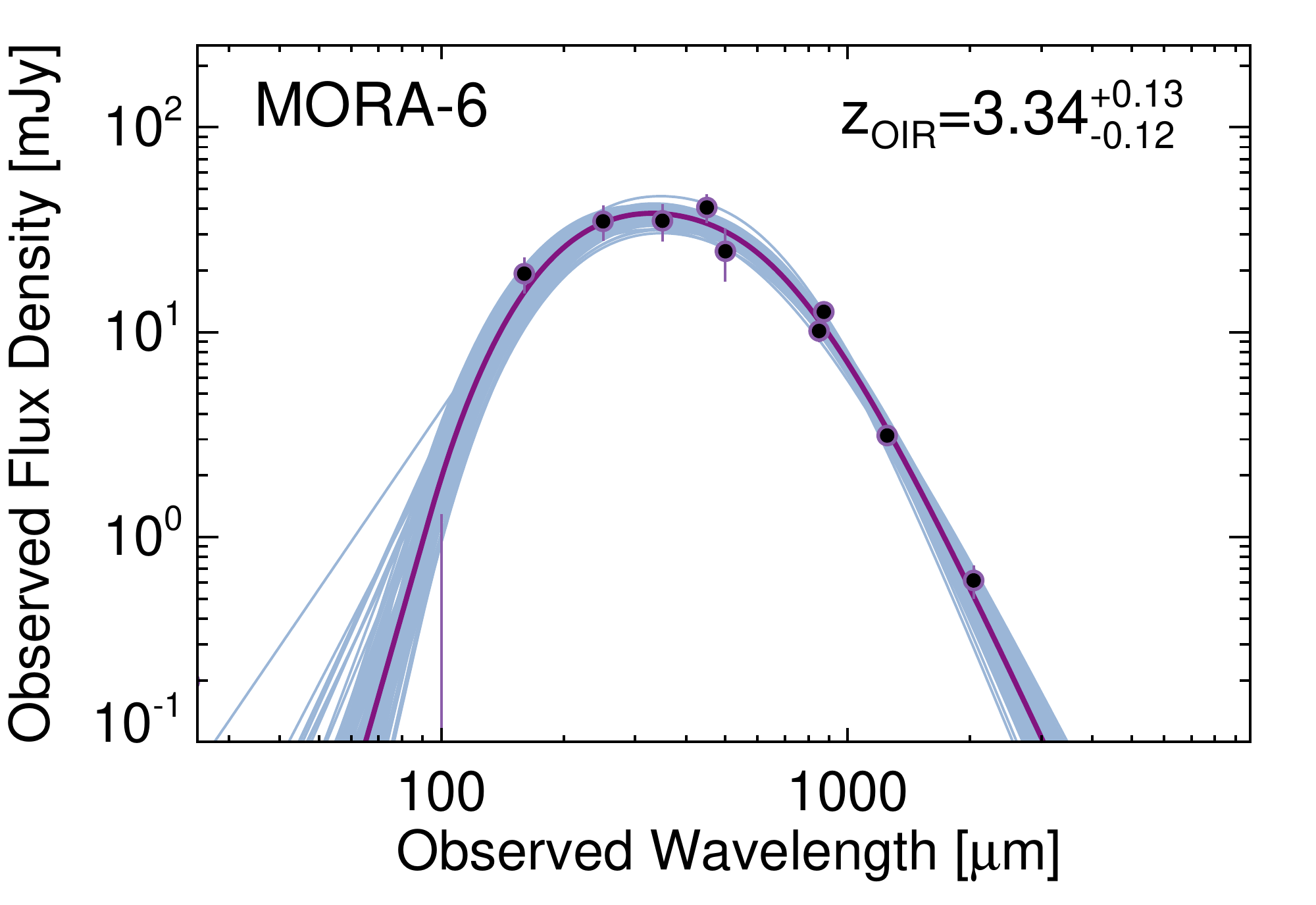}
\includegraphics[width=0.66\columnwidth]{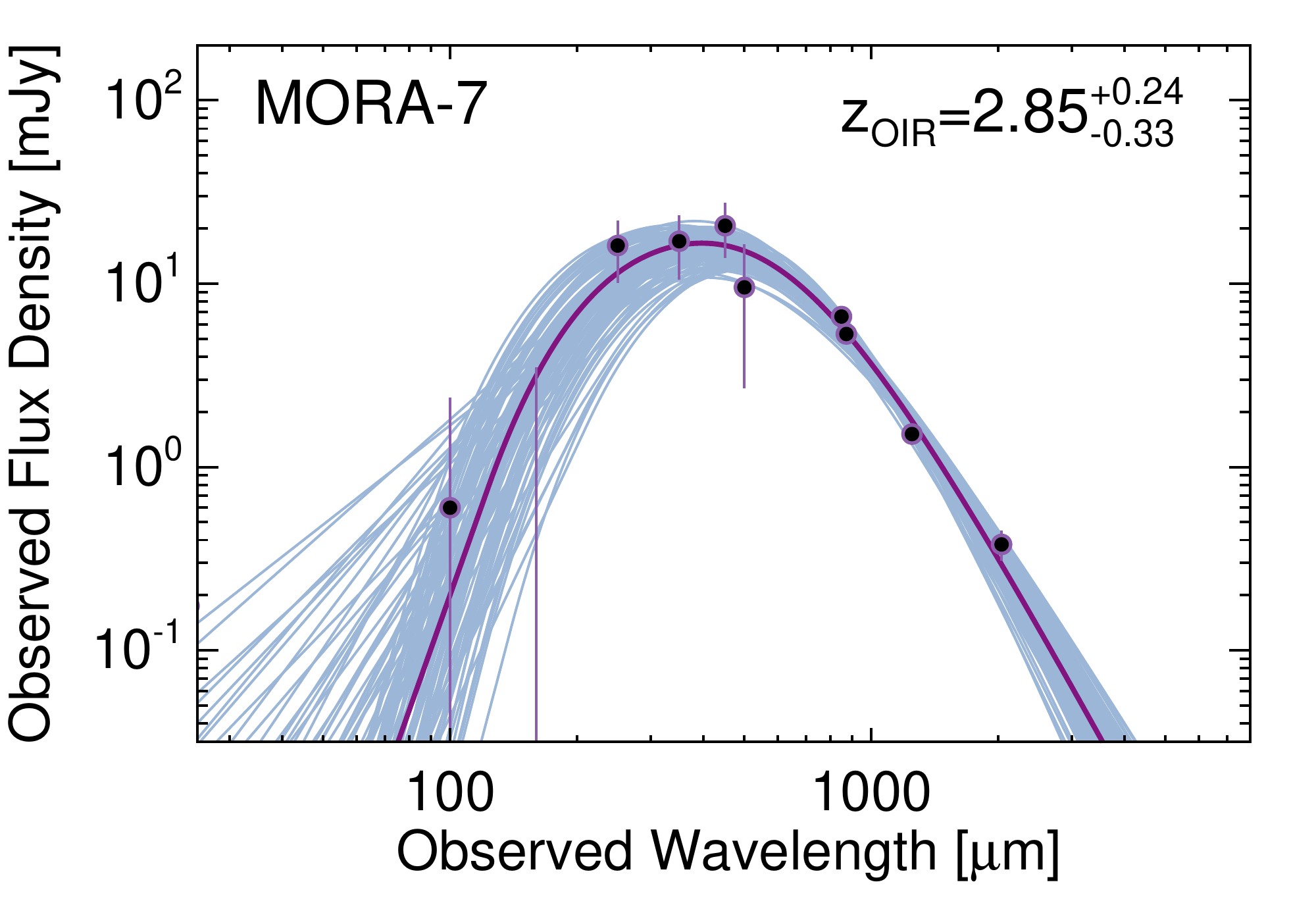}
\includegraphics[width=0.66\columnwidth]{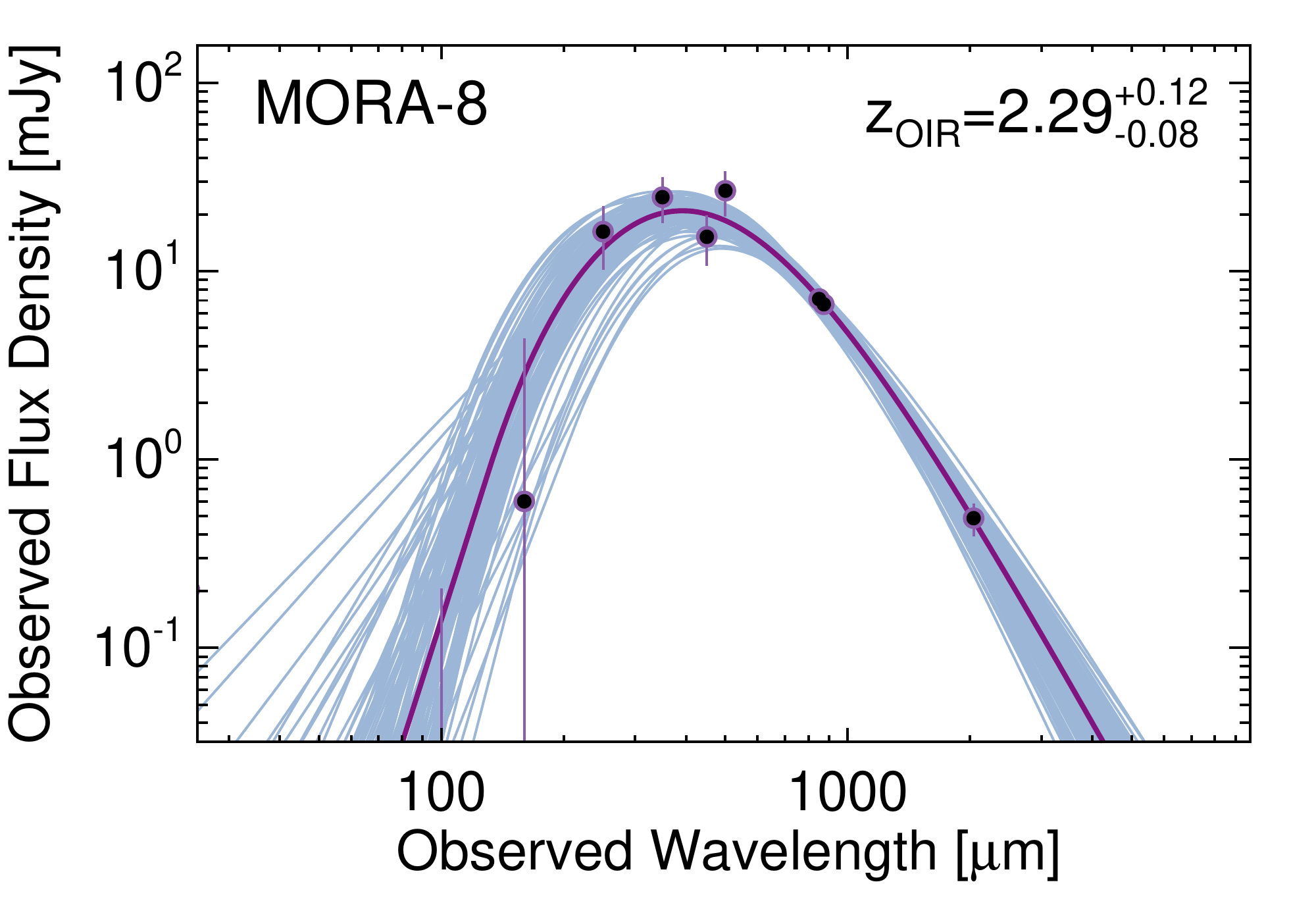}\\

\includegraphics[width=0.66\columnwidth]{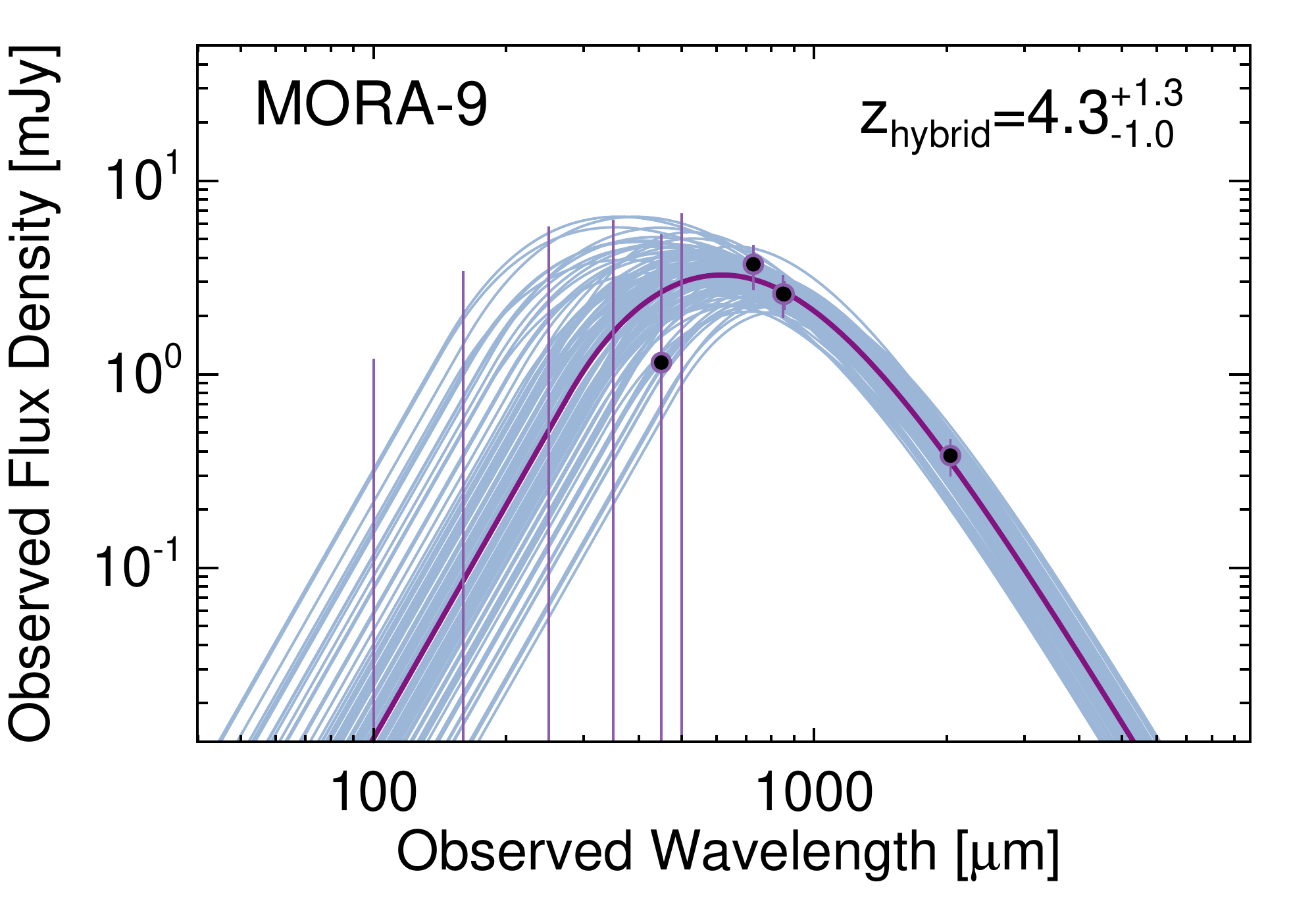}
\includegraphics[width=0.66\columnwidth]{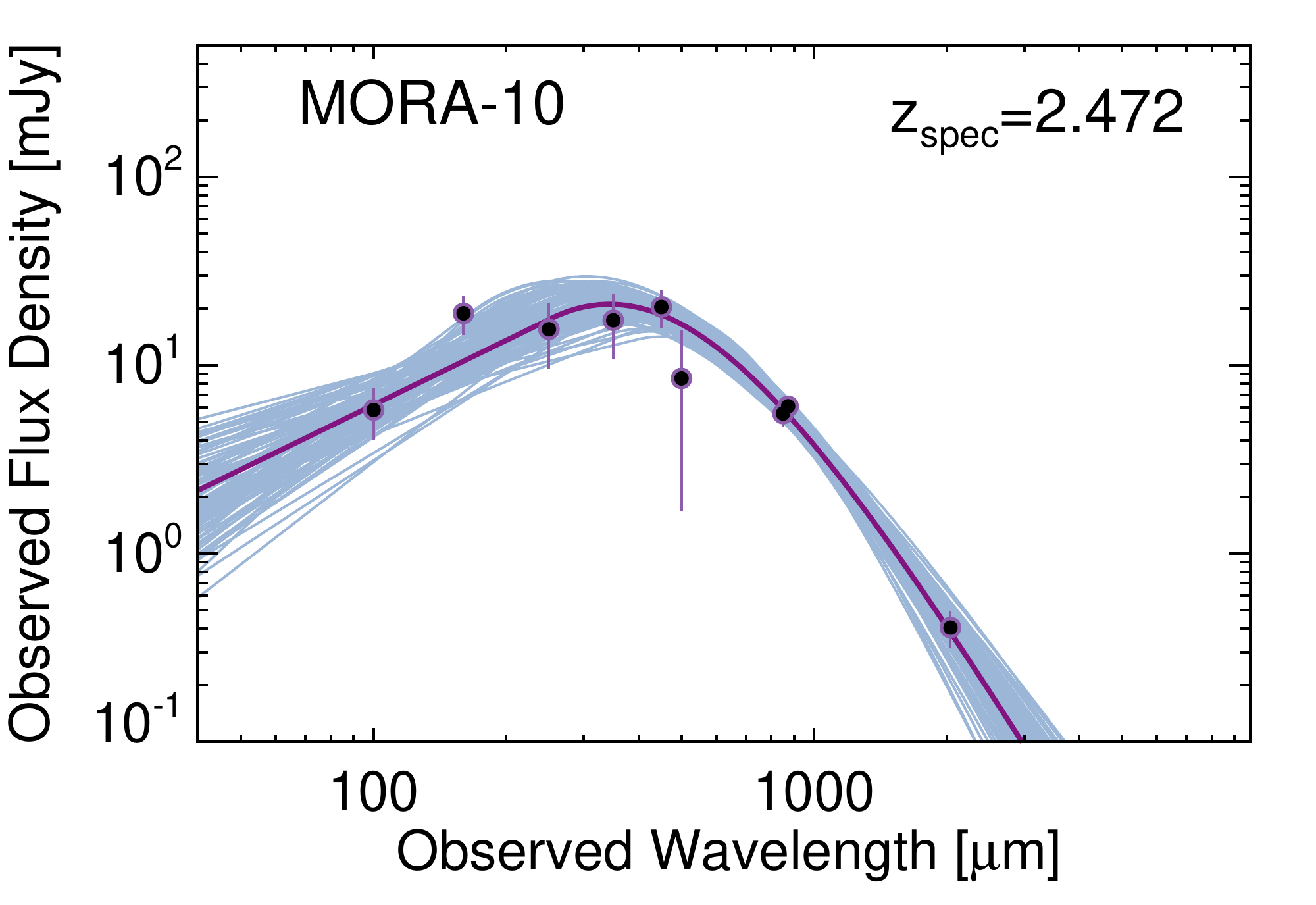}
\includegraphics[width=0.66\columnwidth]{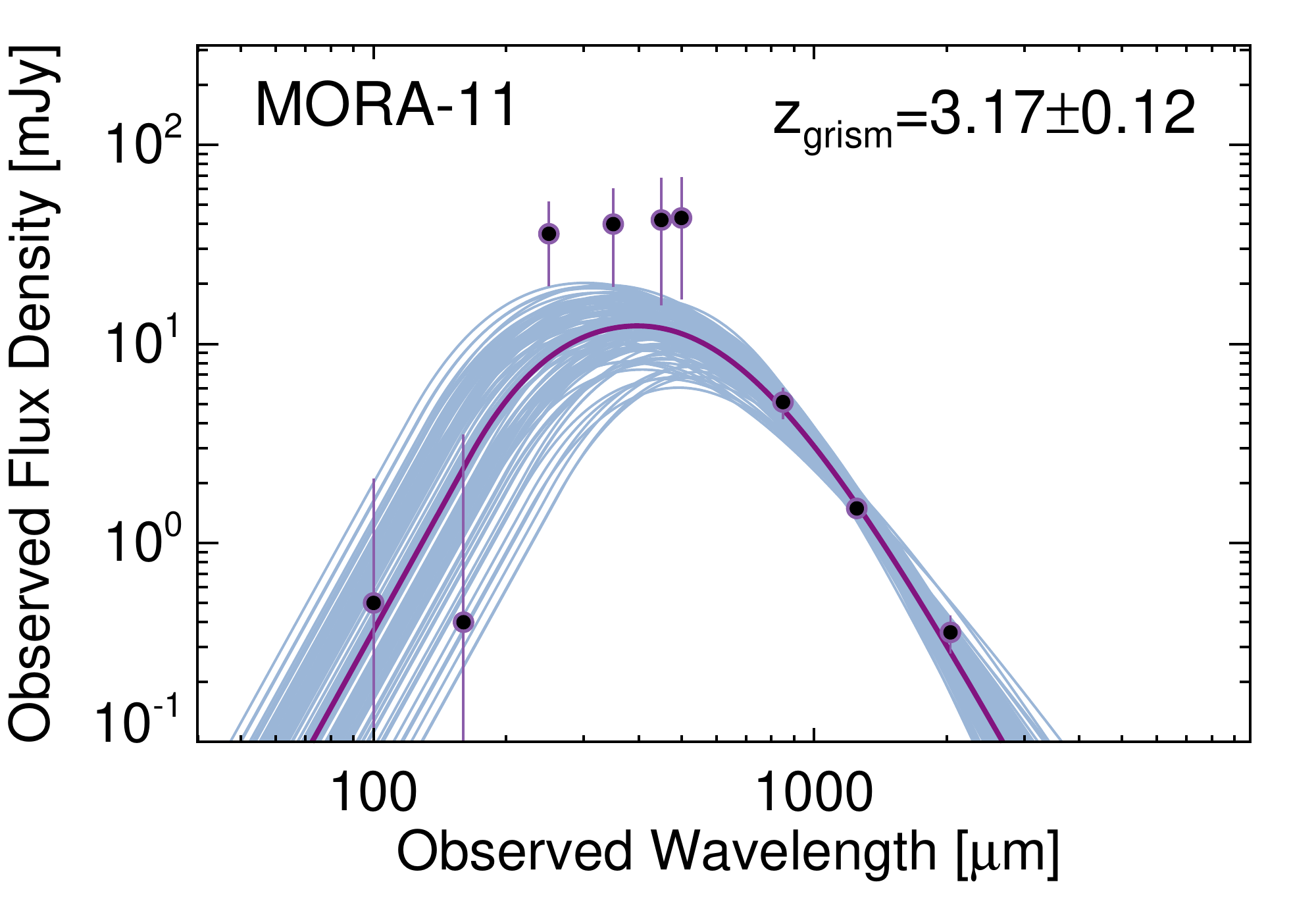}
\caption{The best-fit dust SEDs for the $>$5$\sigma$ 2\,mm-detected
  sample; SEDs are fit as a modified blackbody joined with a
  mid-infrared powerlaw.  The modified blackbody fixes the opacity
  such that $\tau=1$ at $\lambda_0=$\,200\,\um\ near the intrinsic
  peak of the dust SED; this assumption is not meant to be physically
  interpreted, but is rather fixed for convenience since our data
  cannot independently constrain $\lambda_0$.  Fixing $\lambda_0$ has
  no impact on the measured $\lambda_{\rm peak}$.
Four of these sources are detected at 100--160\um\ with {\it
  Herschel}-PACS, leading to a more prominent mid-infrared component
than those that lack rest-frame mid-infrared detections. The
uncertainty of the fit is shown via the light blue SED fits, drawn
randomly from the successful MCMC trials.  Sources without rest-frame
mid-infrared data have fixed mid-infrared powerlaw slopes of
$\alpha_{\rm MIR}\equiv4$ and MORA-9, the only source with limited
photometry on the Rayleigh-Jeans tail, has a fixed $\beta\equiv1.8$.
Measured SED characteristics are given in Table~\ref{tab:physical} and
the sources' photometry is given in Table~\ref{tab:photometry}.}
\label{fig:seds}
\end{figure*}

Figure~\ref{fig:seds} shows the resulting SED fits against measured
source photometry.  We see a variety of SED shapes presented here,
from sources with prominent mid-infrared powerlaws to quite steep
Rayleigh-Jeans tails ($\beta>2$).  The uncertainties are illustrated
by the range of accepted MCMC trial SED solutions, shown as light blue
curves in Figure~\ref{fig:seds}.  Extracted fit characteristics are
given in Table~\ref{tab:physical}.

\begin{table*}
\caption{Measured Physical Characteristics of the 2\,mm-Detected Sample}
\begin{tabular}{cccccccccc}
\hline\hline
{\sc Name} & $z$ & {\sc $z$-Type} & \lir & SFR & \lpeak & $\beta$ & $\alpha_{\rm MIR}$ & M$_{\rm dust}$ & M$_\star$ \\
  & & & [\lsun] & [\sfr] & [\um] & & & [\msun] & [\msun] \\
\hline
MORA-0  & $3.3\pm0.8$ & OIR & (4.7$^{+1.6}_{-1.0}$)$\times10^{12}$ & 690$^{+230}_{-150}$ & 133$^{+15}_{-12}$ & 2.3$^{+0.4}_{-0.1}$ & 3.6$^{+1.1}_{-0.7}$ & ($1.4^{+0.5}_{-0.4}$)$\times10^9$ & (2.2$^{+0.8}_{-0.8}$)$\times10^{10}$ \\
MORA-1  & 3.78$^{+0.27}_{-0.32}$ & OIR & (2.0$\pm0.3$)$\times10^{13}$ & 3000$^{+500}_{-400}$ & 68$\pm$4 & 2.4$^{+0.2}_{-0.4}$ & 5.2$^{+1.2}_{-1.6}$ & ($3.6^{+1.6}_{-1.3}$)$\times10^8$ & (5.8$^{+1.5}_{-1.9}$)$\times10^{10}$ \\
MORA-2  & 3.36$^{+0.60}_{-0.28}$ & OIR & ($9.7^{+4.7}_{-2.7}$)$\times10^{12}$ & 1400$^{+700}_{-400}$ & 84$^{+11}_{-12}$ & 2.0$\pm$0.3 & 5.6$^{+1.8}_{-2.9}$ & ($6.0^{+2.2}_{-2.3}$)$\times10^8$ & (1.4$^{+0.4}_{-0.3}$)$\times10^{11}$ \\
MORA-3  & 4.63 & spec & (1.17$^{+0.18}_{-0.16}$)$\times10^{13}$ & 1730$^{+270}_{-240}$ & 85$^{+4}_{-5}$ & 3.11$^{+0.15}_{-0.26}$ & 4.4$^{+1.3}_{-1.1}$ & $2.1^{+0.5}_{-0.6}$)$\times10^8$ & ... \\
MORA-4  & 5.85 & spec & (3.6$^{+1.1}_{-0.9}$)$\times10^{12}$ & 530$^{+160}_{-130}$ & 96$^{+12}_{-11}$ & 2.1$^{+0.1}_{-0.3}$ & $\equiv4$ & ($6.1^{+1.7}_{-0.7}$)$\times10^8$ & (3.2$^{+1.0}_{-1.5}$)$\times10^{9}$ \\
MORA-5  & 4.3$^{+1.5}_{-1.3}$ & hybrid & (4.1$^{+2.7}_{-1.8}$)$\times10^{12}$ & 610$^{+390}_{-270}$ & 99$\pm$21 & 2.2$^{+0.3}_{-0.5}$ & $\equiv4$ & ($6.2^{+4.2}_{-3.0}$)$\times10^8$ & (1.5$^{+1.0}_{-0.7}$)$\times10^{11}$ \\
MORA-6  & 3.34$^{+0.13}_{-0.12}$ & OIR & (1.43$^{+0.21}_{-0.18}$)$\times10^{13}$ & 2120$^{+310}_{-260}$ & 80$\pm$4 & 2.4$^{+0.2}_{-0.3}$ & 7$\pm$2 & ($5.2^{+1.1}_{-2.0}$)$\times10^8$ & (7.0$^{+1.7}_{-1.4}$)$\times10^{10}$ \\
MORA-7  & 2.85$^{+0.24}_{-0.33}$ & OIR & (3.5$^{+1.5}_{-0.9}$)$\times10^{12}$ & 510$^{+220}_{-130}$ & 105$\pm$13 & 2.2$^{+0.4}_{-0.3}$ & 6$^{+3}_{-1}$ & ($4.4^{+1.6}_{-1.6}$)$\times10^8$ & (2.5$^{+0.5}_{-0.5}$)$\times10^{11}$ \\
MORA-8  & 2.29$^{+0.12}_{-0.08}$ & OIR & (2.2$\pm$0.5)$\times10^{12}$ & 330$^{+70}_{-70}$ & 131$^{+12}_{-10}$ & 1.8$^{+0.3}_{-0.2}$ & 6.6$^{+2.0}_{-2.2}$ & ($1.0^{+0.4}_{-0.2}$)$\times10^9$ & (1.1$^{+0.2}_{-0.3}$)$\times10^{11}$ \\
MORA-9  & 4.3$^{+1.3}_{-1.0}$ & hybrid & (1.2$^{+1.0}_{-0.4}$)$\times10^{12}$ & 180$^{+150}_{-60}$ & 123$^{+26}_{-23}$ & $\equiv1.8$ & $\equiv4$ & ($8.2^{+5.1}_{-3.8}$)$\times10^8$ & (4.1$^{+1.8}_{-1.4}$)$\times10^{10}$ \\ 
MORA-10 & 2.472 & spec & (1.1$^{+0.1}_{-0.1}$)$\times10^{13}$ & 1570$^{+160}_{-130}$ & 102$^{+10}_{-13}$ & 1.7$^{+0.4}_{-0.2}$ & 1.3$^{+0.15}_{-0.14}$ & ($7.1^{+2.2}_{-2.3}$)$\times10^8$ & (7.3$^{+2.0}_{-2.0}$)$\times10^{10}$ \\
MORA-11 & 3.17$\pm$0.05 & OIR & (3.4$^{+1.7}_{-1.1}$)$\times10^{12}$ & 500$^{+250}_{-170}$ & 99$^{+13}_{-14}$ & 2.1$^{+0.4}_{-0.4}$ & $\equiv4$ & ($3.5^{+2.5}_{-0.9}$)$\times10^8$ & (1.7$^{+0.5}_{-0.3}$)$\times10^{11}$ \\
\hline\hline
\end{tabular}
\label{tab:physical}

{\bf Table Notes.} Derived physical properties of the 2\,mm-detected
MORA Sample. Variables which have been fixed are denoted with $\equiv$
notation.  Estimates for ISM masses (not given in this table) can be
obtained by multiplying the dust masses in this table by a factor of
125 \citep{remy-ruyer14a}.  A stellar mass estimate is not available
for MORA-3 (a.k.a. AzTEC-2) due to blending of near-infrared imaging
with foreground galaxies.  All uncertainties in this table indicate
the inner 68\%\ minimum credible interval of posterior distributions.
\end{table*}

\begin{figure*}
\centering
\includegraphics[width=0.99\columnwidth]{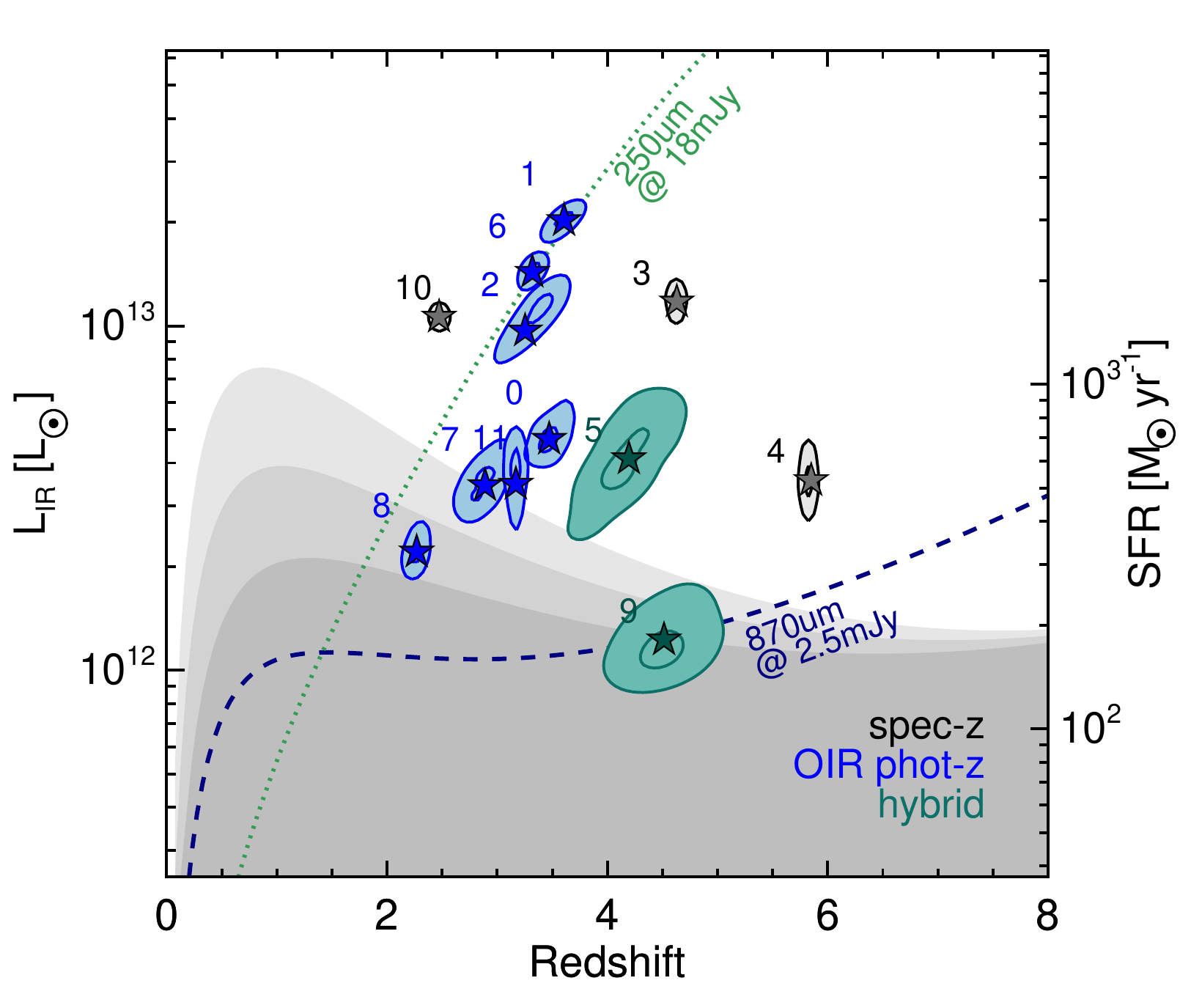}
\includegraphics[width=0.99\columnwidth]{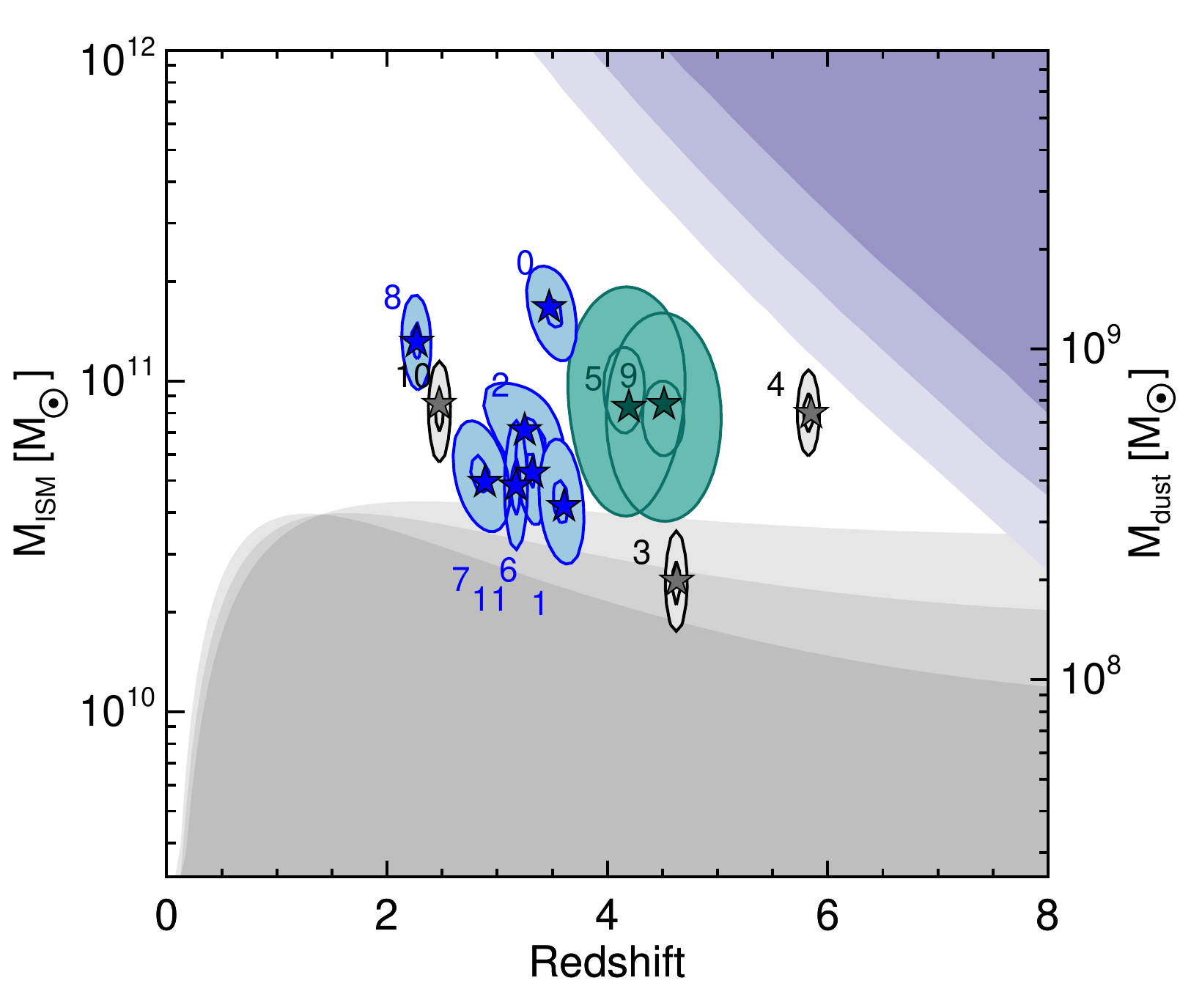}\\
\includegraphics[width=0.99\columnwidth]{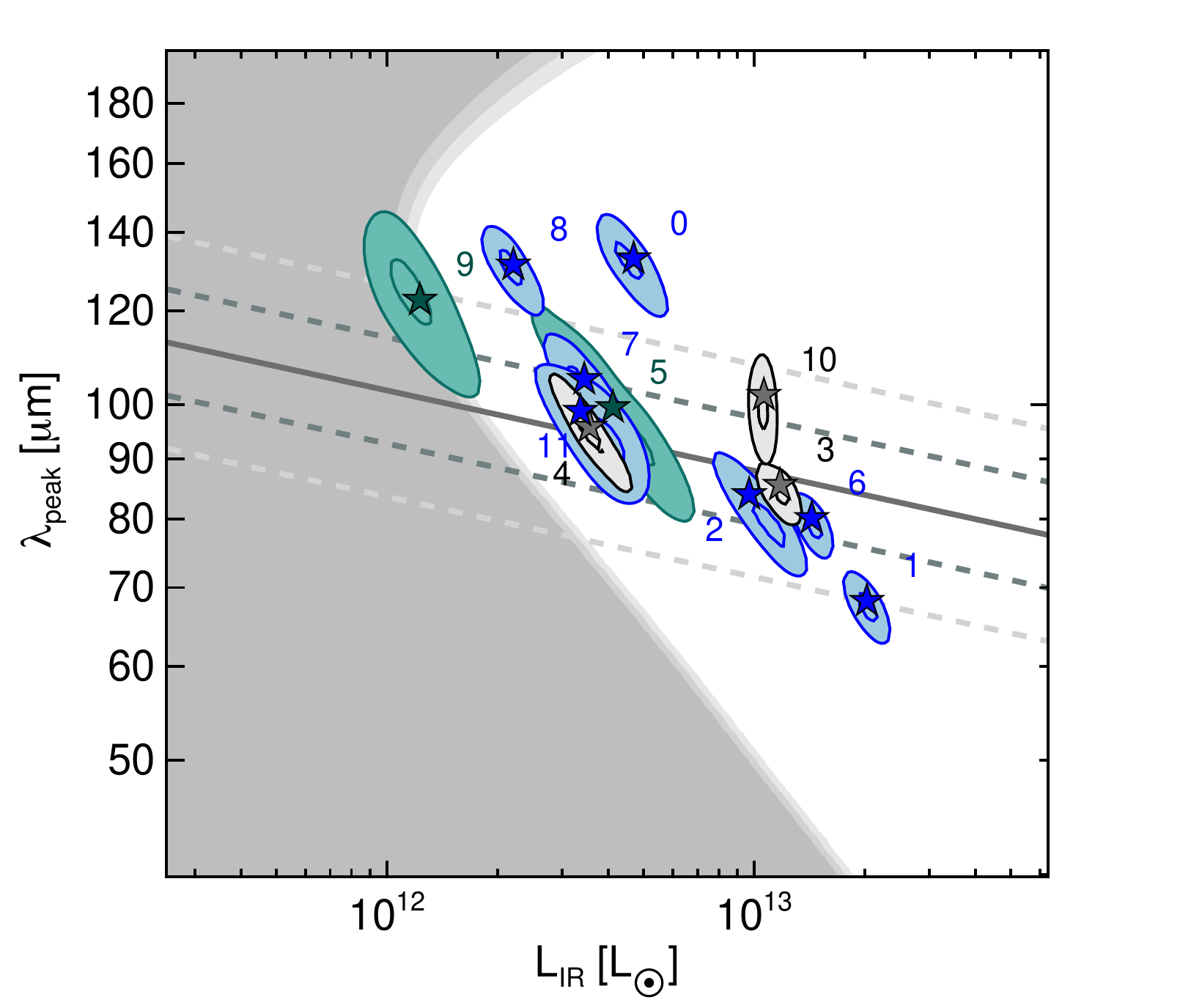}
\includegraphics[width=0.99\columnwidth]{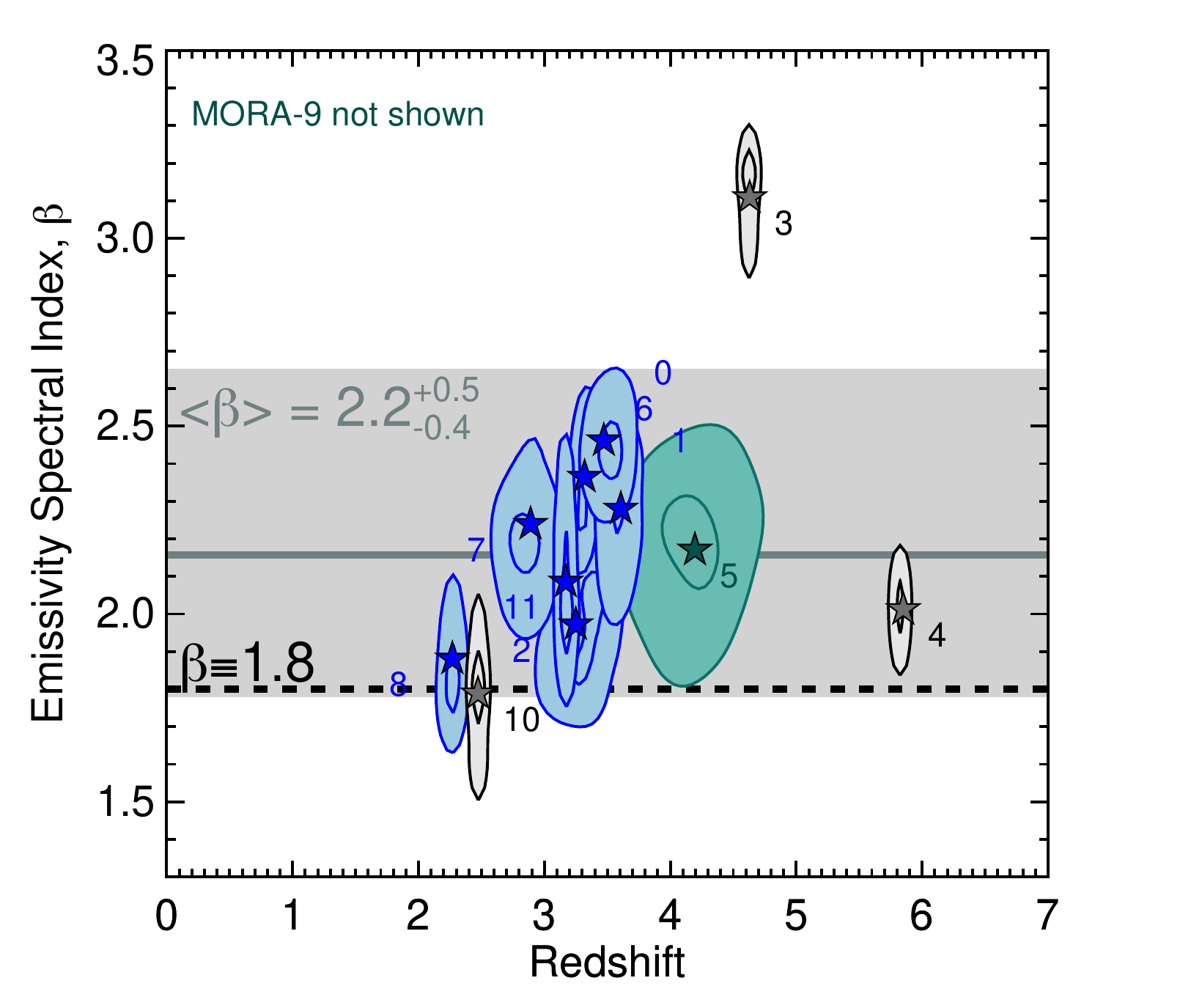}

\caption{The distribution of the 2\,mm-detected sample in derived
  physical parameter space.  Sources are color coded by the origins of
  their their redshift constraints: gray (spectroscopic), blue (OIR
  photometric) and teal (hybrid OIR/mm photometric).  Contours denote
  sources' 1 and 2$\sigma$ confidence intervals for the measured
  parameters, highlighting parameter covariance for the given set of
  photometry.  Regions of the planes inaccessible to the MORA survey
  are shaded in gray in the first three panels: the cutoffs depend
  precisely on parameters of the SED fit.  For example, in the
  \lir$-z$ plane (top left), the gray curves trace out the detection
  limit of a source detected at 5$\sigma$ (1$\sigma$=70\uJy\ at 2\,mm)
  with \lpeak$\,\equiv\,100$\um\ and $\beta=1.8$, $\beta=2.2$ and
  $\beta=2.6$ (with $\beta=2.6$ corresponding to the flattest curve).
  In the \mism$-z$ plane (top right) the lower detection threshold
  corresponds to fits with $\beta=1.8-2.6$ (where the steepest curve
  corresponds to $\beta=2.6$). The light purple shaded regions in the
  \mism$-z$ plane also shows the theoretical maximum \mism\ mass limit
  as a function of redshift based on the \citet{harrison13a} halo mass
  survey estimator; the halo mass is scaled down to \mism\ by a factor
  of 20. The \lir-\lpeak\ plane (bottom left) shows the aggregate,
  non-evolving \lir-\lpeak\ relation from \citet{casey18a} (gray line
  with dashed lines enclosing 1$\sigma$ and 2$\sigma$ scatter) and
  agreement with the MORA sample SEDs.  The gray region marks the
  detection limit for a $z=3$ system (the curves shift in \lir\ with
  $z$ following the gray curves in the \lir-$z$ panel). The
  distribution of sources' emissivity spectral indexes ($\beta$) is
  shown against redshift (bottom right) with the canonical value of
  $\beta=1.8$ marked with the black dashed line. No evidence for
  redshift evolution of $\beta$ is seen in this (small) sample; the
  average value is $\langle \beta\rangle=2.2^{+0.5}_{-0.4}$. }
\label{fig:lirz}
\end{figure*}

\subsection{Distribution in \lir, \lpeak, $M_{\rm ISM}$, and $\beta$}

Figure~\ref{fig:lirz} shows the distribution of the full sample of
twelve galaxies in \lir, \lpeak, $M_{\rm ISM}$, and $\beta$,
demonstrating the relative heterogeneous nature of the sample.
Sources are color coded by quality of their redshift constraints, and
contours denote 1--2$\sigma$ confidence intervals in the given
parameter space.

The dynamic range in MORA sources' IR luminosities is about a factor
of $\sim$25, from the most extreme, MORA-1, topping
2$\times10^{13}$\,\lsun\ to the least extreme, MORA-9,
$\sim$9$\times10^{11}$\,\lsun\ (the corresponding star-formation rates
range from 140--3100\,\sfr).  Given the selection of these sources on
the Rayleigh-Jeans tail of dust blackbody emission at 2\,mm, the
dynamic range in ISM masses is much more narrow, mirroring the
somewhat narrow dynamic range in 2\,mm flux densities (both a factor
of $\sim$3).  The second panel of Figure~\ref{fig:lirz} shows the
measured ISM masses of the sample, scaled up by a factor of 125 from
SED-inferred dust mass (to account for the total mass of gas in the
ISM), which is consistent with the expected gas-to-dust ratio for near
solar metallicity galaxies \citep{remy-ruyer14a}.  Note that a maximum
ISM mass, as a function of redshift, is set in our survey by the size
of the survey itself given a $\Lambda$CDM Universe (purple shaded
regions in the $M_{\rm ISM}-z$ panel of Figure~\ref{fig:lirz}, at
1--3$\sigma$ confidence intervals); this peak is determined by the
maximum halo mass as a function of redshift and survey volume from
\citet{harrison13a}, using an ISM-to-halo mass ratio of 1/20 \citep[as
  is done in][]{marrone18a}.

While the nominal 2\,mm detection limit in \lir\ with redshift follows
the strong negative K-correction (and thus is more sensitive to
sources at $z\simgt3$ than $z=1-2$; denoted by gray bands), the
detection limit in $M_{\rm ISM}$ is approximately flat. These
detection curves are not absolute limits as they are sensitive both to
galaxies' luminosity-weighted dust temperature and the emissivity
spectral index.  For example, a galaxy of ISM mass
$\sim2\times10^{10}\,$\msun\ at $z=4$ may only be detectable in MORA
if its emissivity spectral index is lower than $\beta\simlt2.2$, or a
$\sim2\times10^{12}$\,\lsun\ system at $z=3$ may only be detectable in
MORA if its rest-frame peak wavelength falls within
$95\simlt\lambda_{\rm peak}\simlt175$\,\um.

The galaxies' distribution in rest-frame peak wavelength, \lpeak\ (a
proxy for the luminosity-weighted dust temperature) is shown in the
third panel of Figure~\ref{fig:lirz} relative to the aggregate DSFG
population fit found in \citet{casey18a}, with intrinsic scatter of
1--2$\sigma$ shown.  The MORA galaxies are largely in agreement with
the global DSFG trend, with two sources appearing to be somewhat
anomalously cold (MORA-0 and MORA-8).  For a sample this small, we
would only expect at most one source to sit $>$2$\sigma$ offset from
the global trend given that galaxies are distributed in a Gaussian
about the mean \lir-\lpeak\ relation.
The cold SED for MORA-0 is somewhat uncertain given the significant
uncertainty on the source's redshift; if this source is indeed at the
higher redshift end of its redshift PDF, its SED would be better
aligned with the overall trend for DSFGs' peak wavelengths.
Nevertheless, it is worth highlighting that 2\,mm selection itself may
skew the temperatures of the sample a bit cold; the gray region in the
third panel of Figure~\ref{fig:lirz} shows the parameter space beyond
reach of the MORA survey at a fixed redshift of $z=3$.  If the survey
had a flat selection in \lir\ around $\sim$3$\times10^{12}$\,\lsun,
MORA-9 and MORA-8 may not have made the cut, as they're the least
luminous sources in the sample; excluding them, only one source,
MORA-0, is anomalously cold, in line with expectation for a sample
this size.

The average value of the emissivity spectral index of MORA galaxies in
the sample is $\langle\beta\rangle=2.2^{+0.5}_{-0.4}$, which is skewed
toward higher values than are often typically assumed for galaxies in
the absence of direct measurements.  The fact that the MORA sample is
2\,mm selected would potentially impact the average measured $\beta$.
However, one would assume it would do so in the opposite sense, by
more efficiently identifying sources with shallower emissivity
spectral indices (or $\beta<1.8$) due to higher relative 2\,mm flux
densities for a given \lir.  While nominally substantial heating from
the CMB at high-$z$ ($z\simgt5$) could effectively steepen the
Rayleigh-Jeans slope (and artificially increase $\beta$ as discussed
in \citealt{jin19a}), here we account for CMB heating directly and
find its effect negligible.  There is no evidence in the MORA sample
of evolution in $\beta$; because the sample size is relatively small,
the mean could also not be precisely constrained.  Our interpretation
of $\beta$, relative to other literature samples and other DSFGs
falling within the MORA footprint, is offered in \S~\ref{sec:beta}.

\begin{figure}
\includegraphics[width=0.99\columnwidth]{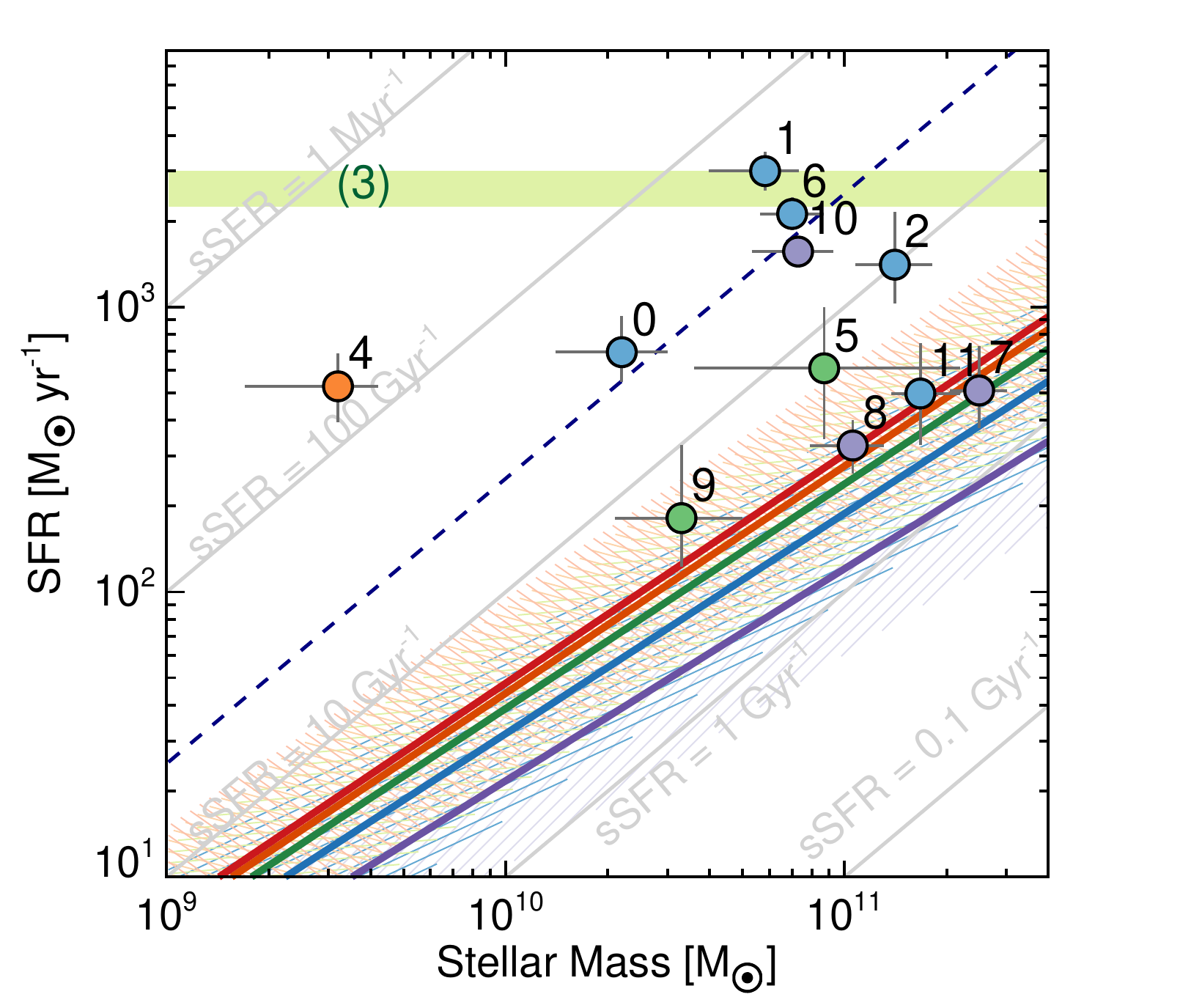}
\caption{Placement of MORA 2mm-selected sample in the galaxy
  M$_\star$-SFR relation, or galaxy ``main sequence.'' Colored lines
  and $\pm$0.3\,dex shaded regions represent the $z=2$ (purple), $z=3$
  (blue), $z=4$ (green), $z=5$ (orange), and $z=6$ (red) average
  relations from \citet{speagle14a}.  The dashed navy line represents
  the lower envelope for starbursts as defined in
  \citet{caputi17a,caputi21a}. Galaxies are labeled by their MORA IDs,
  and color coded according to the $\Delta z=1$ interval in which they
  most likely sit (e.g. MORA-0 with $z_{\rm OIR}=3.3$ is blue to
  denote it sits at $3<z<4$).  All sources except MORA-3
  (a.k.a. AzTEC-2) have stellar mass estimates, and thus that source's
  SFR is noted as a horizontal green band with unconstrained stellar
  mass; MORA-3 lacks an estimate due to severe blending of IRAC
  photometry with foreground galaxies.  With respect to the main
  sequence, the MORA sample splits roughly in half: 5/12 galaxies
  sitting on the galaxy main sequence while 7/12 are significantly
  elevated, the most extreme of which is MORA-4
  \citep[a.k.a. MAMBO-9][]{casey19a} which is (likely) the highest
  redshift source of the sample.}
\label{fig:mainsequence}
\end{figure}

\subsection{Stellar masses and the SFR-M$_\star$ relation}

Stellar masses are derived using the suite of OIR COSMOS photometric
constraints available for each source (Weaver, Kauffmann \etal,
submitted).  While Weaver \etal\ use {\tt LePhare}
\citep{arnouts02a,ilbert06a} and a range of 19 stellar population
templates from \citet{bruzual03a}, we only make use of their posterior
redshift probability density distributions for the subset of MORA
sources detected in the COSMOS2020 catalog.  We choose to remodel the
galaxies' stellar populations using a wider range of templates
inclusive of extreme starbursts.  This refitting is carried out using
the {\sc Magphys} energy balance code \citep{da-cunha08a} and
\citet{bruzual03a} stellar population synthesis templates and an
updated, wider range of star formation histories compatible with DSFGs
\citep{da-cunha15a}.  Redshift uncertainties are accounted for by
iteratively sampling the sources' redshift PDFs; we find that stellar
masses are largely insensitive to redshift uncertainties.  Similarly,
the stellar masses derived from {\sc Magphys} show no systematic
offset with those reported by Weaver \etal, though significant
uncertainty on all derived stellar masses remains.

There are a few exceptions where our technique is approached with
even more caution: MORA-3, MORA-4, MORA-5, and MORA-9.  The first,
MORA-3 (a.k.a. AzTEC-2), lacks any OIR counterpart shortward of
2.2\,\um\ and at longer wavelengths is spatially confused with
foreground galaxies, thus rendering any stellar mass constraint
impossible without future spatially-resolved {\it JWST} observations.
Though the {\sc Magphys} predicted stellar mass for MORA-4
(a.k.a. MAMBO-9) is reasonable on a standalone basis, \citet{casey19a}
present a detailed argument as to why it is most likely an
overestimate given the measured gas mass and implied halo mass.
Therein they provide an alternate stellar mass estimate of MAMBO-9
following the methodology outlined in \citet{finkelstein15a}, which
uses stellar population modeling plus the contribution from nebular
emission.  Note that the necessity of this approach is thought to be
unique to that source, given that MAMBO-9 is likely the highest
redshift source in the sample and intrinsically massive, thus the
survey volume itself sets an upper limit to its mass.

The last two exceptions to the standard {\sc Magphys} stellar
mass-fitting approach are sources MORA-5 and MORA-9, which are the two
OIR-dark sources explored in more detail in Manning \etal, as they
lack good redshift constraints.  Their stellar masses are fit using
the {\sc Magphys}{\tt +photo-z} code \citep{battisti19a}, which is a
further update to {\sc Magphys} allowing for redshift uncertainty;
this was deemed necessary in the case of MORA-5 and MORA-9 in
particular, given the significant uncertainties on both sources'
photometric redshifts.  The resulting best-fit stellar masses are
reported in Table~\ref{tab:physical}.

Figure~\ref{fig:mainsequence} shows the distribution of the MORA
sample in the galaxy SFR--M$_\star$ plane relative to observed trends
for the bulk galaxy population at various redshifts (i.e. the galaxy
``main sequence'').  The trend lines are drawn from \citet{speagle14a}
at fixed redshifts and sources are color coded by redshift, and the
dashed navy line demarcates the lower envelope for ``starbursts''
\citep[sSFR$\,>\,$25\,Gyr$^{-1}$;][]{caputi17a,caputi21a}.
What we observe is a rather heterogeneous sample: 7/12 (58\%)
MORA-selected sources are significantly elevated above the `main
sequence' with specific star formation rates in excess of ${\rm
  sSFR}\,>\,10\,{\rm Gyr}^{-1}$ (with 5/12, 42\%\ above the
25\,Gyr$^{-1}$ threshold)\footnote{Note that this likely includes
MORA-3, which doesn't have a stellar mass constraint, but is unlikely
to have a stellar mass in excess of $\sim$3$\times10^{11}$\,\msun,
which would be required for it to dip below this threshold.  See
\citet{jimenez-andrade20a} for more details on this source.}.  The
remaining 5/12 (42\%) are embedded in the galaxy main sequence itself,
with $1\,$Gyr$^{-1}<{\rm sSFR}<10\,$Gyr$^{-1}$.  There does not appear
to be a strong correlation between sources' redshifts and whether or
not they have elevated sSFRs.  The fact that the sample is somewhat
heterogeneous in the SFR--M$_\star$ plane is not surprising given that
the 2\,mm selection (and submillimeter selection more broadly) roughly
corresponds to a SFR cut, modulo variation due to SED dust temperature
and emissivity.  Later in \S~\ref{sec:quiescent} we discuss some of
the implications of the sample's measured stellar content in relation
to the formation of the first massive quiescent galaxies.

With a median stellar mass of 7$\times10^{10}\,$\msun, we note that
the median dust-to-stellar mass ratio in this sample is $M_{\rm
  dust}/M_\star=(7^{+30}_{-3})\times10^{-3}$.  This is nominally a bit
higher than expected for DSFGs, though one would expect a 2\,mm sample
to be slightly biased in that regard, selecting more galaxies rich in
dust per unit stellar mass than selection at shorter wavelengths.

\subsection{2\,mm Characteristics of Known (Sub)Millimeter Sources}\label{sec:scuba2}

\begin{figure*}
\centering
\includegraphics[width=1.65\columnwidth]{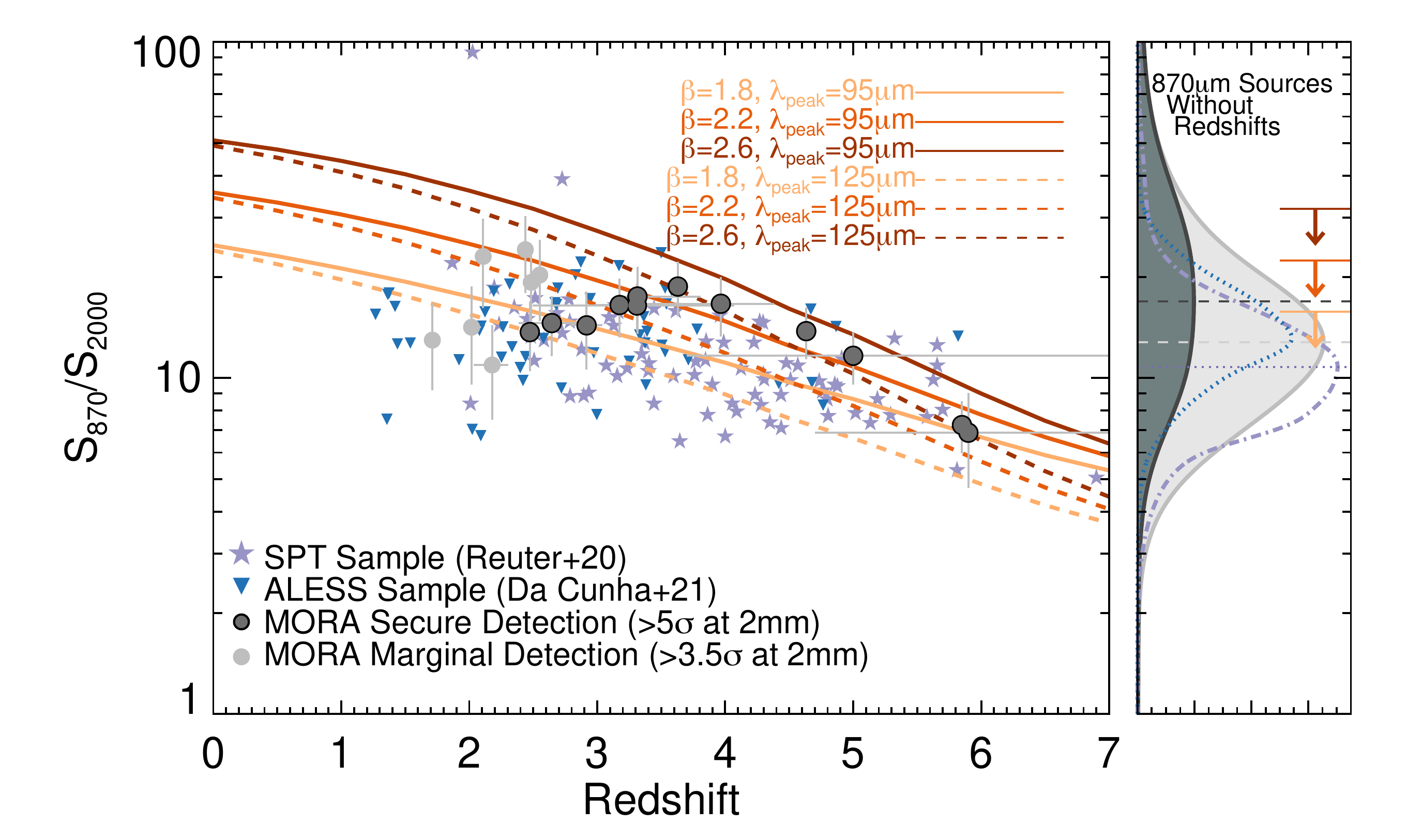}
\caption{ The 870\um\ to 2\,mm flux density ratio against redshift and
  in histogram for DSFGs from various samples. {\it Left:} The $S_{\rm
    870}/S_{\rm 2000}$ ratio for DSFGs that have secure redshift
  constraints, including $>$5$\sigma$ MORA sources (dark gray points)
  and marginal MORA sources (light gray points); we have also
  overplotted two other samples of well-characterized DSFGs in the
  literature: the SPT (1.4\,mm-selected) sample from
  \citet{reuter20a}, and the ALESS (870\um-selected) sample from
  \citet{da-cunha21a}.  Overlaid are tracks of specific SEDs with
  variable dust temperature ($\lambda_{\rm peak}=95$\,\um\ or 125\um,
  solid vs. dashed) and variable emissivity spectral index
  ($\beta=$1.8, 2.2, or 2.6, in increasingly dark shades of orange,
  respectively).  {\it Right:} Histograms of $S_{\rm 870}/S_{\rm 2000}$
  for sources without redshifts that: have 870\um\ ALMA data (dark
  gray), sources with 870\um\ data only from SCUBA-2 (light gray), and
  the full distributions of the SPT sample (purple dot-dashed) and
  ALESS sample (dotted blue).  Downward arrows mark the maximum color
  ratio that could correspond to a given value of $\beta$ at $z=2.5$;
  higher ratios are not allowed for the given $\beta$, as the ratio of
  flux densities only diminishes with redshift.  Our data suggest a
  range of $\beta$ values apply to this sample, from some being fully
  consistent with the often-presumed value of $\beta=1.8$, to some
  being more consistent with steeper slopes, $\beta\approx 2.2-2.6$.}
\label{fig:s2beta}
\end{figure*}

Here we analyze the 2\,mm characteristics of known submillimeter
sources in the MORA footprint, taken from the \citet{simpson19a} deep
SCUBA-2 850\um\ map.  We select sources from that work detected above
a signal-to-noise of 4, roughly corresponding to a flux threshold of
$S_{\rm 850}\simgt$2.3\,mJy (i.e. ${\rm RMS}\approx0.57$\,mJy/beam),
minimizing spurious detections at lower significance.  Only 55/98
($\approx$\,56\%) sources have interferometric follow-up from ALMA other
than our band 4 data, allowing identification of their precise
astrometric positions in the MORA map.  The remainder
(43/98\,$\approx$\,44\%) lack precise position constraints and are not
well known to better than the JCMT beam (14\farcs8).  For those
sources, we search for possible corresponding marginally-detected
2\,mm sources by identifying the highest significance peak within the
FWHM of the JCMT beam.

Out of 98 $>4\sigma$ SCUBA-2 sources overlapping with the main MORA
mosaic, 17/98 (=17.3\%) have 2\,mm peaks detected at $>4\sigma$
(including the 12 sources in our 5$\sigma$ sample).  This compares to
3.7\%\ of randomly placed JCMT beam `apertures' containing
$>4\sigma$ sources.  Most of this excess signal comes from our
$>5\sigma$ sample. However, our data still show an excess between
$4<\sigma<5$ significance with 5/86 (=\,5.8\%) found in total compared
to the random `aperture' rate of 3.7\%.  We find 56/98 (=\,57$\pm$7\%)
SCUBA-2 sources have between $3<\sigma<4$ significance peaks in the
2\,mm map, compared to 48$\pm$2\%\ of randomly placed apertures.  Of
those sources with $3<\sigma<4$ 2\,mm counterparts, the median
850\um\ flux density is 3.4\,mJy, lower than the median for the
$4<\sigma<5$ sample with 6.8\,mJy.  While the excess signal over
random apertures is relatively low significance due to significant
potential for contamination, it is still indicative of real 2\,mm
emission associated with these known 850\um\ sources.

Fifty-five of the 98 SCUBA-2 sources sitting in the MORA footprint
have other ALMA continuum data covered by the archival A3COSMOS
project \citep[described in detail in][]{liu18a} and 45/55 have a
confirmed ALMA counterpart detected at or above 3$\sigma$ in their
corresponding archival ALMA data (of varying observed frequencies,
most at 870\um\ or 1.2\,mm); 7 of the 45 sources have two components
or sources (separated by more than 1$''$) detected by ALMA within the
JCMT beam (and 2/7 have one of their multiples detected at 2\,mm at
$>$5$\sigma$).  The highest SNR 2\,mm detection within the JCMT beam
corresponds to the correct, independently-identified (single) ALMA
counterpart in 24 of 45 cases ($\approx53$\%).

Out of the 98 sources, fifty-five also have some form of redshift
constraint (derived independently from the mm photometry), ranging
from spectroscopic confirmation in the millimeter or
optical/near-infrared, or through an optical/near-infrared photometric
redshift.  These 55 sources with redshift constraints are not the same
55 with independent ALMA data; only 37 sources have both redshifts and
secure ALMA counterparts.

Figure~\ref{fig:s2beta} shows the distribution of millimeter colors
for the sample of known SCUBA-2 sources in the field, i.e. $S_{\rm
  870}/S_{\rm 2000}$, or the ratio of ALMA measured 870\um\ flux
density (or SCUBA-2 850\um) to 2\,mm flux density as measured by MORA.
Sources are split into those with redshift constraints (left panel)
and those without (right panel).  We also include several model tracks
in color space as a function of dust temperature (or $\lambda_{\rm
  peak}$) and emissivity spectral index ($\beta$) as well as
literature samples of DSFGs with well-constrained Rayleigh-Jeaons SEDs,
like the SPT 1.4\,mm-selected sample from \citet{reuter20a} and the
ALESS 870\um-selected sample from \citet{da-cunha21a}.

While naively one might think this 870-to-2\,mm color ratio indicative
purely of $\beta$, the emissivity spectral index, given that both
870\um\ and 2\,mm flux density measurements are likely to sit on the
Rayleigh-Jeans tail of dust blackbody emission, the color ratio has
both a temperature and redshift dependence that can result in
significant deviation from expectation given a fixed value of $\beta$.
For example, the approximation that the flux density on the
Rayleigh-Jeans tail of a blackbody follows $S_{\nu}\propto
\nu^{2+\beta}$ is often used, while in practice the SED only
asymptotically approaches this at rest-frame wavelengths longward of
rest-frame $\sim$2\,mm.  Even at moderate redshifts ($z\sim1$), our
observed bands probe shorter rest-frame wavelengths where this
approximation no longer holds.  This is why there is a redshift
dependence on $S_{\rm 870}/S_{\rm 2000}$ at redshifts well below the
point that either wavelength probes the peak of the SED.  The
temperature dependence may also be somewhat intuitive: at colder
intrinsic temperatures, the evolution toward lower $S_{\rm 870}/S_{\rm
  2000}$ ratios is steeper at lower redshifts, as the 870\um\ band
more quickly probes the peak of the SED than it would for a source
with a hotter SED.

We see that both MORA and literature DSFG samples roughly follow the
model trend evolution of $S_{\rm 870}/S_{\rm 2000}$ with redshift;
however, at higher redshifts, the $>$5$\sigma$ 2\,mm sample (points
enclosed in black circles) also suggest consistency with steeper
values of $\beta$ than the nominal $\beta=1.8$.
Figure~\ref{fig:s2beta} also marks the maximum color value (in $S_{\rm
  870}/S_{\rm 2000}$) allowed for given fixed value of $\beta$, which
is set by the measured color for $z=2.5$ systems (in other words, at
all higher redshifts, the color will be substantially lower). Though
there is a significant degeneracy between $S_{\rm 870}/S_{\rm 2000}$
color and redshift, $\beta$ and \lpeak, as indicated by the tracks on
the figure, it appears that many sources skew toward higher colors
than would be allowable for $\beta=1.8$: both from the sample with
redshift constraints and from the distribution of colors for sources
without redshifts.  This indicates that at least a subsample of the
population likely skews toward steeper values of $\beta$.  This is
consistent with our finding for individual sources in the $>$5$\sigma$
2\,mm sample that have $\langle\beta\rangle=2.2$, described earlier in
\S~\ref{sec:sedfits}.  Further discussion of the implications of
steeper values of $\beta$ are given in \S~\ref{sec:beta}.

\section{Discussion}\label{sec:discussion}

Prior to the MORA survey described herein, constraints on the
prevalence of dust-obscured star forming galaxies beyond $z\simgt3-4$
were relatively weak; refining the measurement of DSFGs' number densities and
characteristics during this epoch was the survey's primary goal.  Here
we discuss the implications of our measurements, from the observed
contribution of DSFGs to the cosmic SFRD beyond $z\simgt3$, to the
buildup of massive galaxies in the first few Gyr, and to the measured dust
emissivity index in early-Universe DSFGs.  We then present the limitations
of the survey, particularly given the survey area and small
sample size of detected sources, as well as various model degeneracies
that could be broken with future millimeter-wavelength observational
campaigns.

\subsection{Contribution to the SFRD}

The question of DSFGs' contribution to the cosmic star-formation rate
density at early times has been difficult to constrain.  However, the
MORA Survey's design is meant to effectively filter out lower redshift
DSFGs, enabling direct measurement of the obscured galaxy contribution
to the SFRD.  For example, the contrast between a ``dust poor''
universe \citep[e.g.][]{koprowski17a,dudzeviciute20a} and a ``dust
rich'' universe \citep[][]{rowan-robinson16a,gruppioni20a} would
result in very different manifestations of the MORA dataset; the
former predicts $\simlt$15 sources with median redshift $z\sim3.4$,
and the latter predicts $\simgt$50 sources with a median redshift of
$z\sim5.4$.  So what does our dataset imply for the SFRD?

\begin{figure*}
\centering
\includegraphics[width=1.5\columnwidth]{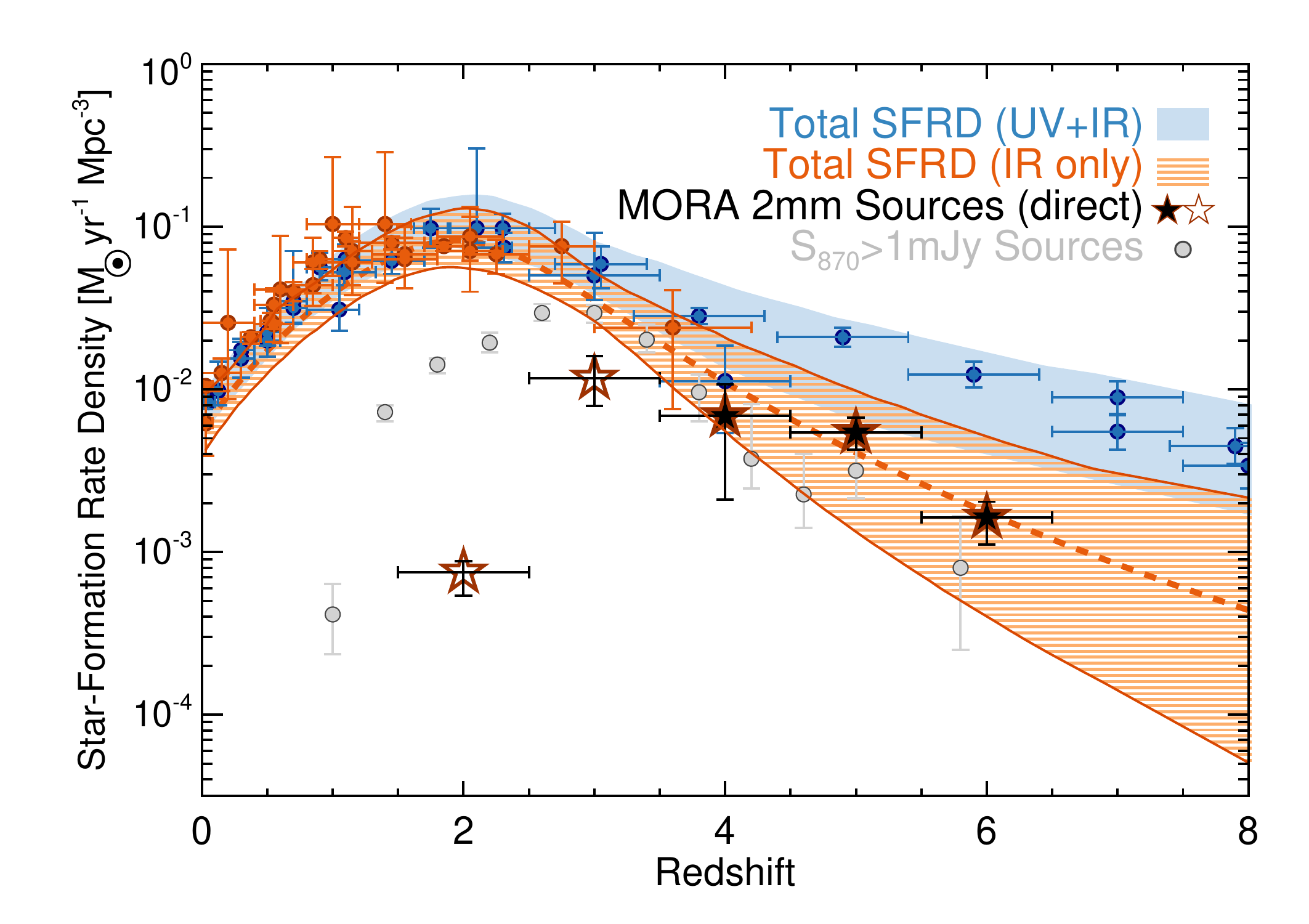}
\caption{The contribution of 2\,mm sources detected in the MORA Survey
  to the cosmic star formation rate density (black stars; open stars
  where incomplete at $z\simlt3$).  Our measurements are shown
  relative to the reported literature values from the \citet{madau14a}
  review (orange points indicating direct IR measurements, and blue
  points indicating rest-frame UV/optical measurements, corrected for
  dust attenuation).  The results from our accompanying paper, Zavala
  \etal\ (Z21), for the integrated IR/obscured component is shown as
  an orange, shaded band out to $z\sim8$ including the impact of
  cosmic variance (the black stars do not). The Z21 band is inferred
  from a joint analysis of 1\,mm, 2\,mm and 3\,mm number counts and
  the empirical model described in \citet{casey18a}.  We compare the
  contribution of this 2\,mm-selected sample with the measured
  contribution from 870\um-selected sources from the AS2UDS Survey
  above 1\,mJy (gray circles) in \citet{dudzeviciute20a}. The
  measurements in this paper (black stars) account directly and {\it
    only} for the eleven sources found above $>5\sigma$ in the MORA
  Survey, and do not represent an extrapolation of a fitted luminosity
  function; the uncertainties in sources' redshifts and SFRs are
  accounted for.  The dearth of sources at $z\simlt3$ is a direct
  consequence of the 2\,mm survey design, meant to efficiently `filter
  out' the majority of DSFGs at $1\simlt z\simlt3$ that dominate
  cosmic star-formation at its peak.}
\label{fig:sfrd}
\end{figure*}

We provide direct estimates of the MORA sample's contribution to
cosmic star-formation in Figure~\ref{fig:sfrd}.  The SFRD contribution
for this sample is estimated in the following way; first, we determine
the area over which each of the 11 sources would be detectable in the
MORA map above $>5\sigma$ significance.  As a reminder, we exclude
MORA-10 for its synchrotron component and MORA-12 as a likely false
positive.  Then we sample both the redshift and SFR probability
density distributions for each source through Monte Carlo trials; this
method accounts for their relative covariance.  Sources' SFRs are
converted to a SFRD by dividing by the appropriate survey volume
corresponding to the area over which a given source is accessible.  A
total SFRD, binned by $\Delta z=1$ intervals, is then measured for
each Monte Carlo trial, and the average and 68\%\ minimum confidence
interval of all trials is used to infer the MORA source contribution
to the SFRD, as shown in Figure~\ref{fig:sfrd}.  Though redshifts in
our sample only nominally span $z=2.2-5.9$, our SFRD estimates span
$z=1.5-6.5$ from the tails in the redshift PDFs for some of the less
well-constrained sources.

Overall, we find the prevalence of $z<2$ sources and $z>6$ sources in
this 2\,mm-selected sample is quite low compared to the $2<z<6$
sample.  Here our $z<3$ SFRD measurements are lower than measured from
other DSFG samples due to 2\,mm selection serving as an effective
filter for $1<z<3$ DSFGs, thus those points are shown as open stars to
contrast with filled stars at $z>3$, where the 2\,mm sensitivity is
more complete for $\simgt$2$\times$10$^{12}$\,\lsun\ DSFGs, as shown
in Figure~\ref{fig:lirz}. 

The total contribution of MORA 2\,mm-detected DSFGs at $3<z<6$ appears
to be roughly $\sim$30\%\ of the total cosmic SFRD at these epochs,
shown as the blue band in Figure~\ref{fig:sfrd}.
Note that what is shown in Figure~\ref{fig:sfrd} represents a {\it
  direct accounting} for the $>5\sigma$ MORA-detected sample only, not an
extrapolation from a luminosity function.  This is clearly seen in the
apparent deficit of the SFRD at $z<3$, where we know the MORA survey
was designed to filter sources out.

Our accompanying paper, Z21, provides a detailed analysis on MORA
constraints on the infrared luminosity function (IRLF) and the implied
contribution of obscured galaxies to the SFRD out to $z\sim7$. This is
done using a joint analysis of 1\,mm, 2\,mm, and 3\,mm number counts,
plus our empirical model \citep[described at length
  in][]{casey18a,casey18b,zavala18a} to measure the redshift evolution
of the IRLF.  In particular, the free parameters of the model are the
evolution of $\Phi_{\star}$ beyond $z\sim2$, the faint-end slope
$\alpha_{\rm LF}$ (assumed not to evolve with redshift for
simplicity), and the average emissivity spectral index of DSFGs
$\beta$ (assumed not to evolve for simplicity).  The data used to
constrain the model are all aggregate (sub)mm number counts, with
particular emphasis on 2\,mm and 3\,mm in their ability to capture the
high-$z$ redshift evolution of $\Phi_{\star}$.  Z21 finds that
$\Phi_{\star}\propto(1+z)^{\psi_2}$, where
$\psi_2=-6.5^{+0.8}_{-1.8}$, and the resulting {\it total}
contribution of obscured emission to the SFRD is shown by the orange
band in Figure~\ref{fig:sfrd}.

What is of particular note in our SFRD estimate is that our direct
accounting of MORA-detected galaxies agrees rather well with the best
estimate of the integrated IRLF inferred by number counts.  This
effectively means that the sources found in our map are the {\it only}
obscured sources to be found at these high redshifts, and there is not
likely to be a population of fainter sources lurking just below the
detection limit (or at least any population that contributes
significantly to cosmic star-formation at those epochs).  The same
cannot be said of rest-frame ultraviolet luminosity functions (UVLF),
for which there is often a significant discrepancy between the
integrated SFRD contribution between directly detected galaxies and
the inferred contribution from an extrapolation down the luminosity
function \citep{finkelstein15a,finkelstein16a}.  The key difference
between the IRLF and UVLF is, of course, their faint-end slopes; while
the UVLF has a steep faint-end slope, suggestive of a large population
of low-mass, UV-luminous galaxies, the IRLF has a very shallow
faint-end slope \citep[see also][for measurement of the faint-end
  slope of the IRLF from the ASPECS
  survey]{gonzalez-lopez20a,popping20a}.  This effect also manifests
in most of the CIB having been resolved into individual, bright point
sources \citep{bethermin12a}.  Our findings -- agreement between
direct accounting for MORA-detected sources to the SFRD and by
integrating the extrapolated IRLF -- are thus consistent with
$\simgt10^{12}\,$\lsun\ sources being the dominant source of obscured
emission.
This is functionally equivalent to most dust luminosity living in
massive galaxies, if indeed high-mass galaxies roughly correspond to
high-luminosity galaxies.

Also of note in our SFRD estimates is the relative agreement between
an estimate based {\it solely} on number counts (that from Z21) and
our results here, that incorporate both sources' inferred luminosities
(thus SFRs) as well as redshift constraints.  It is plausible that the
additional redshift information introduced here would lead to some
discrepancies with the Z21 model, either showing a flatter (or
steeper) SFRD contribution with redshift, but the agreement of the Z21
model with these data holds.  However, it should not be entirely
surprising that the two agree so well across $3<z<6$ given the root
assumption behind the Z21 model: that there are unlikely to be abrupt
kinks or changes in the IRLF with redshift.  While we nominally do not
find any $z>6$ DSFGs in the MORA survey, we know they exist even if
rare \citep{zavala18a,marrone18a}.  Measuring their volume density
will require larger area 2\,mm campaigns, discussed further in
\S~\ref{sec:cosmicvariance}.

We also draw comparison with the 870\um-selected DSFGs from the AS2UDS
survey in \citet{dudzeviciute20a}, who present the characteristics of
870\um\ SCUBA-2-selected DSFGs across a 1\,deg$^2$ area survey and
followed up with ALMA.  The contribution of sources above 1\,mJy at
870\um\ are shown in Figure~\ref{fig:sfrd} as gray points.  A direct
comparison of the 870\um-selection vs. 2\,mm selection technique is
aligned with expectation: 870\um\ efficiently selects DSFGs above
$z\simgt1.5$, with a high-redshift tail extending to $z\sim6$, while
2\,mm selection is weighted towards the higher redshifts only ($z>3$).
Both the AS2UDS and MORA results are well aligned with findings from
the {\sc Shark} model that predicts 2\,mm-selected galaxies contribute
between $\sim15-28$\%\ to cosmic star-formation at $3<z<6$
\citep{lagos20a}.

Lastly, it is worth noting that our measurements are in disagreement
with recent constraints on the $z>4$ IRLF from the ALPINE survey, who
infer an integrated SFRD contribution from dust obscured sources a
factor of $\sim$10 higher than found in this paper; ALPINE used
serendipitous detections in very deep ALMA pointings of $z\sim4-5$
galaxies (the primary goal being the measurement and characterization
of galaxies' [CII] characteristics) to place constraints on the
high-$z$ IRLF \citep{gruppioni20a} and cosmic star-formation rate
density more broadly \citep{loiacono21a,khusanova21a}.  This
discrepancy could be due to a bias in survey area used -- whereby
sources are physically associated with the ALPINE primary targets,
even if that association is not directly known -- or alternatively, could be due
to increased cosmic variance in the relatively small area of the
survey ($\sim$25\,arcmin$^2$ vs. 184\,arcmin$^2$ in MORA).

Recent findings from a similar higher-redshift line
  survey, the REBELS program, led to the discovery of two
  SFR$\sim$70\,\sfr\ systems at $z\sim7$ \citep{fudamoto21a}, whose
  OIR emission is completely obscured. \citeauthor{fudamoto21a}
  estimate their volume density and contribution to the SFRD,
  attempting to correct for the bias introduced by the targeted survey
  approach.  They predict a much higher contribution of lower IR
  luminosity ($\log$\lir$\sim$11.5--12.2) sources to the obscured SFRD,
  of order the same contribution as we estimate exists for higher
  luminosity sources ($\log$\lir$\simgt$12.2).  Indeed, the Z21 model
  would predict that such lower IR luminous sources
  ($\log$\lir$\sim$11.5--12.2) would have an order of magnitude lower
  contribution to the SFRD than measured by \citeauthor{fudamoto21a}.
  The origin of this discrepancy is unclear and cannot be resolved
  without a more extensive census of the lower luminosity regime.
  Nevertheless, detection of such ``OIR-dark'' obscured galaxies
  within the Epoch of Reionization represents a significant leap,
  indicating that even modest SFR galaxies may indeed be dust rich at
  $z>7$.

  Because ALPINE and REBELS are intrinsically targeted surveys and
  not blank field work, we omit their estimates from
  Figure~\ref{fig:sfrd} for clarity.

\subsection{Progenitors of first Massive, Quiescent Galaxies?}\label{sec:quiescent}

An important consequence of our measurement of the prevalence of DSFGs
at $z>3$ is their implications for the formation of the first massive
quiescent galaxies, already well-established three billion years after
the Big Bang \citep[at $z=2$, placing the formation redshift of their
  stellar mass at $z>3$;
  e.g.][]{kriek09a,toft14a,schreiber18b,merlin19a,marsan20a,santini21a}.
Furthermore, the discovery of a substantial population of
$M_\star\sim10^{11}\,$\msun\ galaxies at $z\sim4$
\citep{straatman14a,marsan20a,sherman20a,valentino20a,stevans21a} with
a volume density of (1.8$\pm$0.7)$\times10^{-5}$\,Mpc$^{-3}$ requires
a population of star-forming progenitors at $z\sim5$ with
SFRs\,$\sim$\,100\,\sfr.  Though constraints on the quiescent
population's volume density are yet uncertain, recent observational
works have constrained the number density to a
few$\times10^{-6}$ to $10^{-5}\,$Mpc$^{-3}$ \citep[see recent compilation
  of quiescent galaxy number densities in][]{valentino20a}.  The
population of DSFGs selected at 2\,mm \citep[or similarly, 3\,mm,
  see][]{zavala18a,williams19a} are such star-forming systems; does
their volume density, as a function of redshift, match the quiescent
galaxy population? While not all DSFGs will quench in time to
transition and become quiescent by $z\sim3-3.5$, a comparison of their
measured volume densities serves as a useful benchmark.

The raw volume density of MORA-detected galaxies (with a rough
luminosity cut of $\simgt2\times10^{12}$\,\lsun\ or 300\,\sfr) at
$4<z<6$ is $\sim$(3.8$^{+1.3}_{-0.7}$)$\times10^{-6}$\,Mpc$^{-3}$.
This requires a correction for the DSFGs' relatively short duty cycle,
$\sim$100\,Myr \citep{ivison11a,bothwell13c,swinbank14a}, shorter than
the period of the redshift interval over which the volume density is
calculated ($\sim$600\,Myr from $4<z<6$).  The corrected volume
density is $\sim$(2.2$^{+0.8}_{-0.4}$)$\times10^{-5}$\,Mpc$^{-3}$.
Note that this roughly agrees with other previous constraints on the
$z>4$ DSFG volume density
\citep[e.g.][]{da-cunha15a,michaowski17a,miettinen17a}, though here
the constraints are somewhat more precise.  Note that it is, however,
unclear if {\it all} DSFGs have universally short duty cycles, as the
star formation histories of quiescent galaxies suggest some long DSFG
lifetimes \citep[e.g.][]{schreiber18b,forrest20a}.

Within uncertainties, the volume density of MORA 2\,mm-detected DSFGs
is well aligned with constraints on $z\sim3.5$ quiescent galaxies.
Improved statistics from larger field 2\,mm surveys will substantially
reduce the uncertainty on the DSFG progenitor volume density at these
epochs though there remains substantial relative uncertainty in the
quiescent population.  This may be due, in large part, to
discrepancies between what galaxies qualify as quiescent or
star-forming.
For example, the density of $z\sim4$ quiescent
systems from \citet{straatman14a} presumes that all color-selected
$z\sim4$ sources {\it not} detected in the FIR with {\it Herschel} or
{\it Spitzer} are, indeed, quiescent.  If some additional fraction of
their sample are star-forming, the
MORA-inferred progenitor population would fall into alignment with
observed volume densities.  This is likely, as both {\it Herschel} and
{\it Spitzer} have shallow sensitivity at $z=4$, whereas longer
wavelength selection (starting at 850\um) can more effectively
segregate between quiescent and star-forming systems through FIR/mm
detection, particularly at $z>2$.
Further, well-studied identified quiescent systems
\citep{glazebrook17a} can reveal low levels of hidden star formation
at the sensitivity of deep ALMA observations
\citep{simpson17b,schreiber18c}.
Indeed, alternative estimates of the
quiescent population number density
\citep{muzzin13a,davidzon17a,girelli19a} are quite a bit lower
($\sim$few$\times10^{-6}$\,Mpc$^{-3}$), depending on the selection
criteria.

\begin{figure}
\includegraphics[width=0.99\columnwidth]{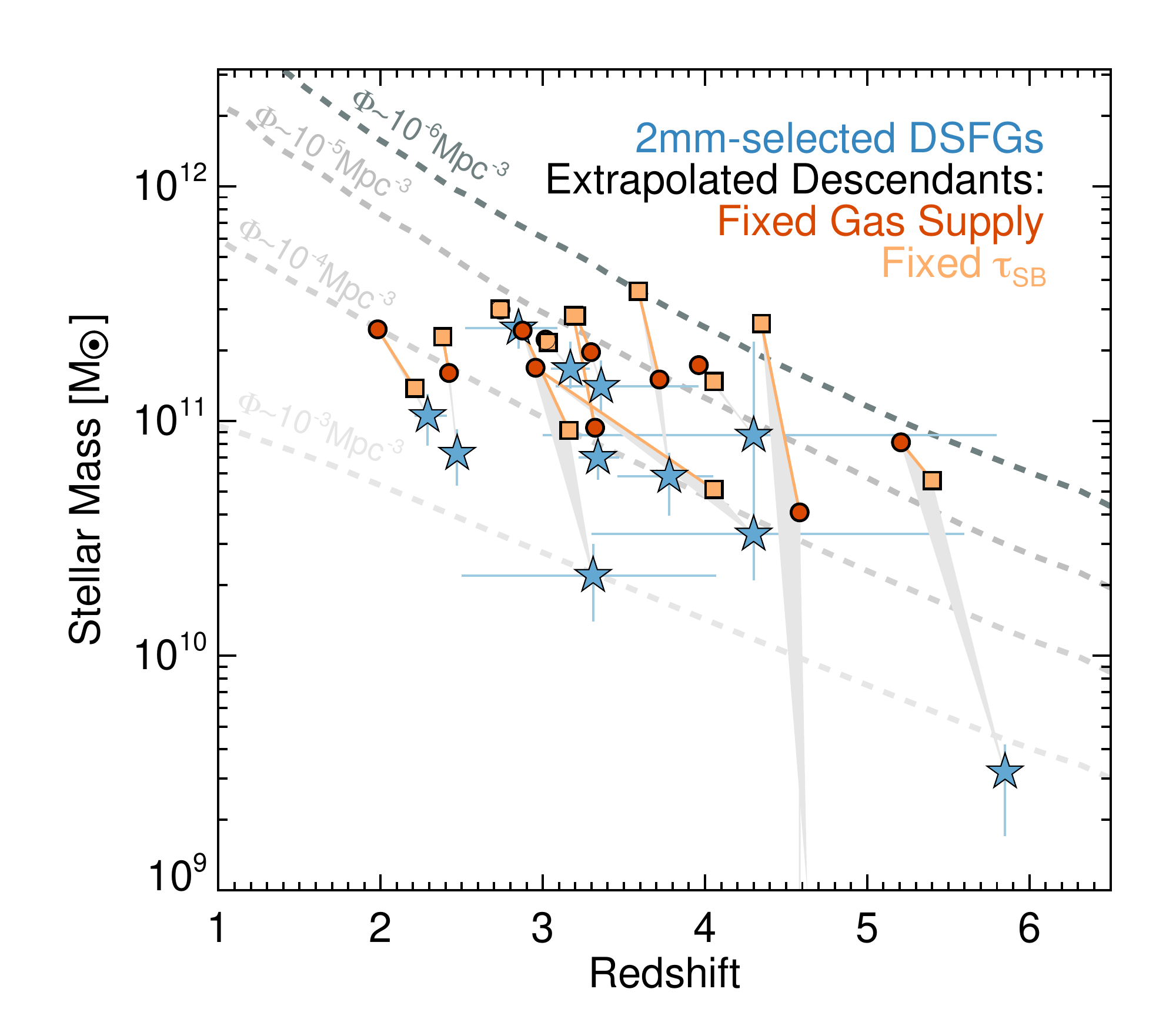}
\caption{Stellar mass with redshift for both the MORA sample of
  2mm-selected DSFGs (blue stars) and the extrapolated descendants.
  Two hypothetical descendants are shown per MORA source: one with
  fixed `starburst' duration of 100\,Myr where star-formation proceeds
  at the observed rate (light orange squares), and one with fixed gas
  supply, such that the current star-formation proceeds at the
  observed rate until the existing gas supply is depleted (dark orange
  circles).  Both descendants for each source are connected with an
  orange line, and back to the MORA progenitor by a gray triangle.
Gray dashed lines show the mass growth
  of halos of different rarities \citep[spanning
    10$^{-3}$ to 10$^{-6}$\,Mpc$^{-3}$;][]{behroozi18a} where halo
  masses have been scaled to stellar masses assuming a
  stellar-mass-to-halo-mass ratio of 2\%.  Note that the stellar mass
  of MORA-3 (a.k.a. AzTEC-2) at $z=4.6$ is unknown and here assumed to
  be extremely low, for lack of other information.  Overall we find
  that MORA descendant galaxies sit at or above 10$^{11}$\,\msun\ and
  are well established at those masses before $z=2$. }
\label{fig:descendants}
\end{figure}

Aside from the simple comparison of volume densities between MORA
DSFGs and quiescent galaxies at high-$z$, we can more closely consider
the established stellar masses of MORA-selected galaxies, their
potential for future star formation, and their likely descendant
population.  In other words, what stellar masses do we expect
MORA-descendant galaxies to have after eventually ceasing their star
formation? And at what redshift are those descendants fully formed?
Figure~\ref{fig:descendants} shows the extrapolated masses and
redshifts for two hypothetical quiescent descendant populations from the MORA sample of DSFGs.  
The first assumes the lion's share of the ISM mass (in the form of
H$_2$) is converted into stars at the current (observed) SFR until
that gas is fully depleted without replenishment\footnote{While the
  conversion of gas to stars is far from 100\%\ efficient
  \citep{evans09a}, the timescale of individual SF episodes is much
  shorter than the duration of DSFGs in their elevated SFR phase, thus
  such substantial conversions are reasonable to presume over long
  timescales \citep[see also][]{walter20a}.}. On average,
  the gas depletion time for the MORA sample is $\tau_{\rm
    depl}=110^{+30}_{-60}$\,Myr. The second adopts an
average starburst timescale of $\sim$100\,Myr \citep[in line with
  measurement of other DSFG samples, as well as the MORA sample
  herein, e.g.][]{bothwell13c} and presumes star formation will
continue at the observed high rate for that fixed period of time
before ceasing.
The redshifts of the descendant population of quiescent systems is
offset from the measured MORA redshifts by the measured or adopted gas
depletion time.
Note that there is no systematic offset of one
population of hypothetical descendants from the other; half of the
sample has higher masses and lower redshifts calculated with one
method\footnote{The median redshift and mass of the first (fixed gas
  supply) is $\langle z\rangle=3.31\pm0.02$ and
  (1.9$\pm$0.2)$\times10^{11}$\,\msun\ while the second (fixed
  starburst timescale) is $\langle z\rangle=3.24\pm0.03$ and
  (2.2$^{+0.3}_{-0.2}$)$\times10^{11}$\,\msun.}.
Depending on the existing stellar
reservoir, we find that anywhere between $\approx$\,20--100\%\ of the
eventual stellar mass in MORA descendants will form in the observed
star forming episode.  The final stellar masses of these systems is
then $\sim$2$\times10^{11}$\,\msun, with the entire MORA 2\,mm
sample producing galaxies with stellar masses in excess of
$>4\times10^{10}$\,\msun; this is well aligned with the measured mass
limits and redshift regimes of high-$z$ quiescent galaxy surveys.

\subsection{Is the emissivity spectral index steep, and/or does it evolve?}\label{sec:beta}

The dust emissivity spectral index, or $\beta$, governs the frequency
dependence of the emissivity of dust grains per unit mass, such that
$\kappa_{\nu}=\kappa_{0}(\nu/\nu_{0})^{\beta}$.  The mass absorption
coefficient, $\kappa_{\nu}$, traces the chemical and optical
properties of dust grains.  A lower value of $\beta$ observationally
manifests in a shallower Rayleigh-Jeans fall off to the SED in the
millimeter, while higher values result in steeper Rayleigh-Jeans fall
offs.

Data on high-redshift galaxies -- where the FIR/mm SED is spatially
unresolved and poorly sampled -- is of limited use in constraining the
intrinsic dust emissivity spectral index, which relates to fundamental
dust composition and likely varies on the scale of molecular clouds.
For example, \citet{arendt19a} measure an average value of
$\beta=2.25$ for the central molecular zone (CMZ) in the Galactic
plane of the Milky Way using new 2\,mm GISMO observations; this is
broadly in line with findings from {\it Planck}, which revealed
$\beta\approx1.6$ at high Galactic latitudes but increasingly steep
values, $\beta=1.8-2.0$ toward the inner Galactic plane
\citep[$|b|\ge10^{o}$][]{planck-collaboration14c}.
Despite our inability to constrain the underlying physical quantity
$\beta$, measurement of the slope of the Rayleigh-Jeans fall off, as
captured by the galaxy-integrated $\beta$ calculated herein, is useful.
Elucidating our view of the average galaxy-integrated $\beta$ is
useful for relating galaxies' dust SEDs to the (sub)millimeter
surveys used to identify DSFGs in the first instance, simulating
galaxies' SEDs and SED fitting to high-$z$ unresolved sources.

Our measurement of $\langle\beta\rangle=2.2^{+0.5}_{-0.4}$ within this
sample is consistent with the often assumed fixed value of
$\beta\equiv1.8$, though also suggestively a bit steeper
\citep[c.f. the prevailing theory based on ISM composition would
  suggest values of $\beta=1-2$;][]{draine11a}.  The analysis of
SCUBA-2 selected sources sitting inside the MORA mosaic (in
\S~\ref{sec:scuba2}) are similarly consistent with a range in $\beta$
that skew a bit higher than $\beta>1.8$, as shown in
Figure~\ref{fig:s2beta}.  How do we interpret these relatively steep
values of $\beta$ relative to shallower values in integrated SEDs of
well constrained local Universe dust \citep{dunne01a,clements10a}?
While this could be suggestive of a higher proportion of large
silicate grains in higher redshift dusty galaxies, it could also mark
different underlying ISM geometries.

Theoretically, low values (e.g. $\beta\approx1$) correspond to dust
comprised of small amorphous carbons with a mix of underlying dust
temperatures from warm to cold; this can be easily understood from the
SED that would result from coadding several blackbodies of different
temperatures yet similar masses: the net unresolved SED would have a
shallower Rayleigh-Jeans tail than a dust distribution of any one
temperature.  Indeed, coadding spatially-distinct SEDs over the scale
of an entire galaxy would always result in a {\it shallowing} of the
Rayleigh-Jeans tail and not a steepening.
One conjecture in the literature \citep[e.g.][]{jin19a} is that the
steepening of the Rayleigh-Jeans tail may be due to CMB heating of the
SED at high redshifts, which would cause a greater reduction in the
observed flux density at longer wavelengths relative to shorter
wavelengths \citep{da-cunha13a}.  However, CMB heating is likely only
a significant effect for galaxies at $z\simgt5$ at very low dust
temperatures ($T<30\,$K) where the CMB temperature is non-negligible;
furthermore, our SEDs do account for this CMB heating (however
negligible), and our derived $\beta$ values still hold after accounting
for this effect.
While steeper values of $\beta$ can also indicate the presence of
large crystalline silicate grains that reach overall lower
temperatures \citep{draine84a,agladze96a,meny07a}, the degeneracies
between spatial distribution of the ISM, dust temperature, opacity and
$\beta$, as well as the relative measurement uncertainties on our
measurements from a dearth of data prevents any direct constraints on
dust composition.  Still it is an interesting puzzle that our
galaxy-integrated $\beta$ seems to skew high relative to prevailing
literature \citep[see also][]{kato18a}, that has, for lack of direct
constraints, fixed the value to either $\beta=1.5$ \citep{paradis09a}
or $\beta=1.8$ \citep{planck-collaboration11a} from measurements of
the Milky Way's ISM in both atomic and molecular states.

Our measurements of $\beta=2.2^{+0.5}_{-0.4}$, along with other recent
direct measurements of $\beta$ in high-$z$ DSFGs \citep[see ][who
  measure $\beta=1.9\pm0.4$ for the ALESS sample via dedicated 2\,mm
  follow-up observations]{da-cunha21a}, suggest that a higher
$\beta\approx2$ is more appropriate than $\beta=1.5$ or $\beta=1.8$ in
instances where a DSFGs' Rayleigh-Jeans tail is not directly
constrained and a choice of $\beta$ is needed to model sources' SEDs.

\subsection{Potential Impact of Cosmic Variance on $z>3$ DSFG sample}\label{sec:cosmicvariance}

\begin{figure}
\includegraphics[width=0.99\columnwidth]{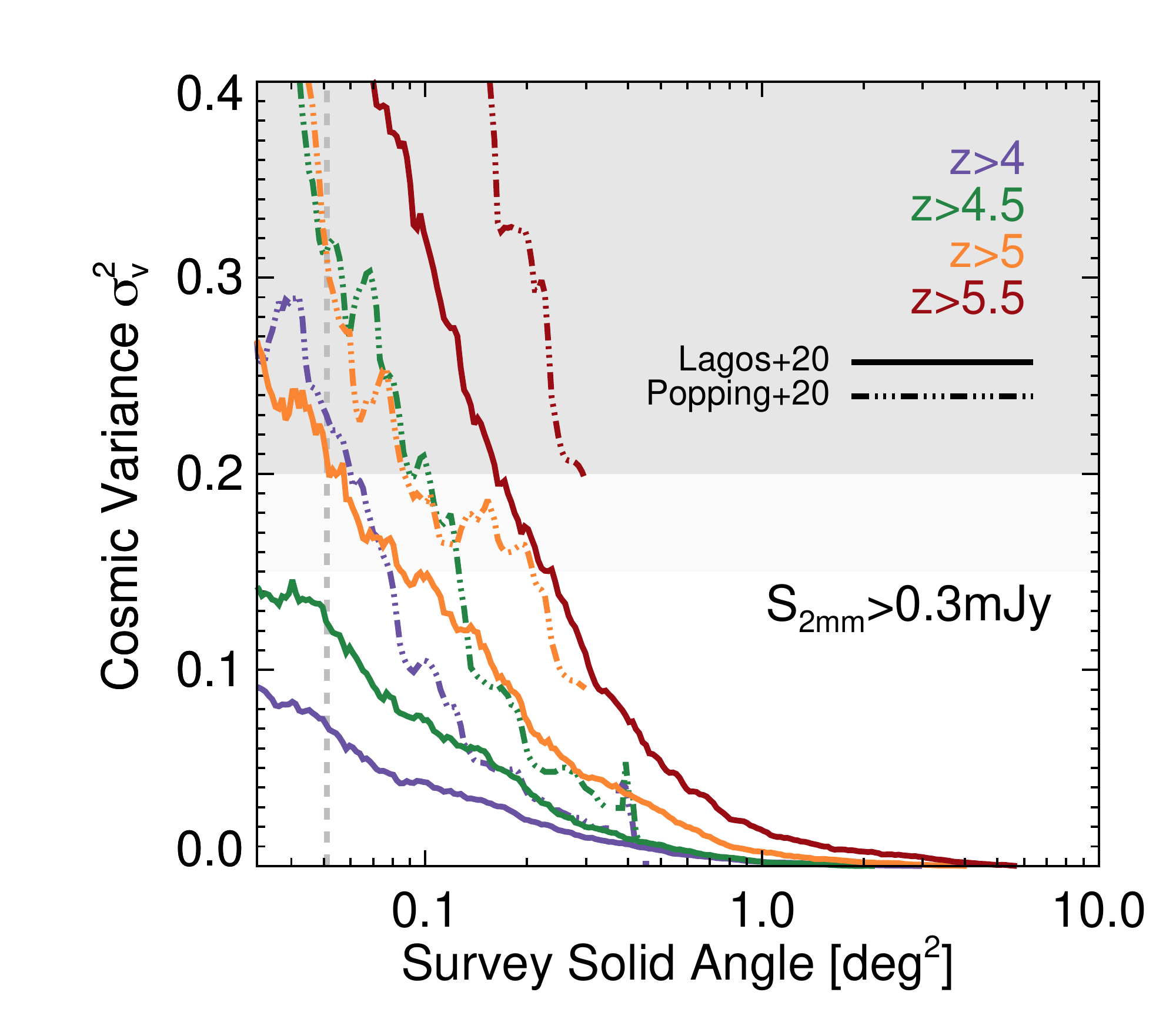}
\caption{Cosmic variance, $\sigma^2_{\rm v}$, as a function of survey
  solid angle for 2mm-detectable galaxies above $S_{\rm 2mm}>$0.3\,mJy
  and a range of redshifts drawn from: the 108\,deg$^2$ {\sc Shark}
  lightcone \citep[solid lines;][]{lagos20a} and the 7.7\,deg$^2$ {\sc
    UniverseMachine} lightcone \citep[dot-dashed lines][]{popping20a}.
  The gray shaded regions denote $\sigma_{\rm v}^2>0.15$ (and
  $>$0.20), a threshold representing significant (severe) cosmic
  variance.  We show cosmic variance for sources above redshifts $z=4$
  (purple lines), $z=4.5$ (green lines), $z=5$ (orange lines), and
  $z=5.5$ (red lines).  Sources at higher redshifts are exceedingly
  rare ($\sim$1\,deg$^{-2}$) in both {\sc Shark} and {\sc
    UniverseMachine} such that direct estimates of cosmic variance
  $\sigma_v^2$ cannot be reliably estimated. The size of the MORA
  survey is indicated with the vertical gray dashed line.  We find
  that 2\,mm sources are roughly a factor of $\sim$2$\times$ more rare
  in the {\sc UniverseMachine} lightcone relative to {\sc Shark}, and
  so the cosmic variance is appreciably higher.
 This work suggests a significant jump in survey area is needed --
 exceeding 0.2-0.3\,deg$^2$ -- to take an accurate census of DSFGs
 beyond $z>5.5$, while surveys of order a few tenths of a square
 degree are adequate at lower redshifts.  It is likely that surveys
 exceeding 1\,deg$^2$ will be needed to sample $z>6$ DSFGs well.}
\label{fig:cosmicvariance}
\end{figure}

Though our MORA survey area is among the largest mosaics stitched
together by ALMA to date, a chief concern of our analysis is that
cosmic variance may impact our measurement of the ubiquity of DSFGs
beyond $z>3$.  In other words, our relatively small area --- small in
comparison to the relative rarity of DSFGs themselves --- may
oversample or undersample the average density of the Universe at any
given epoch.  Our parallel paper, Z21, tests the impact of cosmic
variance using dust emission models applied to galaxies inside large
volume simulations \citep[see Z21, their section 4.4, as well
  as][]{popping20a}. In those simulations, Z21 determines that the MORA
survey volume is robust in the measurement of the number counts of
2\,mm sources out to $z\sim7$.  

Here we extend the cosmic variance analysis of Z21 to analyze the
cosmic variance specifically for the $z>4$ high-redshift tail of the
2\,mm-observed population, and also incorporating predictions from the
{\sc Shark} semi-empirical model \citep{lagos20a}.  The definition of cosmic variance
as given in \citet{moster11a} is:
\begin{equation}
\sigma_v^2 = \frac{\langle N^2\rangle - \langle N\rangle^2 - \langle
  N\rangle}{\langle N\rangle^2},
\end{equation}
where $N$ represents the number of galaxies detectable in a given
survey solid angle above a redshift $z$.  In other words, we generate
Monte Carlo trial mock surveys with increasing area $\Omega$ within
the \citet{popping20a} and \citet{lagos20a} lightcones and assess how
many sources in these lightcones would be 2\,mm detectable (at the
rough sensitivity of the MORA survey) above a given redshift.  At a
fixed survey area, $\Omega$, we generate a distribution of $N$ that
satisfy $S_{\rm 2mm}>0.3$\,mJy and $z>4$ (or other higher redshift
minima); from that distribution of $N$, we are able to easily compute
$\langle N\rangle$ and $\langle N^2\rangle$ from which we calculate
$\sigma_v^2$.

Our results are shown in Figure~\ref{fig:cosmicvariance} where surveys with $\sigma_{\rm v}^2<0.15$ are less prone to the effects of cosmic variance.
We find that, due to the relative rarity of dusty sources in both {\sc
  Shark} and {\sc UniverseMachine}, the MORA survey size
($\sim$0.05\,deg$^2$) is prone to cosmic variance above $z>4$
(according to {\sc UniverseMachine}) and $z>5$ (according to {\sc
  Shark}).  The {\sc UniverseMachine} generated lightcone has roughly
a factor of $2-3\times$ fewer high SFR (SFR$\simgt$100\,\sfr) galaxies
than {\sc Shark} at these redshifts and so the estimated cosmic
variance is higher.  Both models imply that significantly larger areas
are needed to survey the $z>5.5$ Universe, of order 0.2-0.3\,deg$^2$
or greater to suppress cosmic variance below $\sigma_v^2<0.15$.  Given
the dearth of $z>6$ DSFGs in both models, it is hard to characterize
what survey area would be necessary to mitigate cosmic variance at
those epochs, though naturally we may expect it to be
$\simgt$\,1\,deg$^2$.

Note too that the uncertainties for the predicted model redshift
distributions and number counts, shown in Figure~\ref{fig:nz}, account
for the same cosmic variance discussed here by sampling different
regions of either lightcone over the 184\,arcmin$^2$ MORA survey
footprint.  Those estimates account for the heterogeneous noise
characteristics of the MORA survey, while the cosmic variance
estimates in Figure~\ref{fig:cosmicvariance} assume a uniform flux
density threshold of $S_{\rm 2mm}>0.3$\,mJy.

\subsection{Potential Degeneracies in Interpretation}\label{sec:varymodel}

This paper has so far argued that 2\,mm continuum observations can
provide great clarity on the earliest epoch of dust-obscured galaxies
at $z\simgt4$, but how might our conclusions be impacted by
astrophysical unknowns?  The MORA survey represents an important
step in constraining the obscured Universe in this early epoch, but
further refinements will require data to break certain degeneracies,
including unknowns about a possible evolving shape in the IRLF,
galaxies' evolving dust SEDs, and dust emissivities.

For example, there is substantial evidence to suggest that the
faint-end slope of the UVLF evolves, such that it is much steeper at
higher redshifts \citep{finkelstein16a}.  The physical interpretation
of this is that there are far more low mass galaxies at early times,
and with time the distribution of galaxies in the luminosity function
is distributed across a larger dynamic range.  Does the IRLF evolve
similarly, differently, in the opposite sense, or not at all?

As presented in Figure~\ref{fig:sfrd}, our data are consistent with a
non-evolving faint-end slope of the IRLF, though the constraints do
not rule out such an evolution.  Being able to measure such an
evolution will require more painstakingly deep observations (as can be
provided by ALMA with deep 1--2\,mm surveys at $<$1\,mJy depth, like
ASPECS, \citealt{gonzalez-lopez20a}) over significantly larger areas
($\simgt$0.2\,deg$^2$).  Furthermore, source redshifts -- likely most
efficiently obtained through large-area blind searches for millimeter
spectral lines -- will be paramount to this measurement.  At present,
no large field of view millimeter spectrometer exists with the needed
sensitivity.  Once in hand, the combination of a well-measured UVLF
and IRLF, as well as the nature of their evolution, can directly
inform models of early universe dust formation and attenuation.

The question of evolution in galaxies' dust SEDs has come up
frequently in recent literature \citep[e.g.][]{ma19a,liang19a} and may
have significant impact on the results discussed herein and in Z21.
For example, if $z>5$ galaxies contain, on average, substantially
hotter dust than lower redshift DSFGs as such works suggest (with correspondingly lower dust
mass per fixed \lir), then it is likely that 2\,mm dust continuum maps
could miss upwards of half of all DSFGs at early epochs.  In other
words, if the average dust SED is hotter at high redshifts, then those
SEDs peak at shorter rest-frame wavelengths and are likely to have
lower flux densities on the Rayleigh-Jeans tail (2\,mm observed) than
colder DSFGs.  This would lead to a smaller fraction of DSFGs being
detected at 2\,mm, and it would be logical to conclude that our
results have under-estimated the total volume density of early
Universe DSFGs \citep[a point raised as a hypothetical
  in][]{casey18a}.  If that is the case, then some tension may exist
between the inferred volume density of all DSFGs and the measured
volume density of lower redshift quiescent galaxies (as discussed in
\S~\ref{sec:quiescent}).  Nevertheless, with exception of very
luminous lensed systems \citep[e.g.][]{reuter20a} the spectral energy
distributions of DSFGs at $z>4$ are not well constrained as an
aggregate population. Measuring them will provide invaluable insight
into both ISM physics in the first few Gyr as well as further
constraints on the volume density measurements of the population
themselves.

Variation in the dust emissivity index may also impact our
volume density constraints in a similar fashion to variation in the
aggregate dust temperatures of early DSFGs.  With a steeper $\beta$,
flux densities on the Rayleigh-Jeans tail drop; for example, the 2\,mm
flux density of a galaxy at $z=4$ would be three times lower for a
very steep $\beta=3.5$ than for a shallower value $\beta=1.5$,
presuming a fixed value of \lir\ and \lpeak.  As discussed in
\S~\ref{sec:beta}, our sample does show moderate evidence for steeper
emissivity spectral indices across the sample.  If DSFGs on a whole
were to exhibit steeper Rayleigh-Jeans tails it may imply that our
volume density measurements here, and in Z21, are underestimated.

Lastly, it is crucial to point out that many MORA source redshifts are
not yet well constrained.  Given our small sample size, even one
catastrophically mis-identified redshift would shift our integrated
SFRD results by $\sim$10\%, and thus has potential to alter our
perception of the IRLF's evolution beyond $z>3$.

\subsection{Future Surveys of $z\simgt4$ Obscured Sources}

Future 2\,mm surveys of large areas of sky will be crucial to overcome
limitations of small sample sizes and cosmic variance, particularly in
pursuit of rare, dusty starbursts at $z>4$.  In particular, testing
the divergent redshift distribution predictions for bright DSFGs
($S_{\rm 2mm}>1\,$mJy; see Figure~\ref{fig:s2mmz}) in the simulations
will be an essential next step.  While the {\sc Shark} and {\sc
  UniverseMachine} simulations predict that the brightest 2\,mm
sources have a relatively low median redshift ($\langle
z\rangle\sim2.5$),
the SIDES, Casey \etal\ and Zavala \etal\ models
suggest a higher average redshift
($\langle z\rangle\simgt3.5$).  This difference originates with the
fundamental problem of how massive, bright galaxies assemble {\it
  en-masse} so quickly after the Big Bang.  Sampling these bright
sources in statistically large samples well will truly require a
different scale of millimeter continuum survey, of order
$\simgt$5\,deg$^2$, as such bright high-$z$ systems are predicted to
be extremely rare in the models (e.g. SHARK predicts five $z>4$
sources per square degree above 1\,mJy at 2\,mm, and only one such
source above $z>5$ per square degree).  While large-field searches of
{\it Herschel} SPIRE imaging for 500\um-peaking ``red'' sources has
been fruitful \citep{ivison16a,duivenvoorden18a,yan20a}, the population is quite rare
($\sim$10$^{-7}$\,Mpc$^{-3}$) relative to the MORA-detected DSFGs
presented herein, due to the relative shallow sensitivity of {\it
  Herschel} at high redshifts.
Instruments like GISMO, which has pioneered much of the 2\,mm
blank-field mapping work to-date, and TolTEC, which in the future has
the potential to expand 2\,mm sky coverage by orders of magnitude,
will play essential roles in pushing this frontier.  Similarly, the
South Pole Telescope survey has already demonstrated the effectiveness
of $>$1\,mm surveys in recovering an intrinsically high-$z$ population
\citep[see][]{vieira13a,reuter20a}, and the next generation SPT survey
(`SPT3G') will push the depth such that unlensed sources will be
detected at high significance.

One complementary observing strategy to large surveys would be
dedicated 2\,mm--3\,mm continuum follow-up of DSFGs already identified
at shorter wavelengths, allowing a more efficient redshift selection
of the highest-$z$ sources using millimeter colors.  Every source in
the MORA Survey (save one that we attribute to a false noise peak) is
detected above 3.5\,$\sigma$ significance at 850\um, and many at other
(sub)mm wavelengths from existing single dish surveys.  As
Figure~\ref{fig:s2beta} shows, the MORA sources with lower $S_{\rm
  870}/S_{\rm 2000}$ ratios sit at the highest redshifts as would be
expected for sources whose 850\um\ emission is closer to the peak of
dust emission. Though strategic 2\,mm or 3\,mm follow-up of 850\um\ or
1\,mm-selected sources is more biased than the blank field survey
strategy, it is far more observationally efficient for samples that
have already been identified, requiring only a few minutes of
on-source time with ALMA.  It is an effective, and observationally
cheap filter through which needles ($z>4$ DSFGs) can be spotted in the
haystack (dominated by $1<z<3$ DSFGs), complementing 2\,mm blank field
surveys like MORA.

Lastly, extentions of MORA stand to make significant progress; over
the next year, MORA is approved for observations to expand coverage
over 0.2\,deg$^2$ (within the COSMOS-{\it Webb} survey footprinto) to
the same depth of our current mosaics.  These data will likely lead to
the discovery of $\sim$20 DSFGs at $z>4$ and enable more detailed
studies of DSFG clustering.

\section{Conclusions}\label{sec:conclusions}

We have presented the Mapping Obscuration to Reionization with ALMA
(MORA) Survey, the first 2\,mm extragalactic blank field map with ALMA
covering 184\,arcmin$^2$.  Our accompanying paper, \citet{zavala21a},
presents the MORA Survey number counts and number count-constrained
obscured star-formation rate density measurement out to $z\sim7$,
while this paper focuses on what is known about the individual
galaxies that have been detected.  We detect thirteen sources above
$>$5$\sigma$ significance, one of which may have significant
contribution from radio synchrotron emission and another we believe to
be a false-positive noise peak, leaving a total of eleven robust
detections dominated by thermal dust emission.  The redshifts of the
sources span $z=2.2-5.9$ with a median redshift of $\langle z\rangle =
3.6^{+0.4}_{-0.3}$.  Sources span IR luminosities $10^{12}-{\rm
  few}\times10^{13}$\,\lsun, ISM masses ${\rm few}\times10^{10}-{\rm
  few}\times10^{11}$\,\msun, and sit both above and embedded within
the galaxy SFR-stellar mass relation.
 
Our results paint a plausible picture of the build-up of dust-obscured
galaxies at the earliest epochs ($z>4$): overall, DSFGs at this epoch
are rare and contribute a sizable, though non-dominant, fraction
towards cosmic star formation.  These DSFGs are the most likely
progenitors of the Universe's first massive, quiescent galaxies.  Our
primary conclusions are as follows:
\begin{enumerate}
\item ALMA Band 4 (2\,mm) is an effective selection wavelength for
  high-$z$ DSFGs; here we find 77$\pm$11\,\%\ of our 2\,mm-selected sample lies
  above $z>3$ and 38$\pm$12\,\%\ above $z>4$. This contrasts with more
  traditional methods of DSFG selection at 850\um\ or 1\,mm, whose
  proportion of sources at these redshifts is much smaller
  \citep[e.g. 6\%\ of 850\um-selected sources sit at
    $z>4$;][]{dudzeviciute20a}.
\item ``OIR-dark'' sources (i.e. sources lacking counterparts in very deep
  optical and near-infrared data, $<$1.6\,\um) make up a sizable
  fraction of 2\,mm selected sources (4/11$\sim$36\%). Our
  accompanying paper, Manning \etal\ in prep further explores their
  physical characteristics.
\item Several semi-empirical and empirical models from the literature accurately
  reproduce the redshift distribution and volume density of
  2\,mm-detected galaxies.  However, several details are not yet
  resolved by data, namely the epoch when the first DSFGs turn on, and
  how common such systems are at the highest redshifts, $z\simgt5.5$.
  Different cosmological models infer that MORA may be impacted by
  cosmic variance above $z>5$, with the results highly
  dependent on dust radiation prescriptions in early Universe
  galaxies.  Either way, larger area 2\,mm continuum surveys will be
  essential to gather sufficient statistics to constrain the volume
  density of DSFGs at the highest-redshift regimes.
\item Dusty galaxies with star-formation rates in excess of
  300\,\sfr\ contribute $\sim$30\%\ to SFRD between
  $3<z<6$. Furthermore, at these epochs, the IRLF is dominated by
  these sources, with little contribution from galaxies with lower
  levels of obscured star formation ($\simlt$\,100\,\sfr).  This implies that
  the vast majority of cosmic dust lives in such ultraluminous
  galaxies at these epochs, with relatively little to no dust
  impacting the bolometric output of less luminous, less massive
  systems.
\item DSFGs at $z>3$, like those detected in the MORA Survey, are
  often claimed to be the progenitors of the Universe's earliest
  quiescent systems; we find that the number density of quiescent
  galaxies from the literature agrees well with the measured volume
  density of DSFGs measured at $z>3$ ($\sim$10$^{-5}$\,Mpc$^{-3}$).
  Furthermore, the extrapolated characteristics of MORA DSFGs'
  descendants are also well aligned with detected quiescent
  populations, all well established with stellar masses
  $>$10$^{11}$\,\msun\ above $z>2$.
\item Our MORA data hints at a possible higher value for the
  galaxy-integrated dust emissivity index, $\beta$, than is often
  assumed in the literature: the sample has a measured
  $\langle\beta\rangle=2.2^{+0.5}_{-0.4}$ vs $\beta=1.8$.  We
  recommend future works modeling SEDs of high-$z$ DSFGs adopt a value
  of $\beta\approx2$ as needed in lieu of direct constraints.
\end{enumerate}
The final frontier of this work would lead to the detection of the
Universe's first dusty galaxies, the home of the Universe's first
substantial dust reservoirs.  Do such systems exist beyond $z>7$?
$z>8$?  While MORA has taken a tantalizing first step in completing
the census of high-$z$ DSFGs, we know higher redshift DSFGs
(specifically, three with ${\rm SFR}\simgt100$\,\sfr) do exist than
are found in this work
\citep[e.g.][]{cooray14a,zavala18c,strandet17a,marrone18a}.  Yet,
their rarity -- itself not yet well constrained -- implies that
finding more requires large area millimeter surveys sufficiently
sensitive to detect unlensed galaxies, similar conceptually to the
all-sky searches for the highest redshift quasars found by, e.g., the
Sloan Digital Sky Survey \citep{fan01a,fan03a}.  And though rare, the
earliest dusty starbursts yet to be found serve as unique laboratories
for our understanding of the assembly of the first galaxies and,
fundamentally, the formation of dust, on which much of what we know
about the modern day Universe relies.

\acknowledgements

We thank the anonymous reviewer for helpful suggestions which greatly
improved the manuscript.  This paper makes use of the ALMA program
ADS/ JAO.ALMA \#2018.1.00231.S.  ALMA is a partnership of ESO
(representing its member states), NSF (USA) and NINS (Japan), together
with NRC (Canada), MOST and ASIAA (Taiwan), and KASI (Republic of
Korea), in cooperation with the Republic of Chile. The Joint ALMA
Observatory is operated by ESO, AUI/NRAO and NAOJ. The National Radio
Astronomy Observatory is a facility of the National Science Foundation
operated under cooperative agreement by Associated Universities, Inc.
CMC thanks the National Science Foundation for support through grants
AST-1714528, AST-1814034, and AST-2009577 and additionally the
University of Texas at Austin College of Natural Sciences for support.
In addition, CMC acknowledges support from the Research Corporation
for Science Advancement from a 2019 Cottrell Scholar Award sponsored
by IF/THEN, an initiative of Lyda Hill Philanthropies.
The Flatiron Institute is supported by the Simons Foundation.
MT acknowledges the support from grant PRIN MIUR 2017.
ET acknowledges support from CATA-Basal AFB-170002, FONDECYT Regular
grant 1190818, ANID Anillo ACT172033 and Millennium Nucleus NCN19\_058
(TITANs).
MA acknowledges support from FONDECYT grant 1211951,
``CONICYT+PCI+INSTITUTO MAX PLANCK DE ASTRONOMIA MPG 190030'' and
``CONICYT+PCI+REDES 190194.''
The Cosmic Dawn Center (DAWN) is funded by the Danish National
Research Foundation under grant No. 140. ST and JW acknowledge support
from the European Research Council (ERC) Consolidator Grant funding
scheme (project ConTExt, grant No. 648179).
GEM acknowledges the Villum Fonden research grant 13160 “Gas to stars,
stars to dust: tracing star formation across cosmic time,” grant
37440, ``The Hidden Cosmos,'' and the Cosmic Dawn Center.
AWSM acknowledges the support of the Natural Sciences and Engineering
Research Council of Canada (NSERC).
KIC acknowledges funding from the European Research Council through
the award of the Consolidator Grant ID 681627-BUILDUP.

\bibliography{caitlin-bibdesk}

\begin{longrotatetable}
\begin{deluxetable*}{c@{ }c@{ }c@{ }c@{ }c@{ }c@{ }c@{ }c@{ }c@{ }c@{ }c@{ }c@{ }c@{ }c@{ }c@{ }c}
\tablecaption{{\bf Photometry of the $>$5$\sigma$ 2\,mm
  Sample\label{tab:sources}}.}
\tablewidth{700pt}
\tabletypesize{\scriptsize}
\tablehead{
\colhead{\sc Name} & 
\colhead{{\sc Position}} & 
\colhead{{\sc SNR}$_{\rm 2mm}$} & 
\colhead{S$_{\rm 2\,mm}$} &
\colhead{$H$-band} & 
\colhead{S$_{\rm 3.6}$ } & 
\colhead{S$_{\rm 24}$ } & 
\colhead{S$_{\rm 100}$ } & 
\colhead{S$_{\rm 160}$ } & 
\colhead{S$_{\rm 250}$ } & 
\colhead{S$_{\rm 350}$ } & 
\colhead{S$_{\rm 450}$ } & 
\colhead{S$_{\rm 500}$ } & 
\colhead{S$_{\rm 850}$ } & 
\colhead{S$_{\rm 1.2\,mm}$ } & 
\colhead{S$_{\rm 3GHz}$}\\
\colhead{} &
\colhead{ ($\alpha_{\rm J2000}$, $\delta_{\rm J2000}$) } & 
\colhead{} & 
\colhead{[\uJy]} &
\colhead{[AB]} & 
\colhead{[nJy]} & 
\colhead{[\uJy]} & 
\colhead{[mJy]} & 
\colhead{[mJy]} & 
\colhead{[mJy]} & 
\colhead{[mJy]} & 
\colhead{[mJy]} & 
\colhead{[mJy]} & 
\colhead{[mJy]} & 
\colhead{[mJy]} & 
\colhead{[\uJy]}
}{
\startdata
MORA-0  & 10:00:15.617 $+$02:15:49.00 & 7.90 & 818$\pm$103  & --- & 1714$\pm$40 & [21$\pm$27] & [$-$1.5$\pm$1.7] & [0.6$\pm$3.7] & [8.9$\pm$5.8] & [18.7$\pm$6.3] & [10.7$\pm$5.8] & [20.3$\pm$6.8] & 13.56$\pm$0.12$^\dagger$ & 4.13$\pm$0.18 & 19.5$\pm$2.6 \\ 
MORA-1  & 10:00:19.740 $+$02:32:03.80 & 7.74 & 703$\pm$91   & 22.68$\pm$0.02 & 6958$\pm$94 &  189$\pm$13 & [0.0$\pm$1.7] & 31.4$\pm$3.7 & 37.6$\pm$5.8 & 48.4$\pm$6.3 & 29.2$\pm$5.2 & 35.7$\pm$6.8 & 8.67$\pm$0.06$^\dagger$ & 2.67$\pm$0.09 & 74.8$\pm$0.9$^{*}$  \\
MORA-2  & 10:00:10.146 $+$02:13:35.00 & 7.58 & 529$\pm$70   & 24.24$\pm$0.04 & 5223$\pm$100 & [18$\pm$25] & [3.9$\pm$1.7] & 13.1$\pm$3.8 & 15.4$\pm$5.8 & 31.3$\pm$6.3 & [17.2$\pm$9.4] & 27.7$\pm$6.8 & 8.82$\pm$0.71 & 2.66$\pm$0.05 & 30.1$\pm$2.9  \\
MORA-3  & 10:00:08.037 $+$02:26:12.20 & 7.25 & 1220$\pm$168 & ... & ... & ... & [0.3$\pm$1.7] & [1.0$\pm$3.7] & [16.8$\pm$5.8] & 30.8$\pm$6.3 & 23.5$\pm$4.8 & 29.1$\pm$6.8 & 16.75$\pm$0.15$^\dagger$ & 4.62$\pm$0.11 &  15.0$\pm$2.4 \\
MORA-4  & 10:00:26.359 $+$02:15:28.00 & 6.68 & 557$\pm$83   & --- & 87$\pm$29 & [10$\pm$18] & [0.5$\pm$1.5] & [-0.6$\pm$2.8] & [2.9$\pm$5.8] & [2.9$\pm$6.3] & [2.3$\pm$5.8] & [4.9$\pm$6.8] & 5.908$\pm$0.052$^\dagger$ & 2.05$\pm$0.11 & [10.6$\pm$4.1] \\
MORA-5  & 10:00:24.157 $+$02:20:05.40 & 6.63 & 584$\pm$88   & --- & 406$\pm$24 & [-8$\pm$26] & [-0.2$\pm$1.7] & [-0.1$\pm$3.7] & [3.8$\pm$5.8] & [7.3$\pm$6.3] & [10.7$\pm$4.1] & [7.1$\pm$6.8] &  6.80$\pm$0.53 & 2.30$\pm$0.10 & [10.1$\pm$3.4] \\
MORA-6  & 10:00:28.723 $+$02:32:03.40 & 6.37 & 615$\pm$97   & 23.75$\pm$0.03 & 4332$\pm$30 & 198$\pm$13  & [-0.4$\pm$1.7] & 19.3$\pm$3.3 & 34.7$\pm$5.8 & 35.0$\pm$6.3 & 40.6$\pm$5.2 & 24.8$\pm$6.8 & 12.57$\pm$0.15$^\dagger$ & 3.13$\pm$0.10 & 45.5$\pm$0.8$^{*}$ \\
MORA-7  & 10:00:11.574 $+$02:15:05.20 & 6.29 & 378$\pm$60   & 23.52$\pm$0.02 & 11956$\pm$80 & 175$\pm$16 & [0.6$\pm$1.8] & [-0.3$\pm$3.8] & [16.1$\pm$5.8] & [17.0$\pm$6.3] & 20.7$\pm$6.6 & [9.5$\pm$6.8] &  5.31$\pm$0.16$^\dagger$ & 1.51$\pm$0.13 & 12.2$\pm$2.3 \\
MORA-8  & 10:00:25.292 $+$02:18:46.20 & 5.75 & 489$\pm$85   & 23.21$\pm$0.02 & 11156$\pm$79 & 204$\pm$29 & [-1.7$\pm$1.9] & [0.6$\pm$3.8] & [16.2$\pm$5.8] & 24.7$\pm$6.3 & 15.2$\pm$4.3 & 26.8$\pm$6.8 &  7.12$\pm$0.54 & ... & 34.7$\pm$2.9 \\
MORA-9  & 10:00:17.298 $+$02:27:15.80 & 5.55 & 379$\pm$68   & 26.75$\pm$0.30 & 469$\pm$33 & [1$\pm$13] & [-0.5$\pm$1.7] & [-0.3$\pm$3.7] & [0.0$\pm$5.8] & [0.0$\pm$6.3] & [1.1$\pm$4.1] & [0.0$\pm$6.8] &  2.59$\pm$0.37$^\dagger$ & ... & [4.3$\pm$2.4] \\
MORA-10 & 10:00:16.578 $+$02:26:38.00 & 5.21 & 405$\pm$78   & 22.77$\pm$0.01 & 10766$\pm$66 & 890$\pm$17 & 5.8$\pm$1.7 & 18.9$\pm$4.0 & [15.6$\pm$5.8] & [17.4$\pm$6.3] & 20.4$\pm$4.1 & [8.5$\pm$6.8] &  6.07$\pm$0.07$^\dagger$ & ... & 3212.0$\pm$1.6$^{*}$ \\
MORA-11 & 10:00:12.922 $+$02:12:11.40 & 5.16 & 356$\pm$69   & 22.64$\pm$0.01 & 13322$\pm$199 & [22$\pm$44] &  [0.5$\pm$1.6] & [0.4$\pm$3.1] & 35.7$\pm$15.8 & 40.0$\pm$20.3 & [7.6$\pm$14.4] & 41.9$\pm$25.9 &  5.10$\pm$0.78 & 1.49$\pm$0.07 & 12.1$\pm$2.4 \\
MORA-12 & 10:00:04.713 $+$02:29:55.20 & 5.03 & 950$\pm$190  & --- & [1$\pm$30] & [1$\pm$12] & [0.7$\pm$1.7] & [0.4$\pm$3.7] & [13.8$\pm$5.8] & [14.2$\pm$6.3] & [1.2$\pm$4.8] & [5.7$\pm$6.8] &  [0.22$\pm$0.69] & ... & [1.0$\pm$2.3] \\
\enddata
}
\end{deluxetable*}

\hspace{-\columnwidth}
\begin{minipage}{\textwidth}
{\bf Table Notes.}  Photometry enclosed in brackets represent
$<$3$\sigma$ detections.  Ellipses (...) indicate an absence of data
while long dashes (---) represent a non-detection in the COSMOS2020
catalog.  Note that MORA-3 has highly confused {\it Spitzer} imaging,
blending MORA-3 with two lower redshift foreground sources within the
{\it Spitzer} beam at both 3.6\um\ and 24\um.  A dagger ($\dagger$)
indicates that the flux density in the 850\um\ column has been
replaced by improved 870\um\ photometry from ALMA rather than SCUBA-2.
An asterisk ($^{*}$) indicates that the 3\,GHz flux density comes from
the deeper COSMOS-XS maps of \citet{van-der-vlugt21a} and
\citet{algera20a} rather than the shallower/wider coverage of
\citet{smolcic17a}.  Note that some of the sample have ALMA band 6
data from archival datasets (recorded here at 1.2\,mm).
\end{minipage}

\label{tab:photometry}
\end{longrotatetable}

\startlongtable
\begin{deluxetable*}{c@{ }c@{ }c@{ }c@{ }c@{ }c@{ }c}
\tablecaption{{\bf Positions and Multiwavelength Counterparts for the Marginal $4<\sigma<5$ 2\,mm sample}}
\tablewidth{700pt}
\tabletypesize{\scriptsize}
\tablehead{
\colhead{\sc Name} & 
\colhead{{\sc Position}} & 
\colhead{{\sc SNR}$_{\rm 2mm}$} & 
\colhead{S$_{\rm 2\,mm}$} & 
\colhead{$H$-band} & 
\colhead{S$_{\rm 850}$ } & 
\colhead{S$_{\rm 3GHz}$}\\
\colhead{} &
\colhead{ \hspace{1cm}($\alpha_{\rm J2000}$, $\delta_{\rm J2000}$)\hspace{1cm} } & 
\colhead{} & 
\colhead{[\uJy]} &
\colhead{[AB]} & 
\colhead{[mJy]} & 
\colhead{[\uJy]}
}{
\startdata
MORA-P03-0  & 10:00:42.629 $+$02:14:39.00 & 4.68 & 615$\pm$131 & ... & ... & ... \\
MORA-P03-1  & 10:00:42.843 $+$02:14:59.80 & 4.51 & 493$\pm$109 & ... & ... & ... \\
MORA-P03-2  & 10:00:43.643 $+$02:22:05.20 & 4.37 & 308$\pm$70  & ... & ... & ... \\
MORA-P03-3  & 10:00:42.415 $+$02:27:18.60 & 4.28 & 703$\pm$164 & ... & ... & ... \\
MORA-P03-4  & 10:00:46.019 $+$02:28:29.20 & 4.24 & 506$\pm$119 & ... & ... & ... \\
MORA-P03-5  & 10:00:42.629 $+$02:18:29.20 & 4.19 & 551$\pm$131 & ... & ... & ... \\
MORA-P03-6  & 10:00:42.362 $+$02:26:56.60 & 4.07 & 711$\pm$174 & ... & ... & ... \\
MORA-P03-7  & 10:00:45.685 $+$02:24:26.40 & 4.07 & 378$\pm$93  & ... & ... & ... \\
MORA-P03-8  & 10:00:45.818 $+$02:13:06.20 & 4.07 & 416$\pm$102 & ... & ... & ... \\
MORA-P03-9  & 10:00:44.511 $+$02:15:49.20 & 4.04 & 256$\pm$63  & ... & ... & ... \\
MORA-P03-10 & 10:00:42.936 $+$02:11:51.00 & 4.01 & 408$\pm$102 & ... & ... & ... \\
\hline
MORA-13  &  10:00:24.398 $+$02:21:45.00  & 4.78  &  419$\pm$88  & ... & ... & ... \\
MORA-14  &  10:00:26.119 $+$02:19:07.80  & 4.74  &  397$\pm$84  & ... & ... & ... \\
MORA-15  &  10:00:25.118 $+$02:15:37.20  & 4.68  &  398$\pm$85  & ... & ... & ... \\
MORA-16  &  10:00:25.505 $+$02:20:30.20  & 4.68  &  397$\pm$85  & ... & ... & ... \\
MORA-17  &  10:00:27.761 $+$02:24:43.40  & 4.61  &  406$\pm$88  & 26.74$\pm$0.24 & ... & ... \\
MORA-18  &  10:00:23.103 $+$02:12:27.40  & 4.59  &  401$\pm$88  & 26.19$\pm$0.14 & ... & ... \\
MORA-19  &  10:00:12.000 $+$02:23:09.80  & 4.58  &  283$\pm$62  & 25.00$\pm$0.16 & 4.00$\pm$0.57 & 259.0$\pm$2.9 \\
MORA-20  &  10:00:24.572 $+$02:32:03.00  & 4.58  &  404$\pm$88  & ... & ... & 30.3$\pm$3.9$^{*}$ \\
MORA-21  &  10:00:06.543 $+$02:21:31.60  & 4.54  &  414$\pm$91  & 26.81$\pm$0.50 & ... & ... \\
MORA-22  &  10:00:12.960 $+$02:32:31.00  & 4.54  &  304$\pm$67  & ... & ... & ... \\
MORA-23  &  10:00:12.575 $+$02:14:44.20  & 4.52  &  298$\pm$66  & 21.30$\pm$0.01 & 7.19$\pm$0.64 & ... \\
MORA-24  &  10:00:07.837 $+$02:21:05.20  & 4.52  &  656$\pm$145 & 24.88$\pm$0.08 & ... & ... \\
MORA-25  &  10:00:23.970 $+$02:17:49.80  & 4.52  &  398$\pm$88  & 18.94$\pm$0.01 & 10.48$\pm$0.55 & 96.8$\pm$2.8 \\
MORA-26  &  10:00:22.716 $+$02:25:31.80  & 4.51  &  397$\pm$88  & ... & ... & ... \\
MORA-27  &  10:00:22.636 $+$02:23:04.20  & 4.46  &  392$\pm$88  & ... & ... & ... \\
MORA-28  &  10:00:24.945 $+$02:24:05.60  & 4.46  &  385$\pm$86  & ... & ... & ... \\
MORA-29  &  10:00:23.650 $+$02:21:55.40  & 4.34  &  389$\pm$89  & 23.21$\pm$0.02 & 7.87$\pm$0.55 & 256$\pm$2.4  \\
MORA-30  &  10:00:14.563 $+$02:16:45.00  & 4.34  &  422$\pm$97  & ... & ... & ... \\
MORA-31  &  10:00:24.038 $+$02:29:48.20  & 4.32  &  383$\pm$89  & 26.14$\pm$0.16 & 6.12$\pm$0.62 & ... \\
MORA-32  &  10:00:20.434 $+$02:32:57.80  & 4.29  &  397$\pm$93  & ... & ... & ... \\
MORA-33  &  10:00:05.274 $+$02:23:58.40  & 4.29  &  529$\pm$123 & ... & ... & ... \\
MORA-34  &  10:00:27.748 $+$02:26:45.40  & 4.29  &  377$\pm$88  & ... & ... & ... \\
MORA-35  &  10:00:20.768 $+$02:13:16.20  & 4.28  &  382$\pm$89  & ... & ... & ... \\
MORA-36  &  10:00:27.280 $+$02:20:16.80  & 4.28  &  364$\pm$85  & ... & ... & ... \\
MORA-37  &  10:00:15.870 $+$02:24:46.00  & 4.27  &  424$\pm$99  & ... & 5.38$\pm$0.58 & ... \\
MORA-38  &  10:00:18.219 $+$02:18:51.60  & 4.27  &  311$\pm$73  & ... & ... & ... \\
MORA-39  &  10:00:25.438 $+$02:15:58.80  & 4.27  &  360$\pm$84  & ... & ... & ... \\
MORA-40  &  10:00:05.863 $+$02:16:08.40  & 4.26  &  416$\pm$98  & ... & ... & 6.4$\pm$1.0$^{*}$ \\
MORA-41  &  10:00:05.875 $+$02:26:16.20  & 4.25  &  414$\pm$97  & ... & ... & ... \\
MORA-42  &  10:00:22.783 $+$02:16:07.80  & 4.25  &  373$\pm$88  & ... & ... & ... \\
MORA-43  &  10:00:30.736 $+$02:15:47.00  & 4.25  & 1140$\pm$268 & ... & ... & ... \\
MORA-44  &  10:00:27.560 $+$02:21:58.20  & 4.25  &  368$\pm$87  & ... & ... & ... \\
MORA-45  &  10:00:15.603 $+$02:25:50.40  & 4.24  &  458$\pm$108 & ... & ... & ... \\
MORA-46  &  10:00:20.434 $+$02:28:03.00  & 4.23  &  389$\pm$92  & ... & ... & ... \\
MORA-47  &  10:00:15.964 $+$02:20:27.80  & 4.22  &  403$\pm$96  & ... & ... & ... \\
MORA-48  &  10:00:22.743 $+$02:32:41.00  & 4.21  &  373$\pm$89  & ... & ... & ... \\
MORA-49  &  10:00:30.097 $+$02:31:21.40  & 4.20  &  646$\pm$154 & ... & ... & ... \\
MORA-50  &  10:00:05.394 $+$02:29:44.20  & 4.20  &  487$\pm$116 & ... & ... & ... \\
MORA-51  &  10:00:25.333 $+$02:33:33.20  & 4.19  &  369$\pm$88  & ... & ... & ... \\
MORA-52  &  10:00:04.420 $+$02:28:00.80  & 4.19  & 1090$\pm$261 & ... & ... & ... \\
MORA-53  &  10:00:27.321 $+$02:31:40.60  & 4.19  &  358$\pm$86  & ... & 5.95$\pm$0.68 & ... \\
MORA-54  &  10:00:16.751 $+$02:23:59.40  & 4.18  &  313$\pm$75  & ... & ... & ... \\
MORA-55  &  10:00:10.679 $+$02:21:41.60  & 4.17  &  258$\pm$62  & ... & ... & ... \\
MORA-56  &  10:00:28.108 $+$02:27:10.40  & 4.15  &  377$\pm$91  & ... & ... & ... \\
MORA-57  &  10:00:07.184 $+$02:14:58.20  & 4.14  &  427$\pm$103 & 23.69$\pm$0.03 & ... & ... \\
MORA-58  &  10:00:06.823 $+$02:21:18.60  & 4.13  &  389$\pm$94  & ... & ... & ... \\
MORA-59  &  10:00:29.109 $+$02:29:10.80  & 4.13  &  427$\pm$103 & ... & ... & ... \\
MORA-60  &  10:00:05.637 $+$02:11:56.40  & 4.12  &  434$\pm$105 & ... & ... & ... \\
MORA-61  &  10:00:20.114 $+$02:18:01.40  & 4.12  &  375$\pm$91  & ... & ... & ... \\
MORA-62  &  10:00:30.684 $+$02:31:26.00  & 4.12  & 1030$\pm$251 & 23.21$\pm$0.01 & ... & ... \\
MORA-63  &  10:00:27.614 $+$02:22:32.40  & 4.11  &  358$\pm$87  & ... & ... & ... \\
MORA-64  &  10:00:23.170 $+$02:28:51.40  & 4.11  &  364$\pm$89  & ... & ... & ... \\
MORA-65  &  10:00:13.762 $+$02:14:00.60  & 4.10  &  320$\pm$78  & ... & 2.51$\pm$0.66 & ... \\
MORA-66  &  10:00:28.214 $+$02:19:53.20  & 4.09  &  380$\pm$93  & ... & 3.10$\pm$0.53 & ... \\
MORA-67  &  10:00:28.923 $+$02:32:18.60  & 4.09  &  407$\pm$99  & 23.38$\pm$0.02 & ... & ... \\
MORA-68  &  10:00:16.978 $+$02:28:13.00  & 4.09  &  290$\pm$71  & ... & ... & ... \\
MORA-69  &  10:00:25.332 $+$02:29:31.00  & 4.09  &  352$\pm$86  & ... & ... & ... \\
MORA-70  &  10:00:27.867 $+$02:19:51.80  & 4.08  &  366$\pm$90  & ... & 3.10$\pm$0.53 & ... \\
MORA-71  &  10:00:28.175 $+$02:31:48.20  & 4.08  &  372$\pm$91  & ... & ... & ... \\
MORA-72  &  10:00:07.396 $+$02:23:44.60  & 4.08  &  460$\pm$113 & ... & ... & ... \\
MORA-73  &  10:00:12.934 $+$02:23:45.60  & 4.08  &  275$\pm$68  & ... & 2.64$\pm$0.57 & 93.7$\pm$2.6 \\
MORA-74  &  10:00:14.002 $+$02:23:33.40  & 4.08  &  336$\pm$82  & ... & ... & ... \\
MORA-75  &  10:00:06.104 $+$02:10:50.40  & 4.07  &  754$\pm$185 & ... & ... & ... \\
MORA-76  &  10:00:28.963 $+$02:30:13.60  & 4.07  &  408$\pm$100 & ... & ... & ... \\
MORA-77  &  10:00:05.410 $+$02:12:16.40  & 4.06  &  470$\pm$116 & 19.15$\pm$0.01 & ... & 112.0$\pm$2.8 \\
MORA-78  &  10:00:22.997 $+$02:28:09.40  & 4.06  &  359$\pm$88  & ... & ... & ... \\
MORA-79  &  10:00:09.370 $+$02:32:14.40  & 4.06  &  443$\pm$109 & ... & ... & ... \\
MORA-80  &  10:00:04.500 $+$02:28:58.60  & 4.06  &  963$\pm$237 & ... & ... & 25.7$\pm$3.3$^{*}$ \\
MORA-81  &  10:00:26.492 $+$02:15:27.00  & 4.05  &  338$\pm$83  & ... & 5.90$\pm$0.60 & ... \\
MORA-82  &  10:00:23.170 $+$02:17:38.60  & 4.05  &  357$\pm$88  & ... & ... & ... \\
MORA-83  &  10:00:22.410 $+$02:30:26.20  & 4.04  &  354$\pm$88  & 22.72$\pm$0.01 & 3.88$\pm$0.64 & 42.0$\pm$3.5 \\
MORA-84  &  10:00:08.983 $+$02:32:08.00  & 4.04  &  617$\pm$152 & ... & ... & ... \\
MORA-85  &  10:00:21.902 $+$02:22:46.00  & 4.04  &  355$\pm$88  & ... & ... & ... \\
MORA-86  &  10:00:05.982 $+$02:25:12.40  & 4.04  &  384$\pm$95  & ... & ... & ... \\
MORA-87  &  10:00:07.368 $+$02:33:58.00  & 4.03  &  655$\pm$162 & 26.73$\pm$0.48 & ... & ... \\
MORA-88  &  10:00:30.122 $+$02:20:38.60  & 4.03  &  637$\pm$158 & ... & ... & 57.1$\pm$0.9$^{*}$ \\
MORA-89  &  10:00:29.629 $+$02:24:10.20  & 4.03  &  490$\pm$122 & ... & ... & ... \\
MORA-90  &  10:00:20.341 $+$02:16:49.60  & 4.03  &  366$\pm$91  & ... & ... & ... \\
MORA-91  &  10:00:08.129 $+$02:33:01.60  & 4.03  &  728$\pm$181 & ... & ... & ... \\
MORA-92  &  10:00:29.309 $+$02:25:54.20  & 4.02  &  438$\pm$109 & ... & ... & ... \\
MORA-93  &  10:00:26.773 $+$02:20:44.00  & 4.02  &  336$\pm$84  & ... & ... & ... \\
MORA-94  &  10:00:07.571 $+$02:16:31.40  & 4.01  &  492$\pm$122 & ... & ... & ... \\
MORA-95  &  10:00:25.745 $+$02:18:23.00  & 4.01  &  338$\pm$84  & ... & ... & ... \\
MORA-96  &  10:00:14.482 $+$02:28:48.40  & 4.01  &  385$\pm$96  & 20.81$\pm$0.01 & ... & ... \\
MORA-97  &  10:00:06.729 $+$02:25:36.00  & 4.01  &  373$\pm$93  & ... & ... & ... \\
MORA-98  &  10:00:09.412 $+$02:22:02.20  & 4.00  &  419$\pm$105 & ... & ... & ... \\
MORA-99  &  10:00:19.660 $+$02:31:04.60  & 4.00  &  360$\pm$90  & 23.17$\pm$0.02 & ... & ... \\
 \enddata
}
 \end{deluxetable*}
 
\hspace{-\columnwidth}
\begin{minipage}{\textwidth}
{\bf Table Notes.} Marginal 2\,mm sources and their multiwavelength
counterparts, identified in the COSMOS2020 photometric catalog within
1$''$ of the 2\,mm position (with $H$-band magnitude from UltraVISTA;
Weaver \etal\ submitted), at 850\um\ with SCUBA-2 within 8$''$ of the
2\,mm position \citep{simpson19a}, or at 3\,GHz within 1$''$ of the
2\,mm position \citep{smolcic17a}.  Sources marked with asterisks
($^{*}$) have 3\,GHz flux density measurements from the deeper
COSMOS-XS survey \citep{van-der-vlugt21a,algera20a}.
\end{minipage}

\label{tab:marginal}

\end{document}